\documentclass[a4paper,12pt, nofootinbib,notitlepage,preprintnumbers,floatfix]{revtex4-1}

\usepackage{amsmath}           
\usepackage{amssymb}           
\usepackage{xcolor}             
\usepackage{bbm}               
\usepackage{braket}            
\usepackage{xspace}            
\usepackage{mathtools}         
\usepackage{graphicx}          
\usepackage{enumerate}         
\usepackage{hyperref}          
\usepackage{tabularx}  		   
\usepackage{multirow}		   
\usepackage{enumitem}

\usepackage{etoolbox}
\makeatletter
\patchcmd{\subsubsection}{\itshape}{\itshape\bfseries}{}{}
\makeatother

\renewcommand{\k}{\mathbf{k}}
\newcommand{\p}{\mathbf{p}}
\newcommand{\q}{\mathbf{q}}

\newcommand{\2}{\mathbf{2}}
\newcommand{\3}{\mathbf{3}}

\newcommand{\Bc}{\mathcal{B}}
\newcommand{\Cc}{\mathcal{C}}
\newcommand{\Dc}{\mathcal{D}}
\newcommand{\Fc}{\mathcal{F}}
\newcommand{\Gc}{\mathcal{G}}

\newcommand{\Kc}{\mathcal{K}}
\newcommand{\Lc}{\mathcal{L}}
\newcommand{\Mc}{\mathcal{M}}

\newcommand{\Oc}{\mathcal{O}}
\newcommand{\Pc}{\mathcal{P}}
\newcommand{\Qc}{\mathcal{Q}}
\newcommand{\Rc}{\mathcal{R}}
\newcommand{\Tc}{\mathcal{T}}
\newcommand{\hMc}{\wh{\Mc}}
\newcommand{\hFc}{\widehat{\Fc}}
\newcommand{\hKc}{\widehat{\Kc}}
\newcommand{\hTc}{\widehat{\Tc}}
\newcommand{\hLc}{\wh{\Lc}}
\newcommand{\hRc}{\wh{\Rc}}

\newcommand{\wt}[1]{\widetilde{ #1 }}
\newcommand{\wh}[1]{\widehat{ #1 }}
\newcommand{\bh}[1]{\mathbf{\hat{ #1 }}}

\newcommand{\cf}{cf.\xspace}
\newcommand{\eg}{e.g.\xspace}
\newcommand{\ie}{i.e.\xspace}

\newcommand{\nn}{\nonumber}
\newcommand{\diff}{\textrm{d}}
\newcommand{\df}{\textrm{df}}

\newcolumntype{Y}{>{\centering\arraybackslash}X}
\newcolumntype{s}{>{\centering\arraybackslash\hsize=.2\hsize}X}

\definecolor{wm_green}{HTML}{115740}
\definecolor{wm_gold}{HTML}{B9975B}

\usepackage{mfirstuc} 
\newcommand{\addReviewer}[2]{
  \expandafter\newcommand\csname #1\endcsname[1]{{\sf \color{#2} {#1}:\,##1}}
  \expandafter\newcommand\csname #1cor\endcsname[2]{{\color{#2} {#1}:\,\st{##1}{\sf ##2}}}
  \expandafter\newcommand\csname #1color\endcsname{#2}
}

\usepackage{soul,color}
\definecolor{chromeyellow}{rgb}{1.0, 0.65, 0.0}
\definecolor{DodgeBlue}{rgb}{0.118, 0.565,1.000}
\definecolor{asparagus}{rgb}{0.53, 0.66, 0.42}
\definecolor{cadmiumgreen}{rgb}{0.0, 0.42, 0.24}

\definecolor{jlab_red}{RGB}{192,39,45}
\definecolor{jlab_orange}{RGB}{249,102,0}
\definecolor{jlab_blue}{RGB}{47,122,121}
\definecolor{jlab_green}{RGB}{65,125,10}

\addReviewer{cc}{jlab_green}
\addReviewer{rb}{jlab_blue}
\addReviewer{aj}{jlab_orange}

\hypersetup{
pdftitle     = {Partial-wave projection of the one-particle exchange in three-body scattering amplitudes},
pdfsubject   = {Scattering Theory},
pdfkeywords  = {Scattering, Partial Waves, QCD, Hadron, Physics, Lattice},
pdfauthor    = {Andrew W. Jackura},
pdfnewwindow = {true},
colorlinks   = {true},
linkcolor    = {blue},
citecolor    = {blue},
filecolor    = {blue},
urlcolor     = {blue}
} 

\newcommand{\wm}{Department of Physics, 
William \& Mary, 
Williamsburg, VA 23187, USA}
\newcommand{\ucb}{Department of Physics, 
University of California, 
Berkeley, CA 94720, USA}   
\newcommand{\lbnl}{Nuclear Science Division, 
Lawrence Berkeley National Laboratory, Berkeley, 
CA 94720, USA}

\newcommand{\jlab}{Thomas Jefferson National Accelerator Facility, 12000 Jefferson Avenue, Newport News, Virginia 23606, USA}

\begin{document}
\preprint{JLAB-THY-24-4199}

\title{
Partial-wave projection of relativistic three-body amplitudes
}


\author{Ra\'ul A. Brice\~no}
\email[e-mail: ]{rbriceno@berkeley.edu}
\affiliation{\ucb}
\affiliation{\lbnl}

\author{Caroline S. R. Costa}
\email[e-mail: ]{costa@jlab.org; costa@lbl.gov}
\affiliation{\jlab}
\affiliation{\lbnl}

\author{Andrew W. Jackura}
\email[e-mail: ]{awjackura@wm.edu}
\affiliation{\wm}
\begin{abstract}

We derive the integral equations for partial-wave projected three-body scattering amplitudes, starting from the integral equations for three-body amplitudes developed for lattice QCD analyses. The results, which hold for generic three-body systems of spinless particles, build upon the recently derived partial-wave projected one-particle exchange, a primary component of the relativistic framework proven to satisfy $S$ matrix unitarity. We derive simplified expressions for factorizable short-distance interactions, $\Kc_{3}$, in two equivalent formalisms — one symmetric under particle interchange and one asymmetric. For the asymmetric case, we offer parameterizations useful for amplitude analysis. Finally, we examine toy models for $3\pi$ systems at unphysically heavy pion masses with total isospins $0,1,$ and $2$.

\end{abstract}
\date{\today}
\maketitle

\section{Introduction}
\label{sec:intro}

Presently, there is a community-wide effort to have faithful representations for three-hadron scattering amplitudes. This is being driven by three major thrust areas in nuclear and hadronic physics: {\bf hadron spectroscopy}, {\bf nuclear structure}, and {\bf fundamental symmetries}. Before discussing the details of this program, we briefly discuss some of the needs for three-body scattering amplitudes for the subfields above. Most states in the hadron spectrum are unstable resonances whose existence can only be reconstructed by studying the analytical properties of the amplitudes of its byproducts, which normally involve two and/or more multi-hadron states. Notable examples include the recently observed $T_{cc}$~\cite{LHCb:2021vvq}, the $X(2370)$, which was recently hypothesized to be a glueball candidate by the BESIII collaboration~\cite{BESIII:2023wfi}, and the spin-exotic $\pi_1$ resonance being searched for at the GlueX experiment~\cite{Afzal:2024ulu}. To have an accurate determination of the nuclear spectrum and its response to electroweak probes it is critical to have a robust determination of three-nucleon dynamics~\cite{Hergert:2020bxy}, which would preferably be constrained directly from quantum chromodynamics (QCD). Finally, several electroweak heavy-meson decays to multi-meson states may present signals for possible physics beyond the standard model (BSM). These include recent large CP violations observed in the LHCb experiment~\cite{LHCb:2019xmb,LHCb:2022fpg,LHCb:2014mir,LHCb:2019jta,LHCb:2013lcl} in rare heavy-meson decays to three light hadrons ($\pi$'s and $K$'s). Currently, it is not understood if these are evidence of new sources of CP violation or dynamical enhancement due to final state interactions~\cite{Suzuki:1999uc,Wolfenstein:1990ks,Suzuki:2007je,AlvarengaNogueira:2015wpj,Bediaga:2013ela,Garrote:2022uub}. In other words, the inability to have precise and accurate determinations of the QCD contributions of such decays limits our ability to claim evidence for BSM physics confidently.     

The current effort towards determining three-hadron scattering amplitudes has two parallel tracks. 
One track aims to have representations of scattering amplitudes that satisfy the principles of the S matrix, such as enforcing unitarity and including as much of the correct analytic structure as possible~\cite{Jackura:2018xnx, Hansen:2015zga, Jackura:2020bsk, Jackura:2023qtp,Jackura:2022gib,Mai:2017vot,Jackura:2018xnx,Mikhasenko:2019vhk,Dawid:2020uhn, Feng:2024wyg}.
Such a representation can be achieved up to a class of unknown functions, which can at least be proved to be real for physical energies. Different practitioners give these functions different names, here we will refer to these as K matrices as is common in hadron spectroscopy. 
In parallel, there is an ambitious program towards constraining these K matrices directly from QCD using lattice QCD~\cite{Detmold:2008fn,Beane:2007es,Culver:2019vvu,Alexandru:2020xqf, Hansen:2020otl,Draper:2023boj}. 
Scattering observables can not be determined directly from lattice QCD. As a result, this program relies on the derivation of non-perturbative relations between the lattice QCD observables and the physical K matrices~\cite{Polejaeva:2012ut,Hansen:2014eka,Hansen:2015zga,Briceno:2017tce,Briceno:2018mlh,Briceno:2018aml,Briceno:2019muc,Blanton:2019igq,Hansen:2020zhy,Blanton:2020gha,Hammer:2017uqm,Hammer:2017kms,Meng:2017jgx,Pang:2019dfe,Muller:2021uur,Blanton:2020gmf,Blanton:2021mih,Hansen:2020zhy}~\footnote{These works build from an even large literature that has been dedicated to develop such formalism for two-body systems~\cite{Luscher:1985dn,Luscher:1986n2,Luscher:1990ux,Rummukainen:1995vs, Kim:2005gf,He:2005ey,Hansen:2012tf,Briceno:2012yi,Briceno:2013lba, Briceno:2014oea} as well as performing state of the art lattice QCD calculations ~\cite{Dudek:2010ew,Pelissier:2012pi,Dudek:2012xn,Liu:2012zya,Wilson:2014cna,Dudek:2014qha,Lang:2015hza,Wilson:2015dqa,Dudek:2016cru,Briceno:2016mjc,Moir:2016srx,Bulava:2016mks,Hu:2016shf,Alexandrou:2017mpi,Bali:2017pdv,Wagman:2017tmp,Andersen:2017una,Briceno:2017qmb,Woss:2018irj,Brett:2018jqw,Mai:2019pqr,Woss:2019hse,Wilson:2019wfr,Cheung:2020mql,Rendon:2020rtw,Woss:2020ayi,Horz:2020zvv}. We point the reader to recent reviews on this topic~\cite{Briceno:2017max, Hansen:2019nir}.}

In this work, we start from the relativistic integral equations presented in Refs.~\cite{Hansen:2015zga,Hansen:2020zhy, Jackura:2022gib} and the partial-wave projection of the one-particle exchange diagram presented in Ref.~\cite{Jackura:2023qtp}, to derive a set of non-perturbative integral equations that relativistic three-particle scattering amplitudes with definite parity and arbitrary angular momentum must satisfy. In Sec.~\ref{sec:PWA}, we do this for two equivalent types of formalism, distinguished by whether the K matrix is asymmetric~\cite{Blanton:2020gha,Jackura:2022gib} or symmetric~\cite{Hansen:2015zga} under the interchange of the particles. Details of the derivation are presented in  Appendix~\ref{app:conv}. In Sec.~\ref{sec:isospin} we provide a prescription for any system composed of three spinless particles, including any number of coupled channels as well as the particles having flavor isospin symmetry.  

In Sec.~\ref{sec:separable}, we consider the consequence of parameterizations of the K matrix that are factorizable in terms of the initial/final kinematic variables, and we derive simplified expressions for the integral equations presented in Sec.~\ref{sec:PWA}. In Sec.~\ref{sec:LSZ}, we provide further implications for theories that include two-body bound states. In particular, we show how the Lehmann–Symanzik–Zimmermann (LSZ) formalism can be used to provide two-particle scattering amplitudes with definite parity and arbitrary angular momentum, where one of the particles in the initial and final states is a bound state. In Sec.~\ref{sec:numerical}, we consider the numerical solutions of these integral equations for toy models of $3\pi$ system with total isospin $2,1$ and $0$, which have non-zero angular momentum and involve multiple open channels. After providing a numerical prescription, which is a simple generalization from the prescriptions presented in Refs.~\cite{Jackura:2020bsk, Dawid:2023jrj} for $J=0$ angular momentum, we find that for all models considered the numerical solutions satisfy unitarity below the three-particle threshold, as expected~\cite{Briceno:2024txg,Briceno:2019muc}.

In addition to Appendix~\ref{app:conv}, which discusses details needed for performing the partial wave projection of integral equations, Appendix~\ref{app:shifting} discusses freedom in the definition of the two-body phase space appearing in the integral equations for three-body scattering amplitudes. Reference~\cite{Romero-Lopez:2019qrt} showed that this freedom can be used to generalize previously existing formalism for relating finite-volume spectra of three-particle systems to K matrices~\cite{Hansen:2014eka} to accommodate the presence of two-body bound states and/or resonances. Although such formalism will not be used in this work, in Appendix~\ref{app:shifting}, we explain how the shifts in the phase space can be incorporated in the partial-wave projected integral equations.

\section{Partial-wave projected amplitudes}
\label{sec:PWA}

We consider the elastic scattering of three spinless particles, which we denote as $\varphi$, with no internal quantum numbers. We make a partial generalization to coupled spinless systems, e.g., including flavor isospin, in Sec.~\ref{sec:isospin}. We kinematically describe the reaction as one involving \emph{pairs}, where two of the particles form a dimer of relative angular momentum $\ell$, recoiling against \emph{spectators} with some momentum. With this description, the relative momentum between the pair constituents is removed, and the system is described by the magnitudes~\footnote{The system also depends on the orientations of the initial and final spectator momenta. Since our discussion is concentrated on partial wave amplitudes, we ignore this dependence as it is eventually removed.} of the initial and final spectator momenta $k$ and $p$, respectively, as well as the three-body total center-of-momentum (CM) frame energy, $\sqrt{s}$. Note that $k$ and $p$ are also defined in the three-body total CM frame. The effective reaction is denoted
\begin{align}
    \varphi_k + [\varphi_{k_1} \varphi_{k_2}]_{\ell} \to \varphi_{p} + [\varphi_{p_1} \varphi_{p_2}]_{\ell'} \, .
\end{align} 
The momentum $k$ serves to label the initial spectator while the pair $[\varphi_{k_1}\varphi_{k_2}]_\ell$ has additional labels $k_1$ and $k_2$ to indicate the first and second particle of the pair, respectively. A similar notation holds for the final state. We take the mass of the initial and final state particles to be $m_k$, $m_{k_1}$, $m_{k_2}$, and $m_p$, $m_{p_1}$, $m_{p_2}$, respectively. 

Our goal is to construct and study the amplitude for this reaction projected to definite $J^P$, where $J$ is the total angular momentum of the system and $P$ is its parity. We work in the spin-orbit basis of coupled angular momenta, thus for a given energy $\sqrt{s}$ and spin-parity $J^P$ the amplitude is a matrix in the orbital angular momentum $L$ between the spectator and dimer pair and the intrinsic spin $S$ of the pair. Since we restrict our attention to only spinless particles, we have the trivial identity that $S = \ell$ and $S' = \ell'$. We denote matrix elements of this partial wave projected amplitude as $\Mc_{3; L'S',LS}^{J^P}(p,k)$, leaving the dependence on $s$ implicit throughout this work.

The partial wave projection of this amplitude and its integral equations~\cite{Hansen:2015zga, Jackura:2022gib} have been discussed in the simplest case $J = S = S' = 0$ ~\cite{Hansen:2020otl, Jackura:2020bsk}. Here we show that in general, the partial wave projected $\Mc_3^{J^P}$ can be decomposed into two terms~\cite{Hansen:2015zga,Blanton:2020gha,Jackura:2022gib},
\begin{align}
	\label{eq:M3_def}
	\Mc_3^{J^P}(p,k) = \Dc^{J^P}(p,k) + \Mc_{3,\df}^{J^P}(p,k) \, ,
\end{align}
where $\Dc$ contains three-body physics driven by long-range exchanges between pair-wise interactions and $\Mc_{3,\df}$ is driven by short-distance three-body interactions, \ie, the three-body K matrix.~\footnote{There is a technical detail we omit in this discussion regarding summing over all possible spectators. We comment on this technicality in Sec.~\ref{sec:PWA:asymm}.}  Each of these objects is a matrix in angular momentum space of the given quantum number $J^P$, \ie,
\begin{align*}
	\left[\Mc^{J^P}_3(p,k)\right]_{L'S',LS}  = \Mc^{J^P}_{3;L'S',LS} (p,k) \, .
\end{align*}
For example, consider low-energy $3\pi$ scattering in $J^P = 1^+$ isotensor channel. The dominant low lying waves are associated with $S = 1$, corresponding to the dipion pair being in a resonant $P$ wave state, \ie the $\rho$ resonance. For these quantum numbers, the orbital angular momentum between the spectator pion and dipion pair can be $S$ or $D$ wave, thus the $J^P = 1^+$ amplitude is a $2\times 2$ matrix in $LS$ space. We will revisit the $3\pi$ case in detail in Sec.~\ref{sec:numerical}.

We present the details of the partial wave integral equations for both $\Dc^{J^P}$ and $\Mc_{3,\df}^{J^P}$ in the subsequent subsections. There are two main classes of representation for $\Mc_{3,\df}$, which emerge from whether the three-body K matrix, $\Kc_3$, has or has not been symmeterized over all possible spectators, see for example Refs.~\cite{Hansen:2014eka, Hansen:2015zga, Blanton:2020gha, Jackura:2022gib}. We call these representations the \emph{symmetric} and \emph{asymmetric} representations. These two representations lead to the same physical amplitude, but the paths toward performing analyses differ due to a choice in parameterization of $\Kc_3$. After presenting the partial wave formalism for both representations, we discuss their differences in Sec.~\ref{sec:separable} and present general procedures for data analysis. We comment on generalizations to other systems, \eg, systems with flavor isospin, in Sec.~\ref{sec:isospin}.

\subsection{Partial-wave projection for $\Dc$}
\label{sec:DcPWA}

We first present the amplitude $\Dc^{J^P}$ as it is required for both symmetric and asymmetric representations of $\Mc_{3,\df}$. To project $\Dc$ to the $LS$ basis, we follow the steps presented in Ref.~\cite{Jackura:2023qtp}. We start with the helicity-basis definition of the $\Dc$ amplitude, colloquially called the \emph{ladder} amplitude, which functionally looks identical to what was presented in Ref.~\cite{Hansen:2015zga},~\footnote{Reference~\cite{Hansen:2015zga} quantizes the pair angular momentum along some fixed $z$ axis, while Ref.~\cite{Jackura:2023qtp} discussed the advantage of using helicity quantization for constructing the partial wave projection.}
\begin{align}
\label{eq:Dc_hel}
\mathcal{D}_{\ell'\lambda',\ell\lambda}(\p,\k)  &= -   \Mc_{2,\ell'}(\sigma_p)\,\mathcal{G}_{\ell'\lambda',\ell\lambda}(\p,\k)\,\Mc_{2,\ell}(\sigma_k) 
\nn \\[5pt]
    & \qquad - \Mc_{2,\ell'}(\sigma_p)
    \sum_{\ell'',\lambda''}\int\! \frac{\diff^3 \k'}{(2\pi)^{3}\,2\omega_{k'}}\mathcal{G}_{\ell'\lambda',\ell''\lambda''}(\p,\k')\,\mathcal{D}_{\ell''\lambda'',\ell\lambda}(\k',\k) \, .
\end{align}
Here $\k$ and $\p$ are the initial and final spectator momenta, while $\lambda$ and $\lambda'$ denote the helicity of the initial and final state pair, respectively. The scattering amplitude for the pair, $\Mc_{2,\ell}$, is a diagonal matrix in the $\ell$-space and depends on the squared invariant mass of the pair, $\sigma_k$. In the total three particle CM frame, $\sigma_k$ is related to the spectator momentum via  $\sigma_k = (\sqrt{s} - \omega_k)^2-k^2$, where $\omega_k = \sqrt{k^2+m_k^2}$ and $k = \lvert\k\rvert$.

The remaining quantity to define is the one-particle exchange (OPE) propagator, $\mathcal{G}$. The partial wave projection of this object was the main focus of Ref.~\cite{Jackura:2023qtp}, and we will only review the points necessary from that work. In the helicity basis, the $\mathcal{G}$ propagator for a spinless particle with mass $m_e$ can be written as
\begin{align}
	\label{eq:ope}
	 \Gc_{\ell'\lambda',\ell\lambda}(\p,\k) = 
  \left(\frac{k_p^{\star}}{q_p^{\star}}\right)^{\ell'} 
  \frac{4\pi\,H(p,k)  \,  Y_{\ell'\lambda'}^{*}(\bh{\k}_p^{\star}) \, Y_{\ell\lambda}(\bh{\p}_k^{\star}) }{(\sqrt{s}-\omega_p-\omega_k)^2 - (\p + \k)^2 - m_e^2 + i\epsilon} \left(\frac{p_k^{\star}}{q_k^{\star}}\right)^{\ell} ,
 \end{align}
 where $H$ is a generic cut-off function, that must be equal to unity in the physical region. The coordinate system is described in detail in Ref.~\cite{Jackura:2023qtp}. Three key points relevant for our discussion are the following: (\emph{i}) The vector $\p_k^{\star}$ is equal to the value of $\p$ after boosting it to the CM frame of the pair labeled by $k$. (\emph{ii}) $q_k^{\star}$ is the magnitude of $\p_k^{\star}$ when the exchange particle is placed on shell. Similar definitions hold for $\k_p^\star$ and $q_p^\star$. (\emph{iii}) Finally, the spherical harmonics are defined to have the $z$-axis aligned along the direction of the momentum of the two-particle pair labeled by the subscript of the argument. 
 
The main result of Ref.~\cite{Jackura:2023qtp} shows that the partial wave projection of the OPE to definite $J^P$ takes the form
\begin{align}
	\Gc_{L'S',LS}^{J^P}(p,k) = H(p,k) \left[ \Kc_{\Gc;L'S',LS}^{J^P}(p,k) + \Cc_{L'S',LS}^{J^P}(p,k)\,Q_0(\zeta_{pk})\right] \, ,
    \label{eq:G:OPE}
\end{align}
where $Q_0(z)$ is the zeroth-degree Legendre function of the second kind, and  $\zeta_{pk}$, $\Kc_\Gc^{J^P}$ and $\Cc^{J^P}$~\footnote{In Ref.~\cite{Jackura:2023qtp} the function $\Cc^{J^P}$ was called $\Tc^{J^P}$. We change the notation to avoid confusion with Eq.~\eqref{eq:M3df:jp_asym}.} are known kinematic functions given in Ref.~\cite{Jackura:2023qtp}. The $\zeta_{pk}$ is the same for all partial waves, while the other two must be generated for each specific channel. The $\Kc_\Gc$ and $\Cc$ functions have been tabulated for low-lying spins in Ref.~\cite{Jackura:2023qtp}.

Following the steps in Ref.~\cite{Jackura:2023qtp}, which are outlined in App.~\ref{app:conv}, one can show that the partial-wave projected $\Dc^{J^P}$ must satisfy the integral equation, 
\begin{align}
	\label{eq:ladder_eq}
    \Dc^{J^P}(p,k) = \Dc_{0}^{J^P}(p,k) - \Mc_{2}(\sigma_p)\,\cdot\int_{k'}\!   \,  \Gc ^{J^P}(p,k')\cdot \Dc^{J^P}(k',k) \, ,
\end{align}
where we introduce the notation that the product $A^{J^P}\cdot B^{J^P}$ has the $LS$ space matrix element
\begin{align}
\label{eq:mat_prod}
   [A^{J^P}\cdot B^{J^P}]_{L'S',LS} \equiv \sum_{L'',S''} A^{J^P}_{L'S',L''S''} B^{J^P}_{L''S'', L S} \, .
\end{align}
In $LS$ space, the matrix elment for the $\2\to\2$ amplitude is $[\Mc_2]_{L'S',LS} = \delta_{L'L} \delta_{S'S} \, \Mc_{2,S}$. The driving term $\Dc_0^{J^P}$ for the ladder equation is given by the OPE amplitude
\begin{align}
    \left[\Dc_{0}^{J^P}(p,k)\right]_{L'S',LS} = -\Mc_{2,S'}(\sigma_p) \, \Gc_{L'S',LS}^{J^P}(p,k) \, \Mc_{2,S}(\sigma_k) \, .
\end{align}
Also, we have introduced the compact notation %
\begin{align}
	\int_k \equiv \int_0^\infty \! \diff k \, \frac{k^2}{(2\pi)^2 \, \omega_k} \, ,
 \label{eq:radial_int}
\end{align}
for the integral and measure,  which will be used throughout this work.

Although the integral over $k'$ runs to infinity, the $H$ cut-off function ensures that the argument has only finite support. Given a target $J^P$ and two-body scattering amplitudes, Eq.~\eqref{eq:ladder_eq} represents a set of coupled integral equations in the partial waves. In Sec.~\ref{sec:numerical} we show how these can be numerically solved along with examples from $3\pi$ scattering. For numerical applications, it is convenient to introduce an amputated ladder amplitude, $d^{J^P}$, which removes the singularities associated with $\2\to\2$ sub-processes in the initial and final states. This is defined by
\begin{equation}
    \label{def:D:ladder}
    \Dc^{J^P}(p,k) \equiv \Mc_{2}(\sigma_p) \cdot  d^{J^P}(p,k) \cdot \Mc_{2}(\sigma_k) \,.
\end{equation}
Using this definition and Eq.~\eqref{eq:ladder_eq}, it is straightforward to write an integral equation that $d^{J^P}$ must satisfy, which we explicitly give in Sec.~\ref{sec:numerical}.


\subsection{Partial-wave projected for $\Mc_{\df}$ with asymmetric $\Kc_3$}
\label{sec:PWA:asymm}

Once $\Dc^{J^P}$ is known, we reconstruct $\Mc_{3,\df}^{J^P}$ via a second set of equations~\cite{Hansen:2015zga, Jackura:2022gib}. One aspect we have neglected thus far is the choice of spectators in the amplitudes of Eq.~\eqref{eq:M3_def}. In principle, the full scattering amplitude must include the sum over all spectator-pair combinations. In the literature, \eg Refs.~\cite{Hansen:2014eka, Hansen:2015zga, Blanton:2020gha, Jackura:2022gib}, one distinguishes a specific spectator-pair amplitude as $\Mc_3^{(u,u)}$, where the superscript $(u,u)$ references the ``un-symmetrized" nature of the initial and final states since a particle has been chosen to be the ``spectators'' in both states. Equation~\eqref{eq:M3_def} is technically that of the $(u,u)$ amplitude projected to definite $J^P$. In this work, we will not consider the symmeterization of partial wave projected amplitudes. As a result, to improve readability, we will drop the superscript on the amplitudes. 

For the intermediate amplitude $\Mc_{3,\df}^{(u,u)}$, one has a choice to construct it from a symmetric K matrix $\Kc_3$, as originally presented in Ref.~\cite{Hansen:2014eka,Hansen:2015zga}, or that of an asymmetric K matrix as presented in Refs.~\cite{Blanton:2020gha,Jackura:2022gib}. In order to distinguish the intermediate amplitudes between these two constructions, we will label the amplitude associated with the \emph{symmetric} K matrix as $\Mc_{3,\df}^{(u,u)}$, while the one coming from an \emph{asymmetric} K matrix as $\hMc_{3,\df}^{(u,u)}$. It is important to emphasize that these two intermediate amplitudes, $\Mc_{3,\df}^{(u,u)}$ and $\hMc_{3,\df}^{(u,u)}$, are closely related. Which is chosen in the analysis is a matter of choice, and after symmeterization they are identical~\cite{Blanton:2020gha,Jackura:2022gib}. In this section we consider the integral equations for the amplitudes with asymmetric K matrix~\cite{Blanton:2020gha,Jackura:2022gib}, that is $\hMc_{3,\df}^{(u,u)}$.

As mentioned, we drop the $(u,u)$ superscript since we only consider the partial wave projection for a given spectator-pair amplitude. Therefore, we make the following replacements: $\Mc_{3,\df}^{(u,u)}\to \Mc_{3,\df}$ and $\hMc_{3,\df}^{(u,u)}\to \hMc_{3,\df}$. Similarly, to distinguish the symmetric $\Kc_3$ from its asymmetric counterpart, we will label the latter as  $\hKc_3$.  This notation will be used to distinguish the various intermediate functions that will appear in defining $\hMc_{3,\df}$ and closely related observables. 

Having discussed the notation that will be used, we can proceed to define $\hMc_{3,\df}$. Again, we leave the definition of this in the helicity basis to App.~\ref{app:asym_hel}. Following the steps shown in the appendix one arrives at
\begin{align}
	\label{eq:M3df:jp_asym}
    \hMc_{3,\df}^{J^P}(p,k) = \int_{p'} \int_{k'} \, \hLc^{J^P}(p,p') \cdot \hTc^{J^P}(p',k') \cdot \hRc^{J^P}(k',k) \, ,
\end{align}
which is a matrix in $LS$ space where $\hRc^{J^P}$ and $\hLc^{J^P}$ are initial and final state rescattering functions, 
\begin{align}
	\label{eq:Lcap:Jp_asym}
    \left[\hLc^{J^P}(p,k)\right]_{L'S',LS} & = \left[ \, 1 -  \Mc_{2,S'}(\sigma_p) \, \wt{\rho}(\sigma_p) \, \right] \delta_{L'L}\delta_{S'S} \, \frac{(2\pi)^2\omega_k}{k^2} \, \delta(p-k) \nn \\[5pt]
    & \qquad - \Mc_{2,S'}(\sigma_p) \Gc_{L'S',LS}^{J^P}(p,k) - \Dc_{L'S',LS}^{J^P}(p,k) \, \wt{\rho}(\sigma_k) \nn \\[5pt]
    & \qquad \qquad - \sum_{L'',S''}\int_{k'} \Dc_{L'S',L''S''}^{J^P}(p,k') \,  \Gc_{L''S'',LS}^{J^P}(k',k) \, , 
\end{align}
\begin{align}
  \label{eq:Rcap:Jp_asym}
    \left[ \hRc^{J^P} (p,k)\right]_{L'S',LS} & = \left[ 1 - \wt{\rho}(\sigma_k) \, \Mc_{2,S'}(\sigma_k) \right] \delta_{L'L}\delta_{S'S} \, \frac{(2\pi)^2\omega_p}{p^2} \, \delta(p-k) \nn\\[5pt]
    & \qquad - \Gc_{L'S',LS}^{J^P}(p,k) \Mc_{2,S}(\sigma_k) - \wt{\rho}(\sigma_p) \, \Dc_{L'S',LS}^{J^P}(p,k) \,  \nn\\[5pt]
    & \qquad \qquad - \sum_{L'',S''}\int_{p'} \, \Gc_{L'S',L''S''}^{J^P}(p,p') \, \Dc_{L''S'',LS}^{J^P}(p',k) \, .
\end{align}
Here we have introduced $\wt{\rho}$ as~\footnote{This is the minimal definition of $\wt{\rho}$. In Appendix~\ref{app:shifting}, we discuss a more general definition of $\wt{\rho}$, introduced in Ref.~\cite{Romero-Lopez:2019qrt}.}
\begin{align}
	\wt{\rho}(\sigma_k) = -iH(\sigma_k)\, \rho(\sigma_k) \, ,
 \label{eq:rhotilde}
\end{align}
where $H$ again regulates the improper integral with the condition that it is unity for physical $\sigma_k$, and $\rho(\sigma_k)$ is the two-body phase space factor defined as
\begin{align}
	\rho(\sigma_k) = \frac{\xi q_k^\star}{8\pi\sqrt{\sigma_k}} \, ,
 \label{eq:rho_ps}
\end{align}
where $\xi$ is a symmetry factor of the particles in the pair, for which $\xi = 1/2$ if the particles are identical and $\xi = 1$ otherwise, and $q_k^\star$ is the relative momentum of the particles in the pair in its CM frame. The relative momentum can be expressed in terms of the K\"all\'en triangle function, $\lambda(a,b,c) = a^2 + b^2 + c^2 - 2 ( ab + ac + bc )$, as
\begin{align}
    q_k^\star = \frac{1}{2\sqrt{\sigma_k}} \, \lambda^{1/2}(\sigma_k, m_{k_1}^2, m_{k_2}^2) \, .
    \label{eq:pair_rel_mom}
\end{align}
As a matrix in $LS$ space, $\wt{\rho}$ is proportional to the identity, $[\wt{\rho}(\sigma_k) ]_{L'S',LS} = \delta_{L'L}\delta_{S'S}\,\wt{\rho}(\sigma_k)$.

Physically, the rescattering functions characterize all possible initial and final state interactions of the three particles which do not involve three-body short-distance dynamics. We identify the first line of Eq.~\eqref{eq:Lcap:Jp_asym} with either no rescatterings or the scattering of two of the particles, the second and third lines with two particles scattering through any number of possible exchanges, see the discussion in Ref.~\cite{Jackura:2022gib}.

Finally, the amplitude $\hTc^{J^P}$ contains all information on short-distance three-body interactions. It is the solution of the integral equation
\begin{align}
	\label{eq:T_eq_asym}
    \hTc^{J^P}(p,k) = \hKc_3^{J^P} (p,k) - \int_{p'}\int_{k'}  \, \hKc_3^{J^P}(p,p') \cdot \hFc^{J^P}(p',k') \cdot \hTc^{J^P}(k',k) \, ,
\end{align}
where $\hKc_3^{J^P}$ is the previously discussed $\3\to\3$ K matrix,~\footnote{This has been denoted as $\Kc_{3,\df}^{(u,u)}$ in works such as Refs.~\cite{Blanton:2020gha} and~\cite{Jackura:2022gib}, where `df' stands for `divergence free'. Here we drop this notation as the K matrix must be defined to be free of on-shell singularities in a region near the three-body threshold.} defined to be free of unitarity singularities in a region near the three-body threshold, and $\Fc^{J^P}$ characterizes all intermediate state pair-wise rescatterings,
\begin{align}
	\label{eq:F}
	\hFc^{J^P}(p,k) \equiv \, \wt{\rho}(\sigma_p) \, \hLc^{J^P}(p,k) + \int_{k'} \, \Gc(p,k')  \cdot \hLc^{J^P}(k',k)   \, .
\end{align}
%

\subsection{Partial wave projection for $\Mc_{\df}$ with symmetric $\Kc_3$}
\label{sec:PWA:symm}
The general structure of $\Mc_{3,\df}^{J^P}$ with a symmetric $\Kc_3$ is similar to Eq.~\eqref{eq:M3df:jp_asym},
\begin{align}
    \Mc_{3,\df}^{J^P}(p,k) = \int_{p'} \int_{k'} \, {\Lc}^{J^P}(p,p') \cdot \Tc^{J^P}(p',k') \cdot {\Rc}^{J^P}(k',k),
    \label{eq:M3df:jp_sym}
\end{align}
with the only difference being in the definitions of the different building blocks. The $\Lc^{J^P}$ and $\Rc^{J^P}$  rescattering functions are given by,
\begin{align}
    [\Lc^{J^P}(p,k)]_{L'S',LS} & = \left( \frac{1}{3} -  \Mc_{2,S'}(\sigma_p) \, \wt{\rho}(\sigma_p)\right) \delta_{L'L}\delta_{S'S} \, \frac{(2\pi)^2\omega_k}{k^2} \delta(p-k) - \Dc_{L'S',LS}^{J^P}(p,k) \, \wt{\rho}(\sigma_k) \, ,
    \label{eq:Lcap:Jp_sym}\\[5pt]
    [\Rc^{J^P}(p,k)]_{L'S',LS} & = \left( \frac{1}{3} - \wt{\rho}(\sigma_k) \, \Mc_{2,S'}(\sigma_k) \right) \delta_{L'L}\delta_{S'S} \, \frac{(2\pi)^2\omega_p}{p^2} \delta(p-k) - \wt{\rho}(\sigma_p) \, \Dc_{L'S',LS}^{J^P}(p,k) \, ,
    \label{eq:Rcap:Jp_sym}
\end{align}
and $\Tc^{J^P}$ satisfies the integral equation,
\begin{align}
    \Tc^{J^P}(p,k) = \Kc_3^{J^P} (p,k) - \int_{p'} \int_{k'} \, \Kc_3^{J^P}(p,p') \cdot  \wt{\rho}(\sigma_p') \, {\Lc}^{J^P}(p',k') \cdot \Tc^{J^P}(k',k) \, .
	\label{eq:T_eq_sym}
\end{align}
We leave details associated with the derivation of these expressions to App.~\ref{app:conv}.

It is worth commenting on the close resemblance between these equations and the corresponding ones for the asymmetric functions, Eqs.~\eqref{eq:Lcap:Jp_asym}, \eqref{eq:Rcap:Jp_asym}, and \eqref{eq:T_eq_asym}. Qualitatively, one can understand these differences by remembering that $\wt{\rho}\,\wt{\delta}+\Gc$ is equivalent to $3\wt{\rho}\,\wt{\delta}  $ when acting on a symmetric quantity, like $\Kc_3$, where $\wt{\delta}=\frac{(2\pi)^2\omega_k}{k^2} \delta(p-k)$, see \eg Refs.~\cite{Briceno:2019muc,Blanton:2020gha,Blanton:2020jnm,Jackura:2022gib}. Using this identity, if one replaces $\hKc_3$ in \eqref{eq:T_eq_asym} with $\Kc_3/3$, one would find that $\wh{\Mc}_{3,\rm df}$ is exactly equal to $\Mc_{3,\rm df}$. 


\subsection{Generalizations -- Flavor isospin and multi-channel systems}
\label{sec:isospin}

Here we comment on how the above identities can be generalized, focusing on systems with additional quantum numbers like flavor isospin used in hadronic reactions, and multi-channel systems of three scalar particles. Consider first the incorporation of isospin into the above equations for isosymmetric QCD. Let $I$ be the total isospin of the initial pair, and $I'$ the isospin of the final pair. Since strong isospin is conserved in hadronic processes, the $\2\to\2$ amplitude is a diagonal matrix is isospin, whose diagonal components we will denote as $\Mc_{2,IS}$. However, for the $\3\to\3$ amplitude only total isospin, which we denote as $T$, is conserved. Thus, for a given $T(J^P)$, the amplitude is a matrix in $LSI$ space,
\begin{align}
    \left[\Mc_{3}^{T(J^P)}(p,k)\right]_{L'S'I',LSI} = \Mc_{3;L'S'I',LSI}^{T(J^P)}(p,k) \, .
\end{align}
All building blocks involving two-body systems only will be diagonal in $I$ space. The OPE function is the only object that non-trivially mixes isospin~\cite{VMK, Jackura:2023qtp},
\begin{align}
    \Gc^{T(J^P)}_{L'S'I',LSI} \,  & \equiv \Gc^{J^P}_{L'S',LS} \, \braket{([i_ei_k]I',i_p),T |([i_ei_p]I,i_k) T},
\end{align}
where $\braket{([i_ei_k]I',i_p),T |([i_ei_p]I,i_k) T}$ is the three-body recoupling coefficient, \cf, Ref.~\cite{VMK}. This matrix element relates the coupling of a pair (with constituent isospins $i_p$ and $i_e$) to isospin I, which subsequently couples to the spectator isospin $i_k$ to form a total isospin $T$ in the initial state, to a final state with pair $I'$ (constituents $i_k$ and $i_e$) coupled to spectator $i_p$ to definite $T$.~\footnote{The notation is chosen to coincide with the OPE as presented in Ref.~\cite{Jackura:2023qtp}.} For example, consider $3\pi \to 3\pi$ scattering, such that $i_p = i_k = i_e = 1$ and $I,I' = 0,1,2$ and $T = 0,1,2,3$. Thus the recoupling coefficient is
\begin{align}
  \braket{I' ,T |I ,T} \equiv \braket{([11]I',1),T |([11]I,1) T}
  =
  \sqrt{(2I'+1)(2I+1)}
      \begin{Bmatrix} 
    1 & 1 & I \\
    1 & T & I'
    \end{Bmatrix},
\end{align} 
where $\{\cdots\}$ is the Wigner 6-j symbol. Bose symmetry further constrains which partial wave matrix elements are non-zero, see Table I in Ref.~\cite{Jackura:2023qtp} for the $3\pi \to 3\pi$ example. 

Given this additional factor, we can generalize all the expressions above as follows. First, as previously discussed, we enhance the index space in which the matrices above exist. For simplicity, we use Greek letters to denote the product space of isospin, orbital angular momentum, and spin, e.g. $\alpha = LSI$. Second, we replace the matrix product in Eq.~\eqref{eq:mat_prod}, with
\begin{align}
   [A^{T(J^P)}\cdot B^{T(J^P)}]_{\beta\alpha} \equiv 
    \sum_{\gamma} A^{T(J^P)}_{\beta\gamma} B^{T(J^P)}_{\gamma \alpha} \, ,
\end{align}
where $\alpha = LSI,$ $\gamma = L''S''I''$, and $\beta = L'S'I'$. Third, we introduce a Kronecker-$\delta$ in this space, defined by
\begin{align}
    \delta_{\alpha \beta}\equiv\delta_{LL'}\delta_{SS'}\delta_{II'} \, .
\end{align}
Using these relations, all expressions in the previous sections are easily modified. For example, Eq.~\eqref{eq:Lcap:Jp_asym} becomes
\begin{align}
	\label{eq:Lcap:Jp_asym_iso}
    \left[\hLc^{\,T(J^P)}(p,k)\right]_{\beta\alpha} & = \left[ \, 1 -  \Mc_{2,\alpha}(\sigma_p) \, \wt{\rho}(\sigma_p) \, \right] \delta_{\beta\alpha} \, \frac{(2\pi)^2\omega_k}{k^2} \, \delta(p-k) \nn \\[5pt]
    & \qquad\qquad - \Mc_{2,\beta}(\sigma_p) \Gc_{\beta\alpha}^{T(J^P)}(p,k) - \Dc_{\beta\alpha}^{T(J^P)}(p,k) \, \wt{\rho}(\sigma_k) ,
\end{align}
where we define $\Mc_{2,\alpha} \equiv \Mc_{2,IS}$ as it is independent of $L$. Given that this prescription leads to a simple modification of all the expressions above, we avoid rewriting them in this basis.  

Let us now consider the generalization to multiple three-body scattering channels. We focus strictly on spinless systems, \eg $\pi\pi\pi \to \pi\pi\pi, K\bar{K} \pi$. We have thus far presented the partial wave integral equations for general masses, and  Ref.~\cite{Jackura:2023qtp} shows $\Gc^{J^P}$ for arbitrary masses. Let us still consider each particle has definite isospin, therefore we aim to enlarge the matrix space of the preceding discussion. We introduce the channel space index $a = 1, \ldots N_{\mathrm{ch.}}$ where $N_{\mathrm{ch.}}$ is the number of participating three-body channels. The three particles in a definite pair-spectator combination have masses $\{m_{a,k}, m_{a,k_1}, m_{a,k_2}\}$, where $m_{a,k}$ is the mass of the spectator in the $a$ channel, and $m_{a,k_1}$ and $m_{a,k_2}$ are the masses of the particles composing the pair associated the the spectator with momentum $k$ in channel $a$.

We can define the phase space as a diagonal matrix in this channel space,
\begin{align}
    \left[\rho(\sigma_k)\right]_{ab} =\delta_{ab} \frac{\xi_a {q_k^\star}_a}{8\pi\sqrt{\sigma_k}} \, ,
    \label{eq:rho_ps_cc}
\end{align}
where $\xi_a$ is a symmetry factor associated with the two-particle subsystems. If this pair is composed of identical particles $\xi_a=1/2$, otherwise $\xi_a=1$. If the particles are not necessarily identical, but they have been projected to a definite isospin state, \eg $\pi^+\pi^- \to [\pi\pi]_{I=1}$, then $\xi_a=1$. The relative momenta, ${q_k^\star}_a$, is then
\begin{align}
		{q_k^\star}_a= \frac{\lambda^{1/2}(\sigma_k,m^2_{a,k_1},m^2_{a,k_2})}{2\sqrt{\sigma_k}} \,  \, ,
\end{align}
where $\lambda$ is the K\"all\'en function as before in Eq.~\eqref{eq:pair_rel_mom}. The $\2\to\2$ amplitude is now a dense matrix in channel space,
\begin{align}
    \left[\Mc_{2,IS} \right]_{ba} = \Mc_{2;ba}^{IS} \, ,
\end{align}
and all remaining functions receive a channel index trivially. To simplify the notation, we again write generic matrix indices $\alpha$ as $\alpha = a(LSI)$, where $L$, $S$, and $I$ are the orbital angular momentum, spin, and isospin for channel $a$. Following the same extensions presented for including isospin, we then can write our partial wave integral equations in this enlarged angular momentum, isospin, and channel space. Revisiting the extension for Eq.~\eqref{eq:Lcap:Jp_asym} as an example, we have
\begin{align}
    \left[\hLc^{\,T(J^P)}(p,k)\right]_{\beta\alpha} & = \left[ \, 1 -  \Mc_{2,\beta\alpha }(\sigma_p) \, \wt{\rho}_{\alpha}(\sigma_p) \, \right]  \, \frac{(2\pi)^2\omega_k}{k^2} \, \delta(p-k) \nn \\[5pt]
    & \qquad\qquad  - \sum_{\gamma}\Mc_{2,\beta\gamma}(\sigma_p) \Gc_{\gamma\alpha}^{T(J^P)}(p,k) - \Dc_{\beta\alpha}^{T(J^P)}(p,k) \, \wt{\rho}_{\alpha}(\sigma_k) ,
\end{align}
where the sum on $\gamma$ must exist as $\Mc_2$ can transition to different channels.


\section{Separable Parameterizations}
\label{sec:separable}
In this section, we consider a special class of parameterizations, where the asymmetric K matrix, $\hKc_3$, can be written as a separable function. For such parameterizations, we will show that the integral equations for $\hTc$ in Eq.~\eqref{eq:T_eq_asym} have an algebraic solution. We subsequently comment on separable parameterizations for the symmetric K matrix. We begin this section by first defining this class of parameterizations. As discussed in Sec.~\ref{sec:isospin}, incorporating features like isospin leads to a straightforward modification of all the expressions after performing the partial wave projection. As a result, here we focus in deriving the building blocks in the $LS$-basis for our elastic scattering process, and leave the expressions in an enlarged $\alpha$ introduced in Sec.~\ref{sec:isospin} implicit.

\subsection{Separable K matrix parameterizations}
\label{sec:separableK}

In order to justify the separable parameterizations of $\hKc_3$, we begin by outlining some minimal properties it must satisfy. Above threshold, $\hKc_3$ is a purely real function. Because of the spurious singularities of the spherical harmonic at threshold, the partial wave projected $\hKc_3$ must include barrier factors that exactly cancel these singularities. These barrier factors, $\Bc_{LS}(k,s)$, need to cancel the singularities associated with the threshold of the pair sub-system as well as the threshold associated with the pair-spectator system. Because of this, we define the barrier factors as
\begin{align}
	\Bc_{LS}(k,s) = k^L q_k^{\star\,S} \, .
    \label{eq:barrier}
\end{align}
The remaining part of the K matrix is meromorphic in the remaining energy variables, allowing us to choose the separable parameterization as
\begin{align}
	\label{eq:K3.separable}
	\hKc_{3;L'S',LS}^{J}(p,k) = \Bc_{L'S'}(p,s) f_{L'S'}(p^2) \, \wt{\Kc}_{3;L'S',LS}^{J}(s) \, f_{LS}(k^2) \,\Bc_{LS}(k,s) \, .
\end{align}
This definition ensures that the remainder functions, $\wt{\Kc}_3$ and $ f_{LS}$, are real and free of on-shell singularities in a domain near the three-body threshold of the complex ($s$, $\sigma_k$, $\sigma_p$)-hyperplane. Consequently, one can parameterize these by Taylor expansions in this region. Note, since $\sigma_k = \sigma_k(s,k^2)$, we can freely choose to have the expansion in either $\sigma_k$, $\sigma_p$ variables or in terms of $k^2$, $p^2$, and we have chosen the latter. 

The functions $f_{LS}(k^2)$ include residual sub-channel momentum dependence which we are free to choose, \eg,
\begin{align}
	f_{LS}(k^2) = \sum_{j = 0}^{n} \alpha_{LS}^{(j)} \, k^{2j} \, ,
\end{align}
for some $n$ and real parameters $\alpha_{LS}$. Finally, the reduced K matrix $\wt{\Kc}_3$ is a function of $s$ only, thus we can write it as, \eg,
\begin{align}
	\wt{\Kc}_{3;L'S',LS}^{J^{P}}(s) &= \sum_{j=0}^{n'} \frac{\beta_{L'S',LS}^{J^P \, (j) }}{s_0^{(j)} - s} + \sum_{j=0}^{n''} \gamma_{L'S',LS}^{J^P\,(j)} \, s^j \, ,
\end{align}
for real parameters $\beta_{L'S',LS}^{J^{P}}$ and $\gamma_{L'S',LS}^{J^P}$ for some $n'$ and $n''$. For convenience, we write the matrix elements of Eq.~\eqref{eq:K3.separable} as
\begin{align}
\label{eq:factorized}
    \left[\hKc_3^{J^P}(p,k)\right]_{L'S',LS} = \left[h(p)\right]_{L'S'}\, \left[\wt{\Kc}_3^{J^P}(s)\right]_{L'S',LS} \, \left[h(k)\right]_{LS} \, ,
\end{align}
where the $h$ functions contain both the barrier factors and any residual spectator momentum dependence.

This class of separable parameterizations can be generalized to include a sum over any number of terms of the form of Eq.~\eqref{eq:factorized}, called \emph{degenerate} from Fredholm theory,
\begin{align}
\label{eq:factor_general}
    \left[\hKc_3^{J^P}(p,k)\right]_{L'S',LS} = \sum_{j}\left[h(p)\right]_{L'S';j}\, \left[\wt{\Kc}_3^{J^P}(s)\right]_{L'S',LS;j} \, \left[h(k)\right]_{LS;j} \, .
\end{align} 
This additional index can be absorbed by enlarging the space in which $h$ and $\wt{\Kc}_3$ can be considered matrices.

\subsection{Algebraic solution for separable $\hTc$ for asymmetric representation}
\label{sec:sepT_asymm}
The separable K matrix allows us to solve the integral equation for $\hTc^{J^P}$ analytically, as it factorizes the momentum dependence such that $\hTc^{J^P}$ becomes separable,
\begin{align}
    \left[\hTc^{J^P}(p,k)\right]_{L'S',LS} = \left[h(p)\right]_{L'S'}\, \left[\wt{\Tc}^{J^P}(s)\right]_{L'S',LS} \, \left[h(k)\right]_{LS} \, ,
\end{align}
which leads to a system of algebraic equations. The solution of Eq.~\eqref{eq:T_eq_asym} is then
\begin{align}
	\wt{\Tc}^{J^P}(s) = \frac{1}{1 + \wt{\Kc}_3^{J^P}(s) \cdot \wt{\Fc}^{J^P}(s) } \cdot \wt{\Kc}_3^{J^P}(s) \, , 	
    \label{eq:T_:sol}
\end{align}
where each object is a matrix in $LS$ space. Here we have defined 
\begin{align}
	\wt{\Fc}^{J^P}(s) & \equiv  \int_p\int_k \, h(p) \cdot \wh{\Fc}^{J^P}(p,k) \cdot h(k) \, ,\nn\\[5pt]
	& = \int_p \int_k \, h(p) \cdot \Gamma^{J^P}(p,k) \cdot h(k) \nn \\[5pt]
	& \qquad - \int_p \int_k \int_{k'} \,  h(p) \cdot \Gamma^{J^P}(p,k') \cdot \Mc_2 (\sigma_{k'}) \cdot \Gamma^{J^P}(k',k) \cdot h(k) \nn \\[5pt]
	& \qquad\qquad - \int_p \int_k \int_{p'} \int_{k'} \,  h(p) \cdot \Gamma^{J^P}(p,p') \cdot \Dc^{J^P} (p',k') \cdot \Gamma^{J^P}(k',k) \cdot h(k)\, ,
\end{align}
which can be found by direct substitution of Eq.~\eqref{eq:Lcap:Jp_asym} into \eqref{eq:F} and defining the matrix
\begin{align}
	\Gamma^{J^P}(p,k) = \frac{(2\pi)^2\,\omega_k}{k^2} \, \delta(p - k) \,  \wt{\rho}(\sigma_p) + \Gc^{J^P}(p,k)  \, .
\end{align}
Therefore, for a separable $\hKc_3^{J^P}$, the $\3\to\3$ amplitude is given by Eq.~\eqref{eq:M3_def} with $\Dc^{J^P}$ the solution of the ladder equation~\eqref{eq:ladder_eq}, and $\Mc_{3,\df}^{J^P}$ given by
\begin{align}
	\hMc_{3,\df}^{J^P}(p,k) = \wt{\Lc}^{J^P}(p,s) \cdot \wt{\Tc}^{J^P}(s) \cdot \wt{\Rc}^{J^P}(s,k) \, ,
    \label{eq:M3:df:factorizable}
\end{align}
with
\begin{align}
    \wt{\Lc}^{J^P}(p,s) & \equiv \int_k \, \hLc^{J^P}(p,k) \cdot h(k) \, ,  \nn \\[5pt]
    &  = h(p)  - \int_k \, \Mc_2 (\sigma_{p}) \cdot \Gamma^{J^P}(p,k) \cdot h(k) 
    - \int_k \int_{k'} \Dc^{J^P} (p,k') \cdot \Gamma^{J^P}(k',k) \cdot h(k)\, , 
    \label{eq:Lcap:factorizable}\\[5pt]
    \wt{\Rc}^{J^P}(s,k) & \equiv \int_p h(p) \cdot \hRc^{J^P}(p,k) \, ,  \nn \\[5pt] 
    &  = h(k) - \int_p \, h(p) \cdot \Gamma^{J^P}(p,k) \cdot \Mc_2 (\sigma_{k})   
     - \int_p \int_{p'} h(p) \cdot \Gamma^{J^P}(p,p') \cdot \Dc^{J^P} (p',k)  \, . 
    \label{eq:Rcap:factorizable}
\end{align}
In practice, to reduce the number of computations, it is useful to note that $\wt{\Fc}$ can be obtained from either $\wt{\Lc}$ or $\wt{\Rc}$,
\begin{align}
    \label{eq:F_to_L_and_R}
    \wt{\Fc}^{J^P}(s) = \int_p\int_k \, \wt{\Rc}^{J^P}(s,p)\cdot \Gamma^{J^P}(p,k)\cdot h(k) = \int_p\int_k \, h(p)\cdot \Gamma^{J^P}(p,k)\cdot \wt{\Lc}^{J^P}(k,s) \, .
\end{align}
 We can further reduce these $\wt{\Lc}$ and $\wt{\Rc}$ functions to remove the $\delta$ function contributions from $\Gamma^{J^P}$ and express them in terms of $d^{J^P}(p,k)$, defined in Eq.~\eqref{def:D:ladder},
\begin{align}
    \wt{\Lc}^{J^P}(p,s) & = \left[1 - \Mc_2(\sigma_p) \wt{\rho}(\sigma_p) \right]\cdot h(p) 
    \nn\\
    &\hspace{2cm}+ \Mc_2(\sigma_p) \cdot\int_k d^{J^P}(p,k)\cdot\left[1 - \Mc_2(\sigma_k) \wt{\rho}(k)\right]\cdot h(k) \, , 
    \label{eq:Lcap:tilde}\\[5pt]
    \wt{\Rc}^{J^P}(s,k) & = h(k)\cdot \left[1 - \wt{\rho}(\sigma_k)  \Mc_2(\sigma_k) \right] \nn\\
    &\hspace{2cm}
    + \int_p h(p) \cdot\left[1 - \wt{\rho}(\sigma_p)  \Mc_2(\sigma_p) \right]\cdot d^{J^P}(p,k)\cdot \Mc_2(\sigma_k)
    \label{eq:Rcap:tilde} 
    \, .
\end{align}
Using Eq.~\eqref{eq:F_to_L_and_R} on these expressions, we can get a potentially more efficient form for calculating $\wt{\Fc}$,
\begin{align}
    \wt{\Fc}^{J^P}(s) & = \int_p h(p)\cdot \wt{\rho}(\sigma_p) \cdot\left[1 - \Mc_2(\sigma_p) \wt{\rho}(\sigma_p) \right] \cdot h(p)  \nn \\[5pt]
    & \quad + \int_p\int_k h(p)\cdot \wt{\rho}(\sigma_p)\Mc_2(\sigma_p) \cdot d^{J^P}(p,k)\cdot\left[1 - \Mc_2(\sigma_k) \wt{\rho}(\sigma_k) \right]\cdot h(k)   \nn \\[5pt]
    & \qquad + \int_p\int_k\int_{k'}  h(p)\cdot \Gc(p,k)\cdot \Mc_2(\sigma_k)\cdot d^{J^P}(k,k') \cdot\left[1 - \Mc_2(\sigma_k') \wt{\rho}(\sigma_k') \right]\cdot h(k') \, .
    \label{eq:Ftilde_vf}
\end{align}
%

\subsection{Algebraic solution for separable $\Tc$ for symmetric representation}
\label{sec:sepT_symm}
The expressions derived in Sec.~\ref{sec:sepT_asymm} have assumed a separable parametrization for $\hKc_3$ of the form presented in Sec.~\ref{sec:separableK}. For the symmetric formalism, we cannot immediately impose the class of parametrizations presented in Sec.~\ref{sec:separableK}. Instead, one must construct a parametrization of $\Kc_3$ which is symmetric under the interchange of the particles and then perform a partial wave projection of this. Depending on the parametrization, it should be straightforward for low energies to write the resultant $\Kc_3^{J^P}$ matrix in the form of Eq.~\eqref{eq:factor_general}.

If a given symmetric parameterization can be written as Eq.~\eqref{eq:factor_general}, we can follow the same steps as in the asymmetric case to solve Eq.~\eqref{eq:T_eq_sym} to write $\Tc^{J^P}(p,k) = h(p) \cdot \overline{\Tc}^{\, J^P}(s)\cdot h(k)$. The factorized $\overline{\Tc}$ matrix is given in the same form as Eq.~\eqref{eq:T_:sol}, 
\begin{align}
	\overline{\Tc}^{\,J^P}(s) = \frac{1}{1 + \overline{\Kc}_3^{J^P}(s) \cdot \overline{\Fc}^{J^P}(s) } \cdot \overline{\Kc}_3^{J^P}(s) \, , 	
\end{align}
where $\overline{\Kc}_3^{J^P}$ is the factorized symmetric K matrix, and $\overline{\Fc}^{J^P}$ is defined as 
\begin{align}
    \overline{\Fc}^{J^P}(s) 
    &= \int_{p'} \int_{k'} \, h^{J^P}(p') \cdot \wt{\rho}(\sigma_p') \, \Lc^{J^P}(p',k') \cdot h^{J^P}(k') \, ,
    \nn\\[5pt]
    &= \int_{p'}  \, h^{J^P}(p') \cdot  \wt{\rho}(\sigma_p') \, 
    \cdot
    \left(\frac{1}{3}-\wt{\rho}(\sigma_p') \mathcal{M}_{2}(\sigma_p') \right)
    \cdot h^{J^P}(p')
    \nn\\[5pt]
    &\hspace{3cm}
    -
    \int_{p'} \int_{k'} \, h^{J^P}(p') \cdot\wt{\rho}(\sigma_p') \, {\Dc}^{J ^P}(p',k') \, 
    \wt{\rho}(\sigma_k')  
    \cdot h^{J^P}(k')
    ,
\end{align}
where $\Lc^{J^P}$ has been defined in Eq.~(\ref{eq:Lcap:Jp_sym}).

Assuming a $\Kc_3^{J^P}$ matrix in the form of Eq.~\eqref{eq:factor_general}, one can show that the amplitude $\Mc_{3,\df}^{J^P}$ can be written as
\begin{align}
	\Mc_{3,\df}^{J^P}(p,k) = \overline{\Lc}^{J^P}(p,s) \cdot \overline{\Tc}^{J^P}(s) \cdot \overline{\Rc}^{J^P}(s,k) \, ,
    \label{eq:M3:df:sym}
\end{align}
with
\begin{align}
    \overline{\Lc}^{J^P}(p,s) & \equiv \int_k \, \Lc^{J^P}(p,k) \cdot h(k) \, , \nn \\[5pt]
    &  = \left(\frac{1}{3} - \mathcal{M}_{2,S'}(\sigma_p)\tilde{\rho}(\sigma_p )\right)\cdot h(p)  - \int_{p'} \,  \mathcal{D}^{J^P}(p,p') \cdot \tilde{\rho}(\sigma_{p'} )\, h(p') \, , 
    \label{eq:Lcap:sym} \\[5pt]
    \overline{\Rc}^{J^P}(s,k) & \equiv \int_p h(p) \cdot \Rc^{J^P}(p,k)  \, , \nn \\[5pt] 
    &  = h(k)\cdot\left(\frac{1}{3} - \mathcal{M}_{2,S'}(\sigma_k)\tilde{\rho}(\sigma_k )\right)  - \int_{k'} \, h(k')\rho(\sigma_k')\cdot  \mathcal{D}^{J^P}(k',k)   \, .     
    \label{eq:Rcap:sym}
\end{align}
%

\section{Bound-state spectator amplitudes}
\label{sec:LSZ}

In this section, we consider the case where the two-particle pairs can form a bound state. This is worthwhile for two reasons. First, for sufficiently heavy quark masses, lattice QCD calculations have found that both the $\sigma$ and the $\rho$ appear to be bound~\cite{HadronSpectrum:2009krc, Briceno:2016mjc, Briceno:2017qmb,Rodas:2023gma}. For these unphysical masses, exploratory calculations of scattering processes involving these states have been performed in Refs.~ \cite{Woss:2018irj,Woss:2020ayi}, among others.   
Second, as shown in Refs.~\cite{Jackura:2020bsk, Dawid:2023jrj}, the consideration of two-body bound states provides some of the best checks of the three-body formalism. In Sec.~\ref{sec:numerical}, we use this limiting case to provide some checks on partial wave projection performed in Sec.~\ref{sec:PWA}. As in Sec.~\ref{sec:separable}, we consider only amplitudes in the $LS$ basis. Isospin and multi-channel processes can be incorporated using the steps presented in Sec.~\ref{sec:isospin}.

We define the resulting two-body scattering amplitude as $\mathcal{M}^{J^P}_{ \varphi b }$, where $b$ denotes a bound two-body system and $\varphi$ is the remaining third scalar particle. Since we work with generic scalar particles, each pair combination in principle has a distinct bound state. Thus, the reaction becomes $b_k + \varphi_k \to b_p + \varphi_p$. The amplitude can be written as the sum of two terms, 
\begin{align}
    \mathcal{M}^{J^P}_{ \varphi b }
    =\mathcal{M}^{J^P}_{ \varphi b ,\Dc}+\mathcal{M}^{J^P}_{ \varphi b ,\rm df},
    \label{eq:Mphib}
\end{align}
where the first comes from following the LSZ prescription to $\Dc^{J^P}$ and the second comes from $\Mc_{3,\df}^{J^P}$. We begin by defining the first of these. We make use of Eq.~\eqref{def:D:ladder}, which we rewrite here for convenience
\begin{equation}
\Dc^{J^P}(p,k) = \Mc_{2}(\sigma_p) \cdot  d^{J^P}(p,k) \cdot \Mc_{2}(\sigma_k).
\label{def:D:ladder_v2}
\end{equation}

In the presence of two-body bound states, $\Mc_2$ has poles of the form 
\begin{align}
    \Mc_{2}(\sigma_k) = -\frac{g_{k,b}^2}{\sigma_k - \sigma_{k,b}} + \Oc\!\left((\sigma_k - \sigma_{k,b})^0\right)
    \label{eq:bs_amp}
\end{align}
where $g_{k,b}$ denotes the residue of the scattering amplitude at the bound state pole and $\sigma_{k,b}$ is the pole location. The label $k$ emphasizes that the bound state is one of the pair associated with spectator $k$. The bound state mass is then $\sqrt{\sigma_{k,b}}$, and $g_{k,b}$ can be understood as the coupling between the composite and two-particle scattering states. From Eq.~\eqref{def:D:ladder_v2}, we see that $\Dc^{J^P}$ has these same poles. Assuming there is only one such bound state in each pair, we see that 
\begin{align}
    \Dc^{J^P}(p,k) = \left(-\frac{g_{p,b}}{\sigma_p - \sigma_{p,b}}\right) \, g_{p,b}g_{k,b} \, d^{J^P}(q_{p,b},q_{k,b}) \left(-\frac{g_{k,b}}{\sigma_k - \sigma_{k,b}}\right) + \cdots \, ,
\end{align}
where the ellipses denote terms which are not simultaneous poles in both $\sigma_p$ and $\sigma_k$. We have introduced $q_{k,b}$ as the relative momentum of the bound state and the spectator, given by
\begin{align}
    q_{k,b} = \frac{1}{2\sqrt{s}}\,\lambda^{1/2}(s,m_k^2,\sigma_{k,b}) \, .
\end{align}

Having identified the pole structure of $\Dc^{J^P}$, we can use the LSZ reduction formula to obtain the definition for $\Mc^{J^P}_{\varphi b ,\Dc}$,
\begin{equation}
    \Mc^{J^P}_{\varphi b ,\Dc}
    = \lim_{\substack{\sigma_p\to \sigma_{p,b} \\ \sigma_k\to \sigma_{k,b} }}\frac{(\sigma_p - \sigma_{p,b})(\sigma_k - \sigma_{k,b})}{g_{p,b} g_{k,b}} \, \Dc^{J^P}(p,k),
\end{equation}
or equivalently,
\begin{equation}
    \label{eq:phib_D}
    \Mc^{J^P}_{\varphi b ,\Dc} =  g_{p,b} g_{k,b} \,d^{J^P}(q_{p,b},q_{k,b}) \, .
\end{equation}

So far, we have only applied the LSZ prescription to $\Dc^{J^P}$. Next, we will analyze the prescription to the divergent free part of the amplitude, considering both the symmetric and asymmetric cases.

\subsection{LSZ for asymmetric representation}

In a similar manner, we proceed to derive  the $\hMc_{3,\mathrm{df}}^{J^P}$ contribution to the $b_k + \varphi_k \to b_p + \varphi_p$ amplitude. By examining the defining Eqs.~(\ref{eq:M3:df:factorizable}),~(\ref{eq:Lcap:factorizable}) and~(\ref{eq:Rcap:factorizable}) or, equivalently, Eqs.~(\ref{eq:Lcap:tilde}) and~(\ref{eq:Rcap:tilde}), we observe that the pole structure of $\hMc_{3,\mathrm{df}}^{J^P}$ arises from the poles in $\Mc_2$ and $\Dc^{J^P}$. This leads to the structure
\begin{align}
     \hMc_{3,\mathrm{df}}^{J^P} 
    =  \left(\frac{-g_{p,b}}{\sigma_p - \sigma_{p,b}}\right)\, \widetilde{\Lc}_{\varphi b }^{J^P}(q_{p,b}, s)\cdot\widetilde{\Tc}(s)\cdot \widetilde{\Rc}_{\varphi b }^{J^P}(s, q_{k,b})\left(\frac{-g_{k,b}}{\sigma_p - \sigma_{k,b}}\right) + \cdots \, , 
\end{align}
where again the ellipses denote terms which are not simultaneous poles in both $\sigma_p$ and $\sigma_k$. The matrix $\widetilde{\Tc}$ is given by Eq.~(\ref{eq:T_:sol}), while $\widetilde{\Lc}_{\varphi b }^{J^P}$ and $\widetilde{\Rc}_{\varphi b }^{J^P} $ are given by
\begin{align}
    -\frac{1}{g_{p,b}} \, \widetilde{\Lc}_{\varphi b }^{J^P}(q_{p,b},s) & \equiv \tilde{\rho}(\sigma_{p,b})h^{J^P} (q_{p,b}) + \int_k \,\Gc^{J^P}(q_{p,b},k)\cdot h^{J^P} (k)  
    \nn\\
    &\hspace{3.25cm} + \int_k \, d^{J^P}(q_{p,b},k)\cdot\Mc_2 (\sigma_k)\cdot
    \wt{\rho}(\sigma_k) \cdot h^{J^P} (k)  \, , 
    \nn\\
    &\hspace{3.25cm}  + \int_k \int_{k'} \, d^{J^P}(q_{p,b},k')\cdot  \Mc_2 (\sigma_{k'}) \cdot \Gc^{J^P}(k',k) \cdot h^{J^P} (k) \, ,   
	  \label{eq:Lcphib} \\[5pt]
    -\frac{1}{g_{k,b}} \, \widetilde{\Rc}_{\varphi b }^{J^P}(s,q_{k,b}) & \equiv  h^{J^P}(q_{k,b}) \tilde{\rho}(\sigma_{k,b}) + \int_p h^{J^P}(p) \cdot\Gc^{J^P}(p,q_{k,b})
    \nn\\
    &\hspace{3.25cm} + \int_p \,  h^{J^P}(p)\cdot \wt{\rho}(\sigma_p) \cdot  \Mc_2 (\sigma_p)\cdot d^{J^P}(p,q_{k,b})\, , \nn \\
    &\hspace{3.25cm} + \int_p \int_{p'} \, h^{J^P}(p)\cdot \Gc^{J^P}(p,p')\cdot  \Mc_2 (\sigma_{p'})\cdot  d^{J^P}(p',q_{k,b}) \, .
    \label{eq:Rcphib}
\end{align}
Using the integral equation for $d^{J^P}$ (see Sec.~\ref{sec:numerical}), we can simplify these functions to
\begin{align}
    -\frac{1}{g_{p,b}} \, \widetilde{\Lc}_{\varphi b }^{J^P}(q_{p,b},s) & =  \tilde{\rho}(\sigma_{p,b})h^{J^P} (q_{p,b}) - \,\int_k \, d^{J^P}(q_{p,b},k)\cdot\left[1 - \Mc_2 (\sigma_k)\cdot
    \wt{\rho}(\sigma_k)\right] \cdot h^{J^P} (k)  \, ,    
	  \label{eq:Lcphib_v2}\\[5pt]
    -\frac{1}{g_{k,b}} \, \widetilde{\Rc}_{\varphi b }^{J^P}(s,q_{k,b}) 
     &= \tilde{\rho}(\sigma_{k,b})h^{J^P}(q_{k,b}) - \int_p h^{J^P}(p) \cdot [1 - \wt{\rho}(\sigma_p) \cdot  \Mc_2 (\sigma_p)]\cdot d^{J^P}(p,q_{k,b})\, .
     \label{eq:Rcphib_v2}
\end{align}
From these, one can obtain the $\hMc_{\varphi b ,\mathrm{df}}^{J^P}$ amplitude,
\begin{align}
    \hMc_{\varphi b ,\mathrm{df}}^{J^P} 
    = \lim_{\substack{\sigma_p\to \sigma_{p,b} \\ \sigma_k\to \sigma_{k,b} }}\frac{(\sigma_p - \sigma_{p,b})(\sigma_k - \sigma_{k,b})}{g_{p,b} g_{k,b}} \, \hMc_{3,\df}^{J^P}(p,k) 
    &\equiv \wt{\Lc}_{\varphi b }^{J^P}(q_{p,b}, s)\cdot\widetilde{\Tc}(s)\cdot \wt{\Rc}_{\varphi b }^{J^P}(s, q_{k,b}) \, .
    \label{eq:Mc_phib_asym}
\end{align}
Combining this result with Eq.~\eqref{eq:phib_D} in Eq.~\eqref{eq:Mphib}, we find for the $\Mc_{\varphi b}^{J^P}$ amplitude
\begin{align}
    \label{eq:Mphib_asym_complete}
    \Mc_{\varphi b}^{J^P}(s) = g_{p,b} g_{k,b} \,d^{J^P}(q_{p,b},q_{k,b}) + \wt{\Lc}_{\varphi b }^{J^P}(q_{p,b}, s)\cdot\widetilde{\Tc}(s)\cdot \wt{\Rc}_{\varphi b }^{J^P}(s, q_{k,b}) \, .
\end{align}
%

\subsection{LSZ for symmetric representation}

For the symmetric representation of the divergence-free amplitude, $\mathcal{M}_{3,\mathrm{df}}^{J^P}$, the pole analysis follows directly from Eqs.~(\ref{eq:M3:df:sym}), (\ref{eq:Lcap:sym}), and~(\ref{eq:Rcap:sym}). The result is
\begin{align}
    \Mc_{3,\df}^{J^P}(p,k) 
    &= 
     \left(\frac{-g_{p,b}}{\sigma_p - \sigma_{p,b}}\right) \, \overline{\Lc}^{J^P}_{\varphi b }(q_{p,b},s)\cdot\overline{\Tc}(s)\cdot \overline{\Rc}^{J^P}_{\varphi b }(s, q_{k,b})\,\left(\frac{-g_{k,b}}{\sigma_k - \sigma_{k,b}}\right) + \cdots \, ,
\end{align}
where 
\begin{align}
    -\frac{1}{g_{p,b}} \, \overline{\mathcal{L}}^{J^P}_{\varphi b }(q_{p,b}, s) & \equiv \tilde{\rho}(\sigma_{p,b})h^{J^P}(q_{p,b}) + \int_{p'} \, d(q_{p,b},p')\cdot\mathcal{M}_{2
}(\sigma_{p'}) \wt{\rho}(\sigma_{p'})\cdot
h^{J^P}(p') \, , \\[5pt]
    -\frac{1}{g_{k,b}} \, \overline{\mathcal{R}}^{J^P}_{\varphi b }(s, q_{k,b}) & \equiv h^{J^P}(q_{k,b})\tilde{\rho}(\sigma_{k,b}) + \int_{k'} \, h^{J^P}(k')\cdot\mathcal{M}_2(\sigma_{k'})\tilde{\rho}(\sigma_{k'}) \cdot d(k',q_{k,b}).
\end{align}
As before, we use the LSZ reduction formula to obtain the definition of $\mathcal{M}_{\varphi b ,\mathrm{df}}^{J^P}$, allowing us to write it in the form:
\begin{align}
    \mathcal{M}_{\varphi b ,\mathrm{df}}^{J^P} 
    &= \lim_{\substack{\sigma_p\to \sigma_{p,b} \\ \sigma_k\to \sigma_{k,b} }}\frac{(\sigma_p - \sigma_{p,b})(\sigma_k - \sigma_{k,b})}{g_{p,b} g_{k,b}} \, \Mc_{3,\df}^{J^P}(p,k) 
    \equiv \overline{\Lc}_{\varphi b }^{J^P}(q_{p,b}, s)\cdot\overline{\Tc}(s)\cdot\overline{\Rc}^{J^P}_{\varphi b }(s, q_{k,b}) \, ,
\end{align}
thus the bound-state spectator amplitude Eq.~\eqref{eq:Mphib} is
\begin{align}
    \Mc_{\varphi b}^{J^P}(s) = g_{p,b} g_{k,b} \,d^{J^P}(q_{p,b},q_{k,b}) + \overline{\Lc}_{\varphi b }^{J^P}(q_{p,b}, s)\cdot\overline{\Tc}(s)\cdot\overline{\Rc}^{J^P}_{\varphi b }(s, q_{k,b}) \, .
\end{align}
%

\section{Numerical investigation for three-pion systems}
\label{sec:numerical}

In this section, we explore the consequence of this formalism for studying arbitrary $[\pi\pi]_{\ell}^{I} + \pi \to [\pi\pi]_{\ell'}^{I'} + \pi$ reactions. In particular, we consider a scenario at unphysical pion masses such that the $\sigma$ and $\rho$ resonances are bound states~\cite{HadronSpectrum:2009krc, Briceno:2016mjc, Briceno:2017qmb,Rodas:2023gma}. We focus on energies $\sqrt{s} < 3m_\pi$, and use the framework discussed in Sec.~\ref{sec:LSZ} to construct effective $\sigma\pi$ and $\rho\pi$ scattering amplitudes for some definite isospin and spin-parity $T(J^P)$. While both the asymmetric and symmetric representations yield the same physical amplitude, we focus the presentation on the asymmetric K matrix formalism primarily due to the relative ease at which one can parameterize $\hKc_3$ compared with the symmetric case. Indeed, adopting this approach follows that in two-body analyses, where one is agnostic to the parameterization of the two-body matrix, only requiring it to respect the $S$ matrix principles and choosing forms that are flexible for analyses.

After providing a prescription for solving the key set of integral equations appearing in Secs.~\ref{sec:PWA} and~\ref{sec:separable}, we present results for the asymmetric formalism for models including the lowest-lying partial waves for $T=2,1,0$ three pion systems. It is worth remarking that we have considered many models for both the symmetric and asymmetric formalisms, and for all examples considered we observed that unitary is well satisfied for kinematics below the three-body threshold.

\subsection{Review of numerical technique for coupled-channel systems}

We set to compute $\Mc_{\sigma\pi}^{T(J^P)}$ and $\Mc_{\rho\pi}^{T(J^P)}$ using the results of Sec.~\ref{sec:LSZ}. Our starting point is the extension of Eq.~\eqref{eq:Mphib_asym_complete}, which describes the desired amplitudes assuming factorizable parameterizations, to relations involving isospin,
\begin{align}
\label{eq:Mphib_iso}
    \left[\Mc^{T(J^P)}_{\varphi b }\right]_{\beta\alpha}
    & = \lim_{\substack{\sigma_p\to \sigma_\beta \\ \sigma_k\to \sigma_\alpha }}\frac{(\sigma_p - \sigma_\beta)(\sigma_k - \sigma_\alpha)}{g_\beta g_\alpha}\left[\Mc^{T(J^P)}_3(p,k)\right]_{\beta\alpha} \, , \nn \\[5pt]
     & = g_\beta g_\alpha\,\left[d^{T(J^P)}(q_\beta,q_\alpha)\right]_{\beta\alpha}
    +
    \left[\wt{\Lc}_{\varphi b }^{\,T(J^P)}(q_\beta, s)\cdot\widetilde{\Tc}(s)\cdot \wt{\Rc}_{\varphi b }^{T(J^P)}(s, q_\alpha)\right]_{\beta\alpha},
\end{align}
where $\alpha,\beta = LSI$ and $\wt{\Lc}_{\varphi b }^{\,T(J^P)}$, and $\wt{\Rc}_{\varphi b }^{T(J^P)}$ are simple extensions of $\wt{\Lc}_{\varphi b }^{J^P}$ and $\wt{\Rc}_{\varphi b }^{J^P}$, defined in Eqs.~\eqref{eq:Lcphib} and \eqref{eq:Rcphib}, in the $LSI$-basis. Since all particles are identical, we need not to distinguish species type on the bound state poles and residues and simply label them by $\alpha$, which indicates which two-body partial wave they belong to, \ie, $I = 0$, $S = 0$ belongs to the $\sigma$, while $I = 1$, $S = 1$ belongs to the $\rho$.

As detailed in the preceding sections, all quantities can be written in terms of the amputated ladder amplitude, $d^{\,T(J^P)}$, which is the $LSI$-space extension of Eq.~\eqref{def:D:ladder}. We summarize the key steps needed to find numerical solutions of the integral equation for $d^{T(J^P)}$. The integral equation for $d^{T(J^P)}$ follows from the generalization of Eqs.~\eqref{eq:ladder_eq} and~\eqref{def:D:ladder} to the $LSI$ basis,
\begin{align}
    d^{\,T(J^{P})}(p,k) = -\mathcal{G}^{T(J^{P})}(p,k) - \int_{k'} \Qc^{T(J^{P})}(p,k')\cdot d^{\,T(J^{P})}(k',k) \, ,
\label{eq:d:int}
\end{align}
where the kernel $\Qc^{T(J^P)}$ is
\begin{align}
    \Qc^{T(J^{P})}_{\beta\alpha}(p,k) = \mathcal{G}^{T(J^{P})}_{\beta\alpha}(p,k) \, \Mc_{2,\alpha}(\sigma_k) \, .
\end{align}
We remind the reader that $\mathcal{M}_{2,\alpha} = \mathcal{M}_{2,IS}$ in the $LSI$-basis.

Recall from Eq.~(\ref{eq:G:OPE}) that the OPE propagator includes a cut-off function $H$. This cutoff function effectively sets an upper limit on the integral above, represented by $k'_{\mathrm{max}}$. Therefore, the integral Eq.~\eqref{eq:radial_int} becomes 
\begin{align}
    \int_{k'}\rightarrow \int_{0}^{k'_{\mathrm{max}}} \! \diff k'\, \frac{ k'^2}{(2\pi)^2 \omega_k'} \, . \nn 
\end{align} 
Two commonly used options for defining $H$ are a smooth or a hard cut-off. The smooth cut-off has proven advantageous when considering the formalism for describing finite-volume quantities~\cite{Hansen:2014eka,Hansen:2015zga}. Unfortunately, defining an analytic function that is exactly equal to $1$ in the physical region above the three-particle threshold is impossible. Having an analytic cut-off function is key to be able to deform contours, which is generally needed for solving integral equations for $\Mc_3$~\cite{Dawid:2023jrj}. As a result, such a smooth function breaks the desired analytic properties for scattering amplitudes. 

Alternatively, one can introduce a hard cut-off in momentum space and fix $H=1$ everywhere within this domain. In the following numerical results, we will use this and set the maximum momentum to ensure that none of the parameterizations used for $\Mc_2$ have unphysical singularities. The minimal but not necessarily sufficient criteria for this requires that $\sigma_k\geq 0$.

Equation~\eqref{eq:d:int} is a system of Fredholm integral equations of the second kind, for which algorithms for numerical solutions are well-known, see \eg,~\cite{Delves_Mohamed_1985}. First we note that the kernel has a pole singularity at $k' = q_{\alpha}$ (or $\sigma_k = \sigma_{\alpha}$) corresponding to the bound state of $\Mc_2$. To circumvent this pole, we use Cauchy's theorem to deform the integration path to a contour in the complex $k'$-plane like that presented in Ref.~\cite{Dawid:2023jrj}. We then follow the Nystr\"om method, which approximates the integral in Eq.~\eqref{eq:d:int} by a quadrature rule over discrete momenta in the integration interval. Let $N_k$ be the number of momenta appearing in one channel and $N_c$ be the total number of channels in $LSI$ space.~\footnote{In general, we can choose different contours and number of mesh points per channel. In this work, we choose that each channel has the same contour and mesh.} Choosing a sufficient contour, we replace the integral in Eq.~\eqref{eq:d:int} by a quadrature rule of order $N_k$. For some target $T(J^P)$, this leads to the approximate equation
\begin{align}
    d_{\beta\alpha}^{\,T(J^{P})}(p,k) = -\Gc_{\beta\alpha}^{T(J^{P})}(p,k) - \sum_{\gamma}\sum_{j=0}^{N_k-1} \overline{\Qc}_{\beta\gamma}^{T(J^{P})}(p,k_j)  \, d^{\,T(J^{P})}_{\gamma \alpha}(k_j,k) \, ,
    \label{eq:nystrom_interp}
\end{align}
where $k_j$ are the mesh points in momentum space, and the modified kernel is
\begin{align}
    \overline{\Qc}_{\beta\alpha}^{T(J^P)}(p,k_j) \equiv
    \Qc^{T(J^P)}_{\beta\alpha}(p, k_j)
    \frac{ k_{j}^2}{(2\pi)^2 \,\omega_{k_{j}}}\, \Delta_{j},
    \label{eq:measure}
\end{align}
with $\Delta_j$ absorbing the weights from the quadrature rule and the Jacobian from the chosen contour.~\footnote{We use Gauss-Legendre quadrature, which requires performing a variable transformation to relate the standard weights to the form shown in Eq.~\eqref{eq:measure}~\cite{Dawid:2023jrj}.} Finally, we evaluate $p$ and $k$ in Eq.~\eqref{eq:nystrom_interp} on the momentum partition $\{k_j\}_{j=0}^{N_k-1}$, transforming the integral equation~\eqref{eq:d:int} into a square linear algebraic system of order $N_c\times N_k$ in the combined channel and momentum space. 

Using standard computational linear algebra, we solve the approximate linear system,
\begin{align}
    \mathbf{d} &= -\mathbf{G}  - \mathbf{\bar{Q}}\cdot\mathbf{d} \, , \nn\\[5pt]
   &= -\left[\mathbf{1} + \mathbf{\bar{Q}} \right]^{-1}\cdot \mathbf{G} \, ,
   \label{eq:mat_sol}
\end{align}
where the solution $\mathbf{d}$ is a matrix in the combined momentum-channel space. Specifically,  $[\mathbf{d}]_{mn}$ represents an element of the matrix, where the index $m$ accesses the $i$-th momentum element of the $\beta$ channel, while $n$ maps to the $j$-th momentum of the $\alpha$ channel, \ie,
\begin{align}
    [\mathbf{d}]_{mn} = d_{\beta\alpha}^{T(J^P)}(k_i,k_j) \, .
\end{align}
Similarly, $[\mathbf{G}]_{mn} = \Gc^{T(J^P)}_{\beta\alpha}(k_i,k_j)$, and $[\mathbf{\bar{Q}}]_{mn} = \overline{\Qc}^{T(J^P)}_{\beta\alpha}(k_i,k_j)$ for the same $m,n$ mapping. Given a well-defined contour and quadrature rule, one can compute the solution $\mathbf{d}$ from Eq.~\eqref{eq:mat_sol} for moderate values of $N_k$. We have checked a range of $N_k$ between 30--500 and found convergence to our desired precision for most systems at $N_k\approx 150$. To obtain $d_{\beta\alpha}^{T(J^P)}(p,k)$ for values of $p$ and $k$ not in the momentum partition, as is needed when $p\to q_\beta$ and $k\to q_\alpha$ in Eqs.~\eqref{eq:Lcphib_v2},~(\ref{eq:Rcphib_v2}), and~(\ref{eq:Mphib_iso}), we use Eq.~\eqref{eq:nystrom_interp} as an interpolation formula.

Once a solution for $d^{T(J^P)}$ is obtained, we can compute all the contributions feed into the expression for $\Mc^{T(J^P)}_{\varphi b}$. In the next section, we summarize the parameterizations we use for the two-body sub-processes.

\subsection{Parameterizations considered}
\label{sec:toy_models}

Here we consider a simple class of parameterizations that can generate a two-body bound state for $S=0$ and $1$. We parameterize the amplitude via the phase shift $\delta_{S,I}$,
\begin{align}
    \Mc_{2,IS}(\sigma_k) = \frac{16\pi \sqrt{\sigma_k}}{q_k^\star \cot\delta_{S,I} - iq_k^\star} \, ,
\end{align}
where the symmetry factor of Eq.~\eqref{eq:rho_ps} is $1/2$ since the pions are treated as identical isovector states. As exploited extensively in previous work~\cite{Romero-Lopez:2019qrt, Jackura:2020bsk, Dawid:2023kxu}, an S-wave leading order effective range expansion (ERE) can be used to generate two-body bound states. As a result, we only consider parameterizations for the $S=0$ two-body amplitude defined by 
\begin{align}
    q^\star_k \cot\delta_{0,I} 
    &= -\frac{1}{a_{0,I}},
    \label{eq:ere}
\end{align}
where $a_{0,I}$ is the scattering length in the $I = 0$ or $2$ channel. For $a_{0,I} > 0$, the resulting two-body amplitudes would have a two-body bound state with real value residues.  

For a P-wave bound state, the use of a leading order ERE leads to unphysical residue for bound states in the amplitude.
Partial wave projected amplitudes near threshold must be kinematically suppressed by barrier factors,
\begin{align}
    \Mc_{2,IS}(\sigma_k)\propto
    q_k^{\star\,2S}.
\end{align}
Near the pole, the amplitude is proportional to $g_{\alpha}^2$, $g_\alpha$ being the bound state coupling of the $S=1,I=1$ channel, \cf Eq.~\eqref{eq:bs_amp}. 
In order for these two conditions to be simultaneously satisfied, the couplings must be imaginary,
\begin{align}
    g_\alpha^2\propto
    q_\alpha^{\star\,2S}=(i\kappa_\alpha)^{2S},
    \label{eq:anal_g}
\end{align}
where $\kappa_\alpha$ is the binding momentum, \ie, Eq.~\eqref{eq:pair_rel_mom} evaluated at $\sigma_k = \sigma_\alpha$. A simple exercise shows that this condition would not be respected by a P-wave bound state generated from a leading order ERE parametrization. 

Instead, for the isovector P-wave systems, we consider a Breit-Wigner (BW) parameterization   
\begin{align}
    q^\star_k \cot\delta_{1,1} 
    &=  \frac{(m_{\mathrm{BW}}^2-\sigma_k)}{\sqrt{\sigma_k}\,\Gamma_1^{\mathrm{BW}}(\sigma_k)}, 
    \qquad  \Gamma_1^{\mathrm{BW}}(\sigma_k)= \frac{g^{2}_{\mathrm{BW}}}{6\pi \sigma_k} q^{\star 2}_k.
    \label{eq:BW}
\end{align}
For carefully chosen values of $m_{\mathrm{BW}}$ and $g_{\mathrm{BW}}$, one can ensure that Eq.~\eqref{eq:anal_g} is satisfied for a P-wave bound state. 
It is worth noting that a BW near threshold is equivalent to a next-to-leading order ERE, that is the latter would also have been a reasonable choice.

\subsection{Numerical results}
\label{sec:results}

Building on the previous two sections, we consider a class of toy models for $3\pi$ systems where the $\rho$ and $\sigma$ are both stable. That is, we compute $\Mc_{\varphi b}^{T(J^P)}$ from Eq.~\eqref{eq:Mphib_iso} for both $\rho \pi$ and $\sigma\pi$ systems. As previously mentioned, this is a reasonable jumping-off point for analyses of lattice QCD results at unphysically heavy quark masses, where the $\rho$ and $\sigma$ are bound. Moreover, the techonology for computing $d^{T(J^P)}$ is identical for investigating systems at physical pion masses.

\begin{figure}[t!]
\begin{center}
\includegraphics[width=1\textwidth]{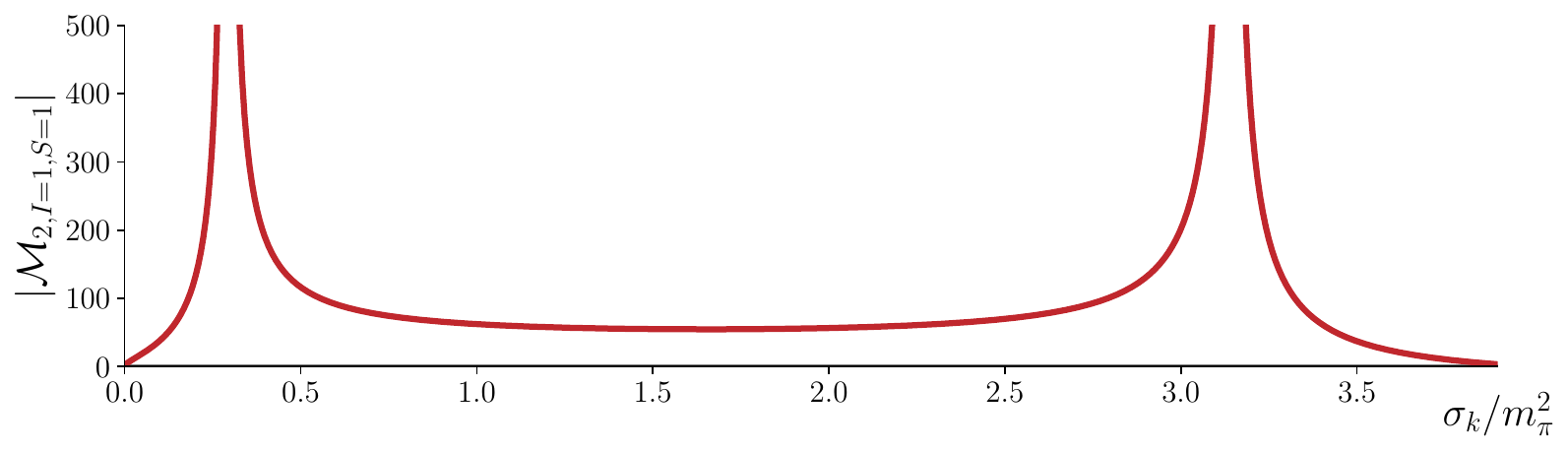}
\caption{Shown is the resulting P-wave amplitude using the Breit-Wigner parametrization, Eq.~\eqref{eq:BW}, for $m_{\mathrm{BW}}=1.8 \, m_\pi$ and $g_{\mathrm{BW}}=5.8$. The pole on the right of the figure is the desired physical $\rho$ pole, while the one on the left is the unphysical pole, which is avoided by introducing a hard cutoff at $\sigma_k= 0.5\,m_\pi^2$.
\label{fig:BW}
}
\end{center}
\end{figure}

We use the BW parametrization for the $I=1$ $\pi\pi$ amplitude, Eq.~\eqref{eq:BW}, and fixing the parameters to $m_{\mathrm{BW}} = 1.8\,m_\pi$ and $g_{\mathrm{BW}}=5.8$. This results in a bound state pole of $\sigma_\rho\approx 3.13 \, m_\pi^2$ with a binding momentum $\kappa_\rho\approx 0.46 \, m_\pi$, and a residue of $g_\rho\approx 4.88 \,i \, m_\pi$, \cf, Eq.~\eqref{eq:bs_amp}. This pole can be seen in Fig.~\ref{fig:BW}, which also shows a deeply bound unphysical pole. To avoid this unphysical state, we fix the hard cut off such that $\sigma_k\geq 0.5 \, m_\pi^2$.

For the $I = 0$ $\pi\pi$ amplitude, we use the leading order ERE in Eq.~\eqref{eq:ere}. For simplicity, we fix the $\sigma$ pole to lie at the same pole location as the $\rho$, that is $\sigma_\sigma = \sigma_\rho \approx 3.13\,m_\pi^2$. This is an arbitrary choice, but it reduces the number of kinematic thresholds to consider when visualizing the results. The associated scattering length, $a_{0,0}$, is determined by fixing the binding momentum of the $\sigma$ to that of the $\rho$, and using the fact that within the leading order ERE, the binding momentum is fixed by the scattering length, $\kappa_\sigma=1/a_{0,0}$. This results in $a_{0,0}\,m_\pi\approx 2.16$, and a residue at the pole of $g_\sigma\approx 18.18\,m_\pi$.

We also use the leading order ERE for the $I=2$ channel. Since this channel is always weakly repulsive regardless of the values of the quark masses~\cite{Dudek:2012gj,NPLQCD:2011htk, Rodas:2023gma}, we only consider negative values of the scattering length, $a_{0,2}$, that have a small magnitude. The first numerical exploration performed below is a demonstration that in the small $a_{0,2}$ limit, the $T=2$ $\rho\pi$ amplitudes are indistinguishable if one solves the coupled set of integral equations with or without the $I=2$ channel present.

In what follows, we limit the orbital momentum $L\leq 2$ and the spin of the dipion $S \leq 1$.  For simplicity, the $h$ vertex functions,  introduced in Eq.~\eqref{eq:factorized}, will be set exactly equal to the barrier factor in Eq.~\eqref{eq:barrier}, 
\begin{align}
      [h(p)]_{LSI}\,  = \Bc_{LS}(k,s)\, ,
\end{align}
which is the minimal requirement for this function. For the remainder piece of $\hKc_3$, we use a simple-pole
\begin{align}
    \left[\wt{\Kc}_3^{T(J^P)}(s)\right]_{\beta\alpha} = -\frac{c_\beta c_\alpha}{s-s_0} \, ,
    \label{eq:K3_param}
\end{align}
where we will vary $c_\beta$, $c_\alpha$ and $s_0$.

For simplicity, we only considers kinematics above the $\rho\pi$ threshold, $s_{\rho \pi}=(m_\rho+m_\pi)^2$ where $m_\rho = \sqrt{\sigma_\rho}$, but below the $3\pi$ threshold, $s_{3\pi}=(3m_\pi)^2$. In this region, S matrix unitarity ensures that 
\begin{align}
    \left[\left(\Mc^{T(J^P)}_{\varphi b}\right)^{-1}\right]_{\beta\alpha}
    =-\delta_{\beta\alpha}\frac{q_\alpha}{8\pi \sqrt{s}}.
    \label{eq:phib_unitarity}
\end{align}
Although this is not shown explicitly, we observe all numerical results satisfy this at sub-percent levels with our solution parameters.

\subsubsection{$T(J^P)=2(1^+)$ channel with stable $\rho$}

As our first example, we compute the $T=2$ channel with $J^P = 1^+$. This system can include with $I =1$ or $2$ $\pi\pi$ processes. If we consider partial waves restricted by $S \leq 1$ and $L \leq 2$, the possible values that $LSI$ can take include $S11$, $D11$, and $P02$, where we have used spectroscopic notation for $L$. If we instead use the more standard ${}^{2S+1}L_J$ notation for these states, the possible channels are ${}^3S_1$, ${}^3D_1$, and ${}^1P_1$, respectively. In this notation, although the isospin of the two-body system is not specified explicitly, it can be readily worked out. The matrix elements of $d^{2(1^+)}$ are then denoted $d({}^{2S'+1}L'_J | {}^{2S+1}L_J) \equiv d^{2(1^+)}_{L'S'I',LSI}$.

First, we consider the dependence of the results on the values of $a_{0,2}$. If $a_{0,2}\neq0$ and we fix $L\leq 2$, we have three open channels. To be more explicit, let us label the $I=2$ channel as $t\pi$, where $t$ refers to the isotensor $\pi\pi$ state. This will not be assumed to be stable. If we then fix the external momenta, we can write the symmetric $d^{\, 2 (1^+)}$ matrix as,
\begin{equation}
    d^{\,2 (1^+)}  =
    \begin{pmatrix}
        d^{}_{\rho \pi,\rho \pi}({}^3S_1|{}^3S_1) & d^{}_{\rho \pi,\rho \pi}({}^3S_1|{}^3D_1)& d^{}_{\rho \pi,t \pi}({}^3S_1|{}^1P_1)  \\
          & d^{}_{\rho \pi,\rho \pi}({}^3D_1|{}^3D_1)& d^{}_{\rho \pi,t \pi}({}^3D_1|{}^1P_1) \\
        && d^{}_{t \pi,t \pi}({}^1P_1|{}^1P_1)
    \end{pmatrix},
\end{equation}
where we are only showing the upper triangle of the symmetric matrix. Each element of this matrix is labeled by the particle content of the in/out state as well as the ${}^{2S+1}L_J$ quantum numbers. 

\begin{figure}[t!]
\begin{center}
\includegraphics[width=1\textwidth]{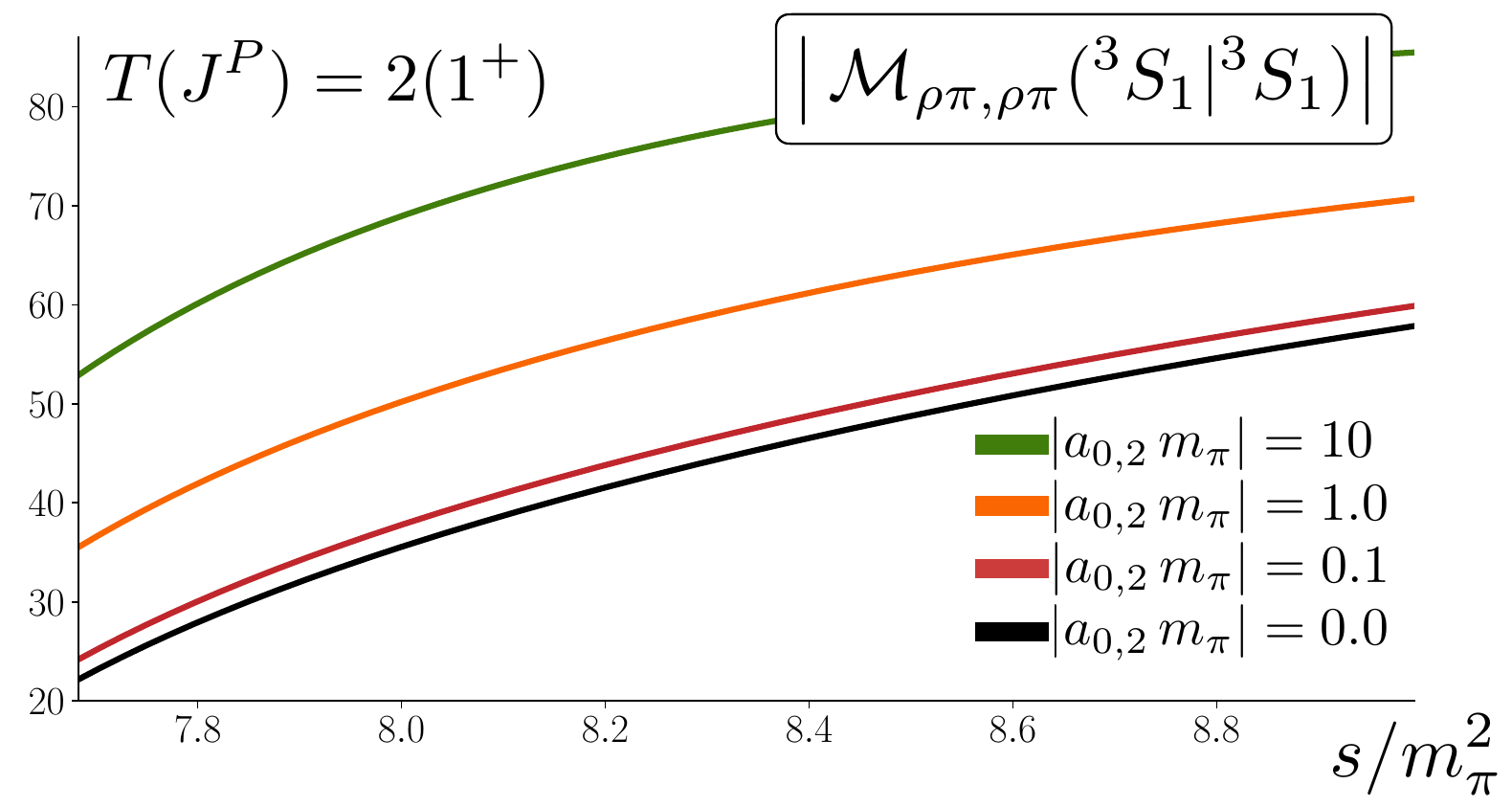}
\caption{Shown is the $\lvert \Mc^{}_{\rho \pi,\rho \pi}({}^3S_1|{}^3S_1)\rvert$ in the $T(J^P) = 2(1^+)$ channel as a function of $s$. The $I=1$ parameters are fixed to those described in the text, and we fix $\wt{\Kc}_3=0$. As labeled in the figure, the different colors represent different values for the scattering length in the isotensor channel, $a_{0,2}=-|a_{0,2}|$.
\label{fig:I2_as_dep}
}
\end{center}
\end{figure}

We obtain $d^{}_{\rho \pi,\rho \pi}({}^3S_1|{}^3S_1)$,  $d^{}_{\rho \pi,\rho \pi}({}^3S_1|{}^3D_1)$, $d^{}_{\rho \pi,\rho \pi}({}^3D_1|{}^3D_1)$ elements in two different ways. The first is by solving the coupled set of integral equations that the three-channel system satisfies for a non-zero value of $a_{0,2}$. The second approach, which holds for the $a_{0,2}=0$ limit, we solve the coupled-integral equations for a system with only $\rho\pi$ channels. In other words, the integral equations the block 
\begin{equation}
    d^{\,2 (1^+)}  =
    \begin{pmatrix}
        d^{}_{\rho \pi,\rho \pi}({}^3S_1|{}^3S_1) & d^{}_{\rho \pi,\rho \pi}({}^3S_1|{}^3D_1)  \\
          & d^{}_{\rho \pi,\rho \pi}({}^3D_1|{}^3D_1)
    \end{pmatrix}.
    \label{eq:T2_sub}
\end{equation}

\begin{figure}[t!]
\begin{center}
\includegraphics[width=.80\textwidth]{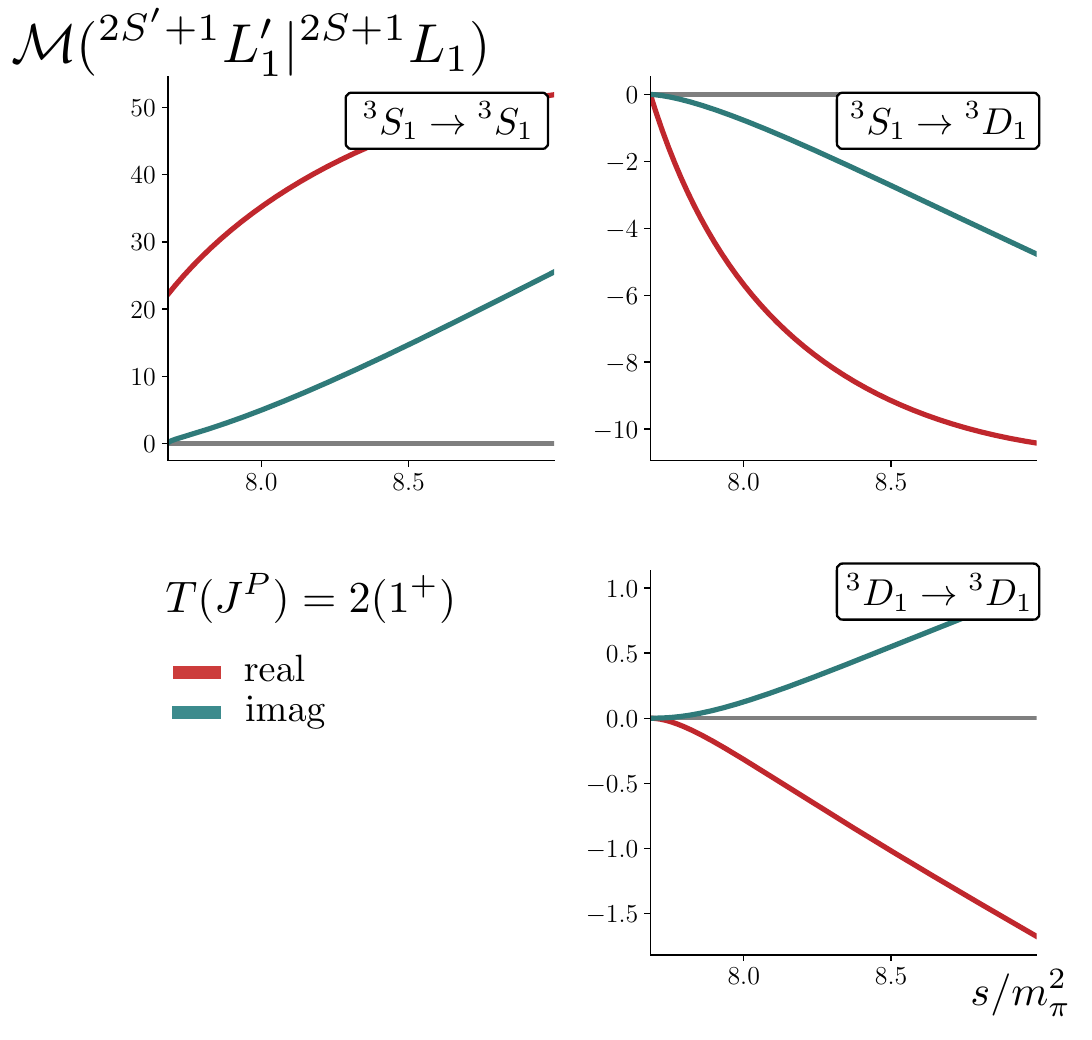}
\caption{Shown are the different matrix elements of the $T=2$ $\Mc_{\rho\pi,\rho\pi}$ amplitude for $\wt{\Kc}_3=0$ and ignoring contributions from the $I=2$ channel. The parameters are as described in the text. In red is the real part of the amplitude, while in dark cyan is the imaginary part of the amplitude.
\label{fig:I2_channel}
}
\end{center}
\end{figure}

In Fig.~\ref{fig:I2_as_dep}, we show the $\Mc^{}_{\rho \pi,\rho \pi}({}^3S_1|{}^3S_1)$ amplitude in the $\wt{\Kc}_3=0$ limit, i.e. $g_\rho^2 d^{}_{\rho \pi,\rho \pi}({}^3S_1|{}^3S_1)$ where we have fixed the spectator momenta to $q_{\rho\pi}= {\lambda^{1/2}(s,m^2_\pi,\sigma_\rho)}/{2\sqrt{s}} $. In the figure, we show results for a range of values of $a_{0,2}$. As can be seen, the results for small $a_{0,2}$ monotonically approach the $a_{0,2}=0$ results. This result is expected because the $t\pi$ contribution to $\rho\pi\to\rho\pi$ will be suppressed by at least one power of $a_{0,2}$.

Given this observation, we fix $a_{0,2}=0$ and consider a smaller channel space, Eq.~\eqref{eq:T2_sub}, throughout this and other cases. By then proceeding to fix $\wt{\Kc}_3=0$, we can predict the full $\Mc^{2 \, 1^+}_{\varphi b}$ matrix,
\begin{equation}
    \Mc^{2 (1^+)}_{\varphi b} =
    \begin{pmatrix}
        \Mc^{}_{\rho \pi,\rho \pi}({}^3S_1|{}^3S_1) & \Mc^{}_{\rho \pi,\rho \pi}({}^3S_1|{}^3D_1)  \\
          & \Mc^{}_{\rho \pi,\rho \pi}({}^3D_1|{}^3D_1)
    \end{pmatrix}.
\end{equation}
The results are shown in Fig.~\ref{fig:I2_channel} as a function of $s$. As previously mentioned, these results satisfy two-body unitary, Eq.~\eqref{eq:phib_unitarity}, in this kinematic region. Furthermore, one can see from the figure that the amplitudes satisfy the expected threshold behavior
\begin{align}
    \Mc^{}_{\rho \pi,\rho \pi}({}^3L'_1|{}^3L_1)\sim q_{\rho\pi}^{L'+L} \, , 
\end{align}
which serves as an additional cross-check for the partial wave projection.

\begin{figure}[t!]
\begin{center}
\includegraphics[width=1\textwidth]{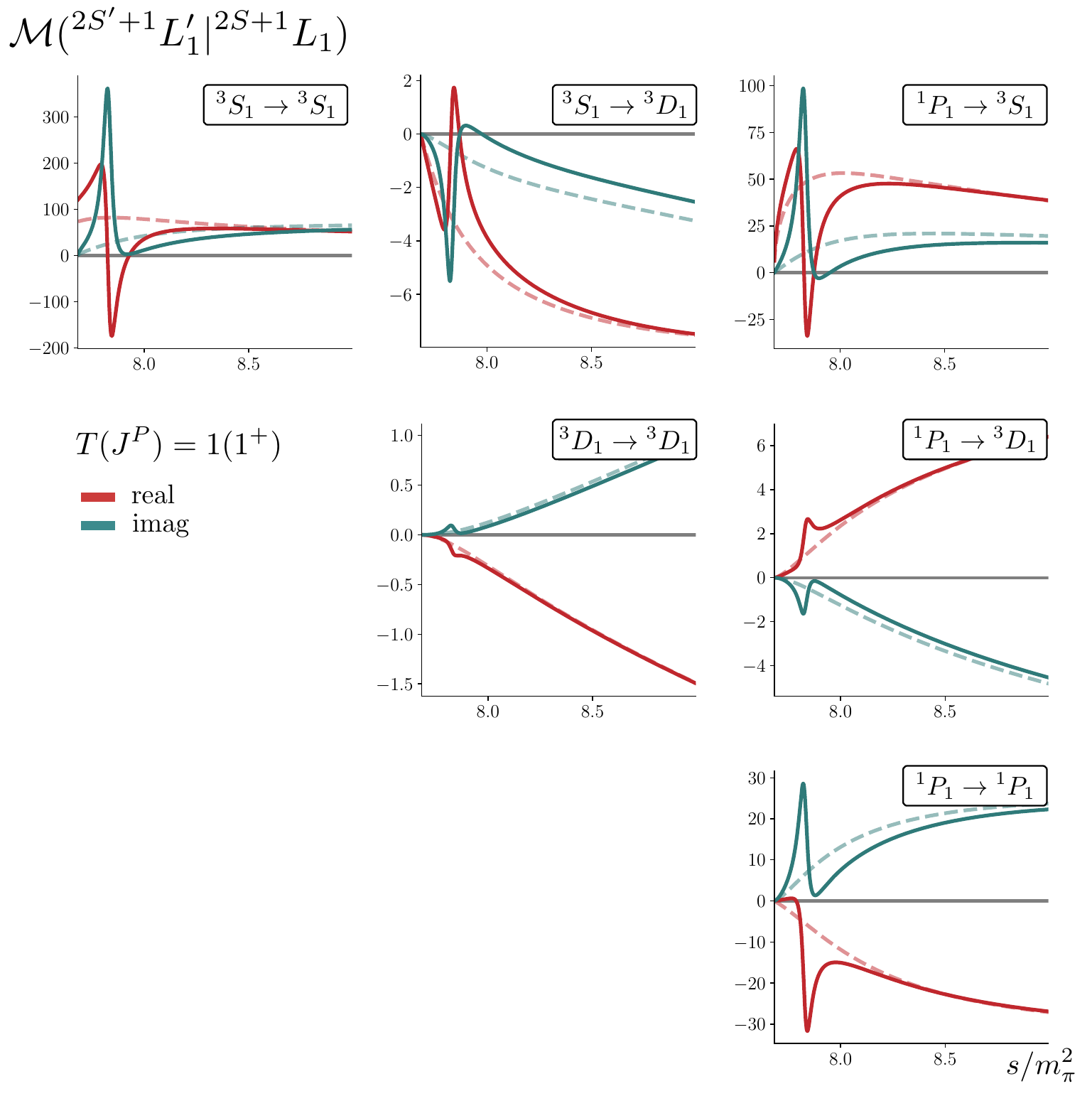}
\caption{Shown are results for the various components of the $T(J^P)=1(1^+)$ amplitude. This amplitude has the quantum numbers of the $a_1$, and the flavor content of the different channels can be found in Eq.~\eqref{eq:Ma1}. The dashed lines denote the amplitude in the limit that $\wt{\Kc}_3=0$, while the solid lines include a pole parametrization for $\wt{\Kc}_3$ described in the body of the text. 
\label{fig:a1_channel}
}
\end{center}
\end{figure}
%

\subsubsection{$T(J^P)=1(1^+)$ channel with stable $\sigma$ and $\rho$}
We examine the $T(J^P)=1(1^+)$ system, where the $a_1$ resonance resides. Ignoring the $t\pi$ contribution and restricting to $L\leq 2$, the $a_1$ can couple to three channels, two $\rho\pi$ and one $\sigma\pi$ partial waves. As a result, the matrix for $\Mc^{1(1^+)}_{\varphi b}$ can be written as
\begin{equation}
    \Mc^{1 (1^+)}_{\varphi b} =
    \begin{pmatrix}
        \Mc^{}_{\rho \pi,\rho \pi}({}^3S_1|{}^3S_1) & \Mc^{}_{\rho \pi,\rho \pi}({}^3S_1|{}^3D_1)& \Mc^{}_{\rho \pi,\sigma \pi}({}^3S_1|{}^1P_1)  \\
          & \Mc^{}_{\rho \pi,\rho \pi}({}^3D_1|{}^3D_1)& \Mc^{}_{\rho \pi,\sigma \pi}({}^3D_1|{}^1P_1) \\
        &&\Mc^{}_{\sigma \pi,\sigma \pi}({}^1P_1|{}^1P_1)
    \end{pmatrix}.
    \label{eq:Ma1}
\end{equation}

In order to mimic the $a_1$ resonance, we parameterize $\wt{\Kc}_3$ with a simple pole according to Eq.~\eqref{eq:K3_param}, with a pole position $s_{\rho\pi}<s_0<s_{3\pi}$. Figure~\ref{fig:a1_channel} shows results where we fix $s_0=8\,m_\pi^2$, $c_{\rho \pi,{}^3S_1}=20\,m_\pi^2$, $c_{\rho \pi,{}^3D_1}=c_{\sigma \pi,{}^1P_1}=5\,m_\pi^2$. For comparison, we show the amplitudes for $\wt{\Kc}_3=0$ as faint dashed lines. In addition to checking that the amplitudes have the right analytic structure and that they satisfy unitarity, we see the canonical behavior of a narrow resonance. In particular, one sees a narrow peak. Naively, one would expect such a peak at $s=8\,m_\pi^2$, since this is the location of the $\wt{\Kc}_3$ pole. However, poles in the K matrices do not coincide with poles in amplitude.~\footnote{Although we do not do the exercise here, using the tools presented in Ref.~\cite{Dawid:2023jrj}, we could analytically continue to the nearest unphysical sheet to find the resonance pole. }

\subsubsection{$T(J^P)=0(1^-)$ channel with stable $\rho$  }

\begin{figure}[t!]
\begin{center}
\includegraphics[width=1\textwidth]{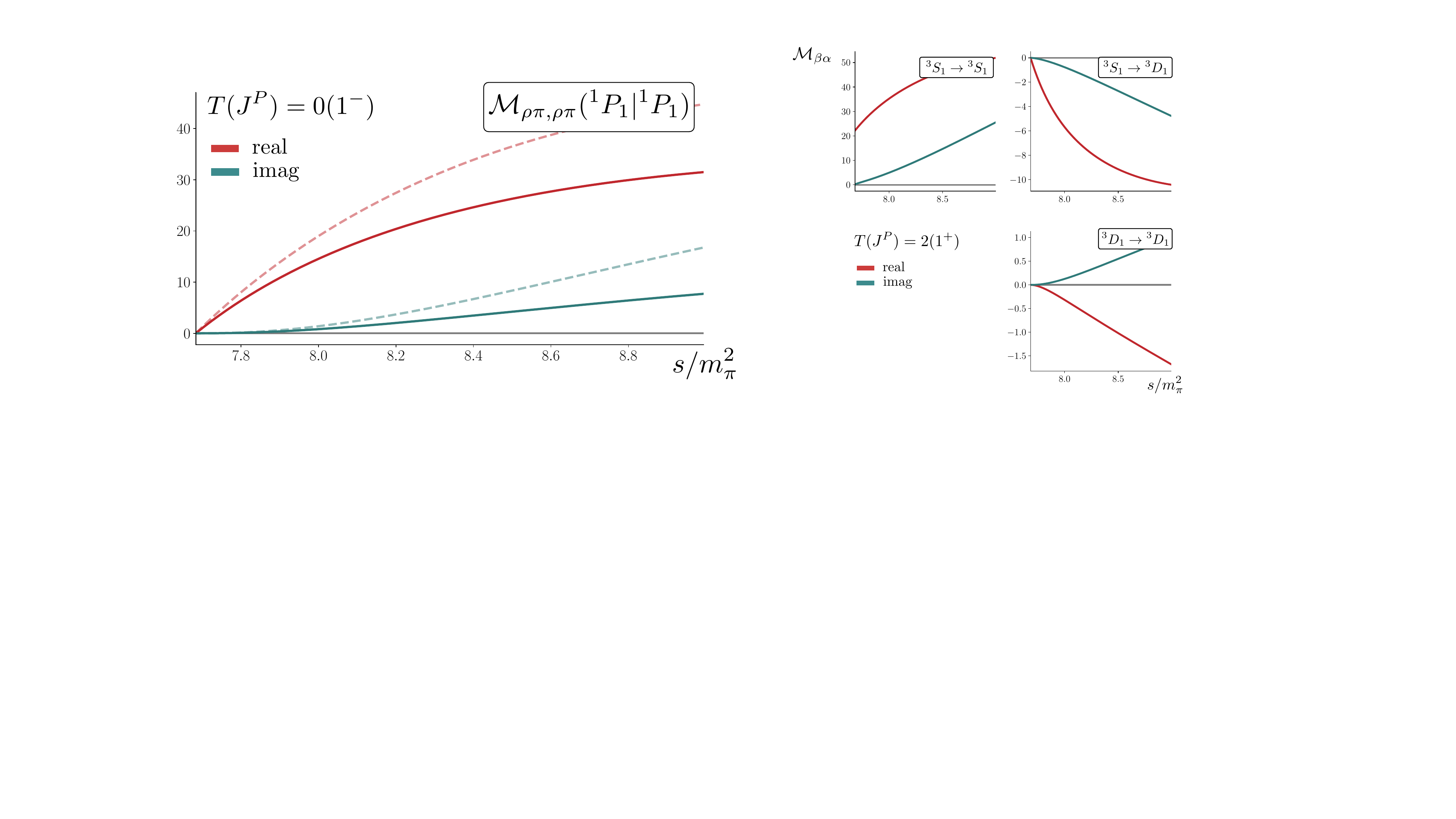}
\caption{ Shown are the real (red) and imaginary (cyan) components of the amplitude in the $T(J^P) = 0(1^-)$ channel. 
As in Fig.~\ref{fig:a1_channel}, dashed and solid lines depict amplitudes with $\wt{\Kc}_3=0$ and $\wt{\Kc}_3\neq0$, respectively. The parameters for $\wt{\Kc}_3\neq0$ are described in the text.  
\label{fig:omega_channel}
}
\end{center}
\end{figure}

Finally, we consider the $T(J^P)=0(1^-)$ channel where the narrow $\omega$ resonance lies. For unphysically heavy quark masses, the $\omega$ is observed to be bound. To mimic this scenario, we use a $\wt{\Kc}_3 $ parameterization with a bound state below the $\rho\pi$ threshold. We use same simple pole parametrization Eq.~\eqref{eq:K3_param} as above but with $s_0<s_{\rho\pi}$.

Assuming the same restrictions in the partial waves as previously discussed, the scattering amplitude in this system is composed of a single channel, the ${}^3P_1$. As a result, we have 
\begin{equation}
    \Mc^{0 (1^-)}_{\varphi b} =
         \Mc^{}_{\rho \pi,\rho \pi}({}^3P_1|{}^3P_1) .
 \end{equation}
In Fig.~\ref{fig:omega_channel}, we show the result for this amplitude for both $\wt{\Kc}_3=0$ (dashed lines) and $\wt{\Kc}_3\neq 0$ (solid lines). For the latter case, we set the pole and coupling of the K matrix to $s_0=7.6\,m_\pi^2$ and $c_{\rho \pi,{}^1P_1}=90\,m^2_\pi$, respectively. Again, we see the expected threshold behavior for a P-wave amplitude, and unitarity is well satisfied for both examples shown.

\section{Summary and outlook}

Using the results derived in Ref.~\cite{Jackura:2023qtp} for the OPE, we have constructed integral equations for partial-wave projected three-body relativistic scattering amplitudes. The integral equations are presented in Sec.~\ref{sec:PWA}, with details presented in Appendix~\ref{app:conv}, for two equivalent formalisms where the three-body K matrix is symmetric or asymmetric under particle interchange. While a pracitioner can choose either framework for analyses, we advocate for the asymmetric formalism due to the relative ease for parameterizing the three-body K matrix, which is illustrated in our numerical applications in Sec.~\ref{sec:numerical}. In particular, a class of flexible parameterizations useful for data analysis are presented in Sec.~\ref{sec:separable}, where $\Kc_3$ is factorizable in the kinematics of the initial and final state allowing one to parameterize $\Kc_3$ in a manner similar to analyses in the two-body sector. In addition, it is shown that factorizable parameterizations for $\Kc_3$ reduce the computational complexity of the integral equations.

In Sec.~\ref{sec:LSZ}, we consider the scenario where one of the two-particle pairs forms a bound state. In particular, we show how the LSZ formalism can be used to reconstruct two-body amplitudes from the three-body amplitudes. Finally, in Sec.~\ref{sec:numerical}, we explore numerical solutions of the three-body amplitudes for toy models for the $3\pi$ channels in total isospin $T=2,1,0$ that include the $\rho$ and $\sigma$ as $\pi\pi$ bound states. Using the expressions derived in Sec.~\ref{sec:LSZ} for the asymmetric formalism, we find that our results satisfy two-body unitarity below the three-body threshold for all models considered, as well as the expected threshold behavior for partial-wave projected amplitudes. 

Given the rapid developments in this line of research, it is worthwhile summarizing some key outstanding problems related to the scattering theory of three-body systems. First, in this work, we considered amplitudes that have not been symmetrized under the interchange of the spectator. Using the notation used in the literature, these are the $\Mc_3^{(u,u)}$. The next step is to symmetrize these and construct amplitudes that may be used to generate Dalitz plots, as was done, for example, in Ref.~\cite{Hansen:2020otl}, for the $T(J^P)=3(0^-)$ lattice QCD calculation. Using formalism presented in, for example, Ref.~\cite{Jackura:2018xnx}, we believe this should be straightforward. As already mentioned, the formalism presented here is built from the partial wave projection of the OPE performed in Ref.~\cite{Jackura:2023qtp}, which only assumed that the particles involved had no intrinsic spin. Lifting such an assumption, although technical, can and will be done. Additionally, including coupled two- and three-particle systems, while attempted in Ref.~\cite{Briceno:2017tce}, requires further investigation. Going beyond these immediate problems, one can envision formulating dispersive representations for the three-body amplitudes. As the connection between scattering theory and lattice QCD matures for few-body systems, preserving S matrix principles such as unitarity and analyticity is vital to exploring the excited QCD spectrum.

\section*{Acknowledgements}

The authors would like to thank S. Dawid, M. Hansen, and F. Romero-L\'opez for their useful comments and discussions. This work was partly supported by the  U.S. Department of Energy (DOE) contract DE-AC05-06OR23177, under which Jefferson Science Associates LLC manages and operates Jefferson Lab. RAB was supported in part by the U.S. Department of Energy, Office of Science, Office of Nuclear
Physics under Awards No. DE-AC02-05CH11231 and No. DE-SC0019229. AWJ acknowledges the support of the USDOE ExoHad Topical Collaboration, contract DE-SC0023598.

\begin{appendix}

\section{Partial wave projection}
\label{app:conv}

We follow the procedure as presented in Refs.~\cite{Jackura:2018xnx,Jackura:2023qtp} and further references therein, namely first projecting to definite helicity amplitudes of definite $J$, then forming definite $J^P$ amplitudes by taking linear combinations to the $LS$ basis. 

Given the three body  helicity  amplitude $\mathcal{M}_{3;\ell'
\lambda',\ell\lambda}$, one can expand it in terms of amplitudes of definite total angular momentum $J$ of the three particle system:
\begin{align}
     \Mc_{3;\ell'\lambda',\ell\lambda}(\p,\k)  
    &=\sum_{J = J_{\min}}^{\infty} (2J+1) \Mc_{3;\ell'\lambda',\ell\lambda}^{J}(p,k) \, d_{\lambda\lambda'}^{(J)}(\theta_{pk}) \, ,
    \label{eq:M3:hel:amp}
\end{align}
where $J_{\min} = \max(\lvert\lambda\rvert,\lvert\lambda'\rvert)$, $d$ is the Wigner matrix elements,~\footnote{Note this $d$ is not to be confused with the amputated ladder amplitude, Eq.~(\ref{def:D:ladder}), that has been defined in Sec.~\ref{sec:DcPWA}.} and $\theta_{pk}$ is the CM frame scattering angle $\theta_{pk}$, defined through
$\cos\theta_{pk} = \bh{p}\cdot \bh{k}$.  Using the orthogonality relation for Wigner $d$ matrices, we use Eq.~(\ref{eq:M3:hel:amp}) to project the helicity amplitudes to definite angular momentum $J$,
\begin{align}
\label{eq:Jbasis}
\Mc_{3;\ell'\lambda',\ell\lambda}^{J}(p,k) = \frac{1}{2}\int_{-1}^{1} \mathrm{d}\cos\theta_{pk}\,d^{J}_{\lambda\lambda'}(\theta_{pk}) \Mc_{3;\ell'\lambda',\ell\lambda}(\mathbf{p},\mathbf{k}).
\end{align}

As discussed in detail in Ref.~{\cite{Jackura:2023qtp}}, one can transform from the helicity-state basis, whose corresponding partial waves do not have definite parity, to the spin-orbit state basis using the spin-orbit coupling coefficients $ \Pc_{\lambda}^{(\ell)}$ to be defined below. This allows one to obtain amplitudes that have both definite angular momentum and parity, which can be written as
\begin{align}
    \Mc_{3;L'S',LS}^{(\ell'\ell),J^{P}}(p,k) = \sum_{\lambda',\lambda} \Pc_{\lambda'}^{(\ell')}(^{2S'+1}L'_J) \, \Mc_{3;\ell'\lambda',\ell\lambda}^{J}(p,k) \, \Pc_{\lambda}^{(\ell)}(^{2S+1}L_J) \, .
    \label{eq:m3:J:LS}
\end{align}
For particles with no intrinsic spin, the spin-orbit coupling coefficient is given by~\cite{Jacob:1959at} 
\begin{align}
    \Pc_{\lambda}^{(\ell)}(^{2S+1}L_J) = \sqrt{\frac{2L+1}{2J+1}} \braket{J\lambda | L 0 , S \lambda} \delta_{S \ell} \, .
    \label{eq:spin_orbit}
\end{align}
The inner product in brackets is the usual Clebsh-Gordan coefficient, which relates different complete sets of the combined system. Note that the spin-orbit couplings are orthonormal,
\begin{align}
    \sum_{\ell,\lambda} \Pc_\lambda^{(\ell)}(^{2S'+1}L'_J) \Pc_\lambda^{(\ell)}(^{2S+1}L_J) = \delta_{L'L} \delta_{S'S} \, .
\end{align}
For spinless particles, we have identically $S = \ell$ as enforced by the Kronecker $\delta_{S\ell}$ appearing in Eq.~\eqref{eq:spin_orbit}, making the superscripts in Eq.~\eqref{eq:m3:J:LS} redundant. Therefore, we adopt the simpler notation, 
\begin{align}
    \Mc_{3;L'S',LS}^{J^{P}}(p,k) = \sum_{\lambda',\lambda} \Pc_{\lambda'}^{(\ell')}(^{2S'+1}L'_J) \, \Mc_{3;\ell'\lambda',\ell\lambda}^{J}(p,k) \, \Pc_{\lambda}^{(\ell)}(^{2S+1}L_J) \, .
    \label{eq:M3:J:LSvf}
\end{align}

In summary, one can use  Eq.~\eqref{eq:Jbasis} to first project to definite $J$ and then Eq.~(\ref{eq:m3:J:LS}) to project the subsequent amplitude to the $JLS$ basis with definite parity. Recall that $\Mc_3$ can be written as the sum of two terms: The ladder amplitude, $\Dc$, and a divergent free amplitude, $\Mc_{3,\mathrm{df}}$ (or the asymmetric $\hMc_{3,\mathrm{df}}$). We now apply the procedure outlined above to write the partial wave projection of these different contributions to the three-particle scattering amplitude.

\subsection{Partial-wave projected $\Dc$}
\label{app:Dc}

We start with the partial wave projection of $\Dc$. For convenience, we repeat here the expression for $\Dc$ in the helicity basis given in Eq.~\eqref{eq:Dc_hel}:
\begin{align}
\mathcal{D}_{\ell'\lambda',\ell\lambda}({\bf p},{\bf k})  &= -   \Mc_{2,\ell'}(\sigma_p)\,\mathcal{G}_{\ell'\lambda,\ell\lambda}(\mathbf{p},\mathbf{k})\,\Mc_{2,\ell}(\sigma_k) 
\nn \\[5pt]
    & \qquad - \Mc_{2,\ell'}(\sigma_p)
    \sum_{{\ell}_1,{\lambda}_1}\int\, \frac{\mathrm{d}^3 \textbf{k}'}{(2\pi)2\omega_{k'}}\mathcal{G}_{\ell'\lambda',\ell_1\lambda_1}(\textbf{p},{\bf k'})\,\mathcal{D}_{\ell_1\lambda_1,\ell\lambda}({\bf k'},{\bf k}). 
\label{eq:int-eq-bigD}
\end{align}
Using Eq.~\eqref{eq:Jbasis}, the partial wave projection of $\Dc$ to definite angular momentum $J$ can be obtained by integrating over the angular dependence
\begin{align}
\mathcal{D}_{\ell'\lambda',\ell\lambda}^{J}(p,k) 
&= \frac{1}{2}\int_{-1}^{1} \mathrm{d}\cos\theta_{pk}\,d^{J}_{\lambda\lambda'}(\theta_{pk}) \mathcal{D}_{\ell'\lambda',\ell\lambda}(\mathbf{p},\mathbf{k})
\\
&= -   \Mc_{2,\ell'}(\sigma_p)\,\mathcal{G}^{J}_{\ell'\lambda,\ell\lambda}(p, k)\,\Mc_{2,\ell}(\sigma_k) 
\nn \\[5pt]
    & \qquad - \Mc_{2,\ell'}(\sigma_p)
    \sum_{{\ell}_1,{\lambda}_1}\int\, \frac{\mathrm{d}k'\, k'^2}{(2\pi)^2\,2\omega_{k'}}\mathcal{G}^{J}_{\ell'\lambda',\ell_1\lambda_1}(p,k')\,\mathcal{D}^{J}_{\ell_1\lambda_1,\ell\lambda}(k',k)
.
\end{align}
The explicit form of $\mathcal{G}^{J}_{\ell'\lambda',\ell\lambda}$ can be found in Eq.~(57) of Ref.~\cite{Jackura:2023qtp}.
Next, Eq.~\eqref{eq:M3:J:LSvf} is used to obtain the desired amplitudes of definite total angular momentum and parity that we are interested in this work,
\begin{align}
    \mathcal{D}_{L'S',LS}^{J^{P}}(p,k) 
    &= \sum_{\lambda',\lambda} \Pc_{\lambda'}^{(\ell')}(^{2S'+1}L'_J) \, \mathcal{D}_{\ell'\lambda',\ell\lambda}^{J}(p,k) \, \Pc_{\lambda}^{(\ell)}(^{2S+1}L_J)
    \nn\\
    &= -   \mathcal{M}_{2,S'}(\sigma_p)\,\mathcal{G}^{J^{P}}_{L'S',LS}(p, k)\,\mathcal{M}_{2,S}(\sigma_k) 
\nn \\[5pt]
    & \qquad - \mathcal{M}_{2,S'}(\sigma_p)
    \sum_{L_1,S_1}\int\, \frac{\mathrm{d}k'\, k'^2}{(2\pi)^2\,2\omega_{k'}}\mathcal{G}^{J^{P}}_{L'S',L_1S_1}(p,k')\,\mathcal{D}^{J^{P}}_{L_1 S_1,LS}(k',k) \, .
\end{align}
Here, $\mathcal{G}^{J^{P}}$ is the partial wave projected OPE explicitly given in Eq.~(\ref{eq:G:OPE}) in Sec.~{\ref{sec:PWA}}.

\subsection{Partial wave projection of asymmetric formalism}
\label{app:asym_hel}

As discussed in the main body of this work, there are two equivalent classes of formalism for `divergent free' part of $\Mc_3$. We first discuss in some detail the partial wave projection of $\wh{\Mc}_{3,\df}(\p,\k)$, which appears in the asymmetric formalism. In the helicity basis~\cite{Jackura:2022gib}, this object can be written as   
\begin{align}
    \wh{\Mc}_{3,\df;\ell'\lambda',\ell\lambda}(\p,\k) = \sum_{\ell_{1},\lambda_{1}}\sum_{\ell '_1,\lambda '_1}\int_{\p'} \int_{\k'}
    \wh{\Lc}_{\ell'\lambda',\ell_{1}\lambda_{1}}(\p , \p') \, 
    \widehat{\Tc}_{\ell_{1}\lambda_{1}\ell '_1,\lambda '_1}(\p' , \k') \,
    \wh{\Rc}_{\ell '_1\lambda '_1,\ell\lambda}(\k' , \k),
    \label{eq:Mcdf_hel}
\end{align}
where we introduce the following compact notation for the three-dimensional integral~\footnote{This is not to be confused with Eq.~\eqref{eq:radial_int}, where only the magnitude of the momentum is being integrated over.}
\begin{align}
\int_{\k}\equiv\int
\frac{\mathrm{d}^3 \textbf{k}}
{(2\pi)^{3}\,2\omega_{k}}.\end{align}
Using this notation, the different building blocks are defined by
\begin{align}
    \wh{\Lc}_{\ell'\lambda',\ell\lambda}(\p,\k) & = (2\pi)^3\,2\omega_k \,\delta^{(3)}(\p - \k) \,  \delta_{\ell'\ell} \delta_{\lambda'\lambda} - \Mc_{2}^{(\ell')}(\sigma_p) \, \Gamma_{\ell'\lambda',\ell\lambda}(\p,\k) \nn\\[5pt]
    & \qquad  - \sum_{\ell_{1},\lambda_{1}} \int_{\k'}\Dc_{\ell'\lambda',\ell_1\lambda_1}(\p,\k')\,\Gamma_{\ell_1\lambda_1,\ell\lambda}(\k',\k),
\label{eq:L:cap_asymm}
\\
\wh{\Rc}_{\ell'\lambda',\ell\lambda}(\p,\k) & = (2\pi)^3\,2\omega_k \,\delta^{(3)}(\p - \k) \,  \delta_{\ell'\ell} \delta_{\lambda'\lambda} -  \Gamma_{\ell'\lambda',\ell\lambda}(\p,\k) \, \Mc_{2}^{(\ell)}(\sigma_k)  \nn\\[5pt]
    & \qquad  - \sum_{\ell_{1},\lambda_{1}} \int_{\p'} \Gamma_{\ell'\lambda',\ell_1\lambda_1}(\p,\p') \, \Dc_{\ell_1\lambda_1,\ell\lambda}(\p',\k),
    \label{eq:R:cap_asymm}
\end{align}
where
\begin{align}
    \Gamma_{\ell'\lambda',\ell\lambda}(\p,\k) = (2\pi)^3\,2\omega_k \delta^{(3)}(\p - \k) \, \delta_{\ell'\ell} \delta_{\lambda'\lambda}\,
    \wt{\rho}
    (\sigma_k) + \Gc_{\ell'\lambda',\ell\lambda}(\p,\k).
    \label{eq:Gamma_hel}
\end{align}
 Finally, the remaining object appearing in Eq.~\eqref{eq:Mcdf_hel} needing to be defined is $\widehat{\Tc}$,
\begin{align}
    \widehat{\Tc}_{\ell'\lambda',\ell\lambda}(\p,\k) &= \widehat{\Kc}_{3;\ell'\lambda',\ell\lambda}(\p,\k) \nn \\[5pt] 
    &\hspace{-.9cm} - \sum_{\ell_{1},\lambda_{1}}\sum_{\ell '_1,\lambda '_1} \sum_{\ell_2,\lambda_2} \int_{\p'} \int_{\q'} \int_{\k'} 
    \widehat{\Kc}_{3;\ell'\lambda',\ell_{1}\lambda_{1}}(\p,\p')
    \Gamma_{\ell_{1}\lambda_{1},\ell '_1\lambda '_1}(\p',\q')
    \wh{\Lc}_{\ell '_1\lambda '_1,\ell_2\lambda_2}(\q',\k') \, 
    \widehat{\Tc}_{\ell_2\lambda_2,\ell\lambda}(\k',\k).
\end{align}
Here we note that there is a freedom in the definition of $\wt{\rho}$. Throughout the main body of this work, we use the minimal definition, Eq.~\eqref{eq:rhotilde}. In Appendix~\ref{app:shifting}, we discuss the possible shifts in the definition $\wt{\rho}$, as well as to why this might be necessary.

We can apply the above mentioned procedure to obtain the partial wave projection of the different building blocks. For a given functional form of $\wh{\Kc}_3$, one can use the analogues of Eqs.~\eqref{eq:Jbasis}, \eqref{eq:M3:J:LSvf}, to project this to definite $J$ and subsequently to the $JLS$ basis
\begin{align}
     \label{eq:K3Jbasis}
\wh{\Kc}_{3;\ell'\lambda',\ell\lambda}^{J}(p,k) &= \frac{1}{2}\int_{-1}^{1} \mathrm{d}\cos\theta_{pk}\,d^{J}_{\lambda\lambda'}(\theta_{pk}) \,  \wh{\Kc}_{3;\ell'\lambda',\ell\lambda}(\mathbf{p},\mathbf{k}) \, ,
    \\
\wh{\Kc}_{3;L'S',LS}^{J^{P}}(p,k) &= \sum_{\lambda',\lambda} \Pc_{\lambda'}^{(\ell')}(^{2S'+1}L'_J) \, \wh{\Kc}_{3;\ell'\lambda',\ell\lambda}^{J}(p,k) \, \Pc_{\lambda}^{(\ell)}(^{2S+1}L_J) \, .
    \label{eq:K3:J:LSvf}
\end{align}

The only element that remains to be partial-wave projected is the term proportional to the $\delta$ function appearing in the two rescattering functions defined in Eqs.~(\ref{eq:L:cap_asymm}) and~(\ref{eq:R:cap_asymm}). To do this, we write the $\delta$ function in the spherical basis,
\begin{align}
    \delta^{(3)}(\p-\k) = \frac{\delta(p-k)}{k^2} \, \delta^{(2)}(\Omega_p - \Omega_k), 
\end{align}
where as defined in Ref.~\cite{Jackura:2023qtp}, $\Omega_k$ is the solid angle of the pair in the CM frame, \ie the vector $-\bh{\k}$. We then use the the completeness relation for Wigner D matrix to rewrite the angular part of the $\delta$ function as
\begin{align}
    \delta^{(2)}(\Omega_p - \Omega_k) 
    &= \sum_{J,m_J,\lambda} \frac{2J+1}{4\pi}\, D_{m_J\lambda}^{(J)}(\Omega_p) D_{m_J\lambda}^{(J)\,*}(\Omega_k) \, ,
    \nn\\
    &= \sum_{J,\lambda} \frac{2J+1}{4\pi}\, d_{\lambda\lambda}^{J}(\theta_{pk}) \, ,
    \label{eq:complete}
\end{align}
where we have used the fact that $ \sum_{m_J = -J}^{J} D_{m_{J}\lambda'}^{(J)\,*}(-\bh{p}) \, D_{m_{J}\lambda}^{(J)}(-\bh{k})  = d_{\lambda\lambda'}^{(J)}(\theta_{pk}) \,$ with $\cos\theta_{pk}=(-\bh{k})\cdot (-\bh{p})$. 
With this, the $\delta$ function term appearing in Eqs.~\ref{eq:L:cap_asymm} and \ref{eq:R:cap_asymm} becomes
\begin{align}
    (2\pi)^32\omega_k \delta^{(3)}(\p-\k) & = (2\pi)^32\omega_k\frac{\delta(p-k)}{k^2} \, \delta^{(2)}(\Omega_p - \Omega_k) \, , \nn\\[5pt]
    & =  \frac{(2\pi)^2 \omega_k}{k^2} \delta(p-k) \sum_{J,\lambda} (2J+1)\, d_{\lambda\lambda}^{J}(\theta_{pk})
    \label{eq:delta_d}
\end{align}

The first equality of Eq.~\eqref{eq:complete} is easy to prove using the orthogonality relation of the Wigner D matrices,
\begin{align}
    D_{m_J \lambda}^{(J)}(\Omega_p) & = \int\!\diff \Omega_k \, \delta(\Omega_p - \Omega_k) \, D_{m_J \lambda}^{(J)}(\Omega_k) \, , \\[5pt]
    & = \int\!\diff \Omega_k \, \sum_{J',m_J',\lambda'} \frac{2J+1}{4\pi}\, D_{m_J'\lambda'}^{(J')}(\Omega_p) D_{m_J'\lambda'}^{(J')\,*}(\Omega_k) \, D_{m_J \lambda}^{(J)}(\Omega_k) \, , \\[5pt]
    & =  \sum_{J',m_J',\lambda'} \frac{2J+1}{4\pi}\, D_{m_J'\lambda'}^{(J')}(\Omega_p) \int\!\diff \Omega_k \, D_{m_J'\lambda'}^{(J')\,*}(\Omega_k) \, D_{m_J \lambda}^{(J)}(\Omega_k) \, , \\[5pt]
    & =  \sum_{J',m_J',\lambda'} \frac{2J+1}{4\pi}\, D_{m_J'\lambda'}^{(J')}(\Omega_p) \, \left( \frac{4\pi}{2J+1} \delta_{JJ'}\delta_{m_Jm_J'}\delta_{\lambda\lambda'} \right) \, , \\[5pt]
    & = D_{m_J\lambda}^{(J)}(\Omega_p) \, ,
\end{align}
where we emphasize the the normalization of the Wigner D matrices is due to the fact that we have integer spin systems and describe the particle orientation by only a polar and azimuthal angle. Using Eq.~\eqref{eq:delta_d}, and the analogues of Eqs.~\eqref{eq:Jbasis}, \eqref{eq:M3:J:LSvf}, we can project the $\delta$ function and consequently the rescattering functions for the $J$ and the $JLS$ basis. For example, the $\Gamma$ function appearing in Eq.~\eqref{eq:Gamma_hel} is projected to these two basis as
\begin{align}
    \Gamma^{J}_{\ell'\lambda',\ell\lambda}(p,k) &= \frac{(2\pi)^2 \omega_k}{k^2} \delta(p-k) \, \delta_{\ell'\ell} \delta_{\lambda'\lambda}\,
    \wt{\rho}
    (\sigma_k) + \Gc^{J}_{\ell'\lambda',\ell\lambda}(p,k) \, ,
    \label{eq:Gamma_Jhel}
    \\
    \Gamma^{J}_{L'S',LS}(p,k) &= \frac{(2\pi)^2 \omega_k}{k^2} \delta(p-k) \, \delta_{L'L} \delta_{S'S}\,
    \wt{\rho}
    (\sigma_k) + \Gc^{J}_{L'S',LS}(p,k) \, .
    \label{eq:Gamma_PW}
\end{align}
Using these results, one then arrives to Eqs.~\eqref{eq:Lcap:Jp_asym}, \eqref{eq:Rcap:Jp_asym}, and \eqref{eq:T_eq_asym} for the partial-wave projected $\wh{\Lc}$,  $\wh{\Rc}$, and  $\wh{\Tc}$ functions.

\subsection{Partial wave projection of symmetric formalism}

The partial wave projection for the symmetric formalism is identical to the asymmetric formalism. This is because the angular-depencence of these two formalisms are encoded in the same building blocks, namely $\Dc$, $\Mc_2$, the three-dimensional $\delta$ function, and a short distance $\Kc_3$. As a result, we will not repeat the steps outlined above for the asymmetric formalism. Instead, we will just write out the expression for $\Mc_{3,\rm df}$, $\Lc$, and $\Rc$ in the helicity basis~\cite{Hansen:2015zga},
\begin{align}
    \mathcal{M}_{3,\mathrm{df};\ell'\lambda',\ell\lambda}({\bf p},{\bf k})
    &=
    \int_{\p'}
    \int_{\k'}
    \mathcal{L}_{\ell'\lambda',\ell_2\lambda_2}({\bf p} , {\bf p}')
    \mathcal{T}_{\ell_2\lambda_2,\ell_1\lambda_1}({\bf p}' , {\bf k}')
    \mathcal{R}_{\ell_1\lambda_1,\ell\lambda}({\bf k}' , {\bf k}),
    \label{eq:Mdf3}
\\
\mathcal{L}_{\ell'\lambda',\ell\lambda}({\mathbf p} , {\mathbf k})
&=
\left(\frac{1}{3}-\wt{\rho}(  \sigma_p) \Mc_{2,\ell}(\sigma_p)
\right)
\delta_{\ell' \ell}
\delta_{\lambda' \lambda}
\, 
2\omega_k\,(2\pi)^3\delta^3(\mathbf{p}-\mathbf{k})
\nn\\
&\hspace{7cm}
-
\mathcal{D}_{\ell'\lambda',\ell\lambda}({\mathbf p} , {\mathbf k})
\wt{\rho}(\sigma_k),
\label{eq:L:cap}
\\
\mathcal{R}_{\ell'\lambda',\ell\lambda}({\mathbf p} , {\mathbf k})
&=
\left(\frac{1}{3}-
\wt{\rho}( \sigma_p) \Mc_{2,\ell}(\sigma_p)
\right)
\delta_{\ell' \ell}
\delta_{\lambda' \lambda}
\,2\omega_p\, 
(2\pi)^3\delta^3(\mathbf{p}-\mathbf{k})
\nn\\
&\hspace{7cm}-
 \wt{\rho}(\sigma_p)
\mathcal{D}_{\ell'\lambda',\ell\lambda}({\mathbf p} , {\mathbf k}),
\label{eq:R:cap}
\end{align}
where the $\mathcal{T}$ function satisfies an integral equation
\begin{align}
\mathcal{T}_{\ell'\lambda',\ell\lambda}({\bf p} , {\bf k})
&=
\mathcal{K}_{3;\ell'\lambda',\ell\lambda}({\bf p} , {\bf k})
\nn\\
&\hspace{1cm}-
    \int_{\p'}
    \int_{\k'}
    \mathcal{K}_{3;\ell'\lambda',\ell_2\lambda_2}({\bf p} , {\bf p}')
    \frac{\wt{\rho}(p')}{2\omega_{p'}}
    \mathcal{L}_{\ell_2\lambda_2,\ell_1\lambda_1}({\bf p}' , {\bf k}')
    \mathcal{T}_{\ell_1\lambda_1,\ell\lambda}({\bf k}' , {\bf k}).
    \label{eq:int-eq-T}
\end{align}
Following the steps discussed for partial wave projecting the asymmetric formalism, one arrives at Eqs.~\eqref{eq:M3df:jp_sym}-\eqref{eq:T_eq_sym} for $\Mc_{3}^{J^P}$ and its various contributions.

\section{Shifting the phase space, $\wt{\rho}$}
\label{app:shifting}

There is freedom in defining the phase space, $\wt{\rho}$, appearing in the definition of both  $\wh{\Mc}_{3,\df}$ and $\Mc_{3,\df}$, in Eqs.~\eqref{eq:Mcdf_hel} and \eqref{eq:Mdf3}, respectively. Throughout this work, we have assumed the simplest definition for $\wt{\rho}$, given by Eq.~\eqref{eq:rhotilde}.  This freedom stems from the fact that the two-body scattering amplitude, $\Mc_{2,\ell}$, must be scheme-independent, but one can modify the phase space and the two-body K matrix $\Kc_{2,\ell}$ simultaneously, in such a way that $\Mc_{2,\ell}$ is invariant~\cite{Jackura:2022gib}. 

To make this more explicit, let us write $\Mc_{2,\ell}$ in terms of two-body K matrix, $\Kc_{2,\ell}$, and the standard two-body phase space, $\rho$, defined in Eq.~\eqref{eq:rho_ps},
\begin{align}
    \Mc_{2,\ell}^{\,-1}
    &=
    \Kc_{2,\ell}^{\,-1} - i \rho \, ,
     \\[5pt]
     &=
    \wt{\Kc}_{2,\ell}^{\,-1} + \wt{\rho} \, ,
\end{align}
where we in the second equality we introduced
\begin{align}
    \wt{\Kc}_{2,\ell}^{\,-1} \equiv \Kc_{2,\ell}^{\,-1}  - i (1-H)\,\rho \, .
\end{align}

This makes it evident that $\Mc_{2,\ell}$ is invariant under a simultaneously shift of $\Kc_{2,\ell}$ and $\wt{\rho}$ of the form, 
\begin{align}
    \wt{\Kc}_{2,\ell}^{\,-1} &\to \wt{\Kc}_{2,\ell}^{\,-1} - \frac{\widetilde{I}_{\rm PV}^{(\ell)} }{q_k^{\star 2\ell}} \, ,
\\[5pt]
\wt{\rho} 
&\to 
\wt{\rho} 
+ \frac{\widetilde{I}_{\rm PV}^{(\ell)} }{q_k^{\star 2\ell}},
\label{eq:rhoshift}
\end{align}
where $\widetilde{I}_{\rm PV}^{(\ell)} $ has to be a real and non-singular function of $\sigma_k$ to ensure unitarity, and the barrier factors have been introduced to guarantee the correct threshold behavior of $\Mc_{2,\ell}$.

Reference~\cite{Romero-Lopez:2019qrt} showed that one can use this freedom to generalize previous finite-volume three-body formalism~\cite{Hansen:2014eka} to describe systems where the $\wt{\Kc}_{2,\ell}$ can have poles. Because $\wt{\Kc}_{2,\ell}$ depends on the cut off, these poles are unphysical, but they can appear for systems where there are two-body bound states and/or resonances. A minimal choice introduced in there is to use $\widetilde{I}_{\rm PV}^{(\ell)} $ to move these poles away from the kinematic region considered.

Because $\Dc$ only depends on $\Mc_{2,\ell}$, as opposed to $\wt{\Kc}_{2,\ell}$ or $\wt{\rho} $, it is unaffected by this shift. Meanwhile, $\Mc_{3, \rm df}$ does depend on $\wt{\rho} $, implying that the functional form does change. For example, the $\Lc$ function within the symmetric formalism, Eq.~\eqref{eq:L:cap}, will be shifted to 
\begin{align}
\mathcal{L}_{\ell'\lambda',\ell\lambda}({\mathbf p} , {\mathbf k})
&\to
\mathcal{L}_{\ell'\lambda',\ell\lambda}({\mathbf p} , {\mathbf k})
-\frac{\widetilde{I}_{\rm PV}^{(\ell)} }{b_p^{\star 2\ell}} \Mc_{2,\ell}(\sigma_p)\,2\omega_p \,(2\pi)^3\delta^3(\mathbf{p}-\mathbf{k})
-
\mathcal{D}^{(u,u)}_{\ell'\lambda',\ell\lambda}({\mathbf p} , {\mathbf k})
\frac{\widetilde{I}_{\rm PV}^{(\ell)} }{q_k^{\star 2\ell}}.
\end{align}
Note, this shift leads to an implicit redefinition of $\Kc_3$ to absorb this modification to $\Mc_3$.

Because this shift is done for partial-wave projected two-body amplitudes, it does not introduce any further subtleties in the partial-wave projecting procedure outlined in Appendix~\ref{app:conv}. In other words, one can use the formalism presented in the main body after making the global replacement
\begin{align}
    \wt{\rho} \, \delta_{S'S}
&\to 
\wt{\rho} \, \delta_{S'S}
+ \frac{\widetilde{I}_{\rm PV}^{(S)} }{q_k^{\star 2S}}.
\end{align}

These artifacts of the finite-volume formalism are not of immediate relevance for the infinite-volume formalism, which is the concern of this work. But, it is important to keep these details in mind, if one is interest in using lattice QCD results for $\Kc_3$, which may require performing these shifts in $\wt{\rho}$ to then determine physical partial wave projected amplitudes. 
\end{appendix}

\bibliography{bibi.bib}

\begin{thebibliography}{97}%
\makeatletter
\providecommand \@ifxundefined [1]{%
 \@ifx{#1\undefined}
}%
\providecommand \@ifnum [1]{%
 \ifnum #1\expandafter \@firstoftwo
 \else \expandafter \@secondoftwo
 \fi
}%
\providecommand \@ifx [1]{%
 \ifx #1\expandafter \@firstoftwo
 \else \expandafter \@secondoftwo
 \fi
}%
\providecommand \natexlab [1]{#1}%
\providecommand \enquote  [1]{``#1''}%
\providecommand \bibnamefont  [1]{#1}%
\providecommand \bibfnamefont [1]{#1}%
\providecommand \citenamefont [1]{#1}%
\providecommand \href@noop [0]{\@secondoftwo}%
\providecommand \href [0]{\begingroup \@sanitize@url \@href}%
\providecommand \@href[1]{\@@startlink{#1}\@@href}%
\providecommand \@@href[1]{\endgroup#1\@@endlink}%
\providecommand \@sanitize@url [0]{\catcode `\\12\catcode `\$12\catcode `\&12\catcode `\#12\catcode `\^12\catcode `\_12\catcode `\%12\relax}%
\providecommand \@@startlink[1]{}%
\providecommand \@@endlink[0]{}%
\providecommand \url  [0]{\begingroup\@sanitize@url \@url }%
\providecommand \@url [1]{\endgroup\@href {#1}{\urlprefix }}%
\providecommand \urlprefix  [0]{URL }%
\providecommand \Eprint [0]{\href }%
\providecommand \doibase [0]{http://dx.doi.org/}%
\providecommand \selectlanguage [0]{\@gobble}%
\providecommand \bibinfo  [0]{\@secondoftwo}%
\providecommand \bibfield  [0]{\@secondoftwo}%
\providecommand \translation [1]{[#1]}%
\providecommand \BibitemOpen [0]{}%
\providecommand \bibitemStop [0]{}%
\providecommand \bibitemNoStop [0]{.\EOS\space}%
\providecommand \EOS [0]{\spacefactor3000\relax}%
\providecommand \BibitemShut  [1]{\csname bibitem#1\endcsname}%
\let\auto@bib@innerbib\@empty
\bibitem [{\citenamefont {Aaij}\ \emph {et~al.}(2022)\citenamefont {Aaij} \emph {et~al.}}]{LHCb:2021vvq}%
  \BibitemOpen
  \bibfield  {author} {\bibinfo {author} {\bibfnamefont {R.}~\bibnamefont {Aaij}} \emph {et~al.} (\bibinfo {collaboration} {LHCb}),\ }\href {\doibase 10.1038/s41567-022-01614-y} {\bibfield  {journal} {\bibinfo  {journal} {Nature Phys.}\ }\textbf {\bibinfo {volume} {18}},\ \bibinfo {pages} {751} (\bibinfo {year} {2022})},\ \Eprint {http://arxiv.org/abs/2109.01038} {arXiv:2109.01038 [hep-ex]} \BibitemShut {NoStop}%
\bibitem [{\citenamefont {Ablikim}\ \emph {et~al.}(2024)\citenamefont {Ablikim} \emph {et~al.}}]{BESIII:2023wfi}%
  \BibitemOpen
  \bibfield  {author} {\bibinfo {author} {\bibfnamefont {M.}~\bibnamefont {Ablikim}} \emph {et~al.} (\bibinfo {collaboration} {BESIII}),\ }\href {\doibase 10.1103/PhysRevLett.132.181901} {\bibfield  {journal} {\bibinfo  {journal} {Phys. Rev. Lett.}\ }\textbf {\bibinfo {volume} {132}},\ \bibinfo {pages} {181901} (\bibinfo {year} {2024})},\ \Eprint {http://arxiv.org/abs/2312.05324} {arXiv:2312.05324 [hep-ex]} \BibitemShut {NoStop}%
\bibitem [{\citenamefont {Afzal}\ \emph {et~al.}(2024)\citenamefont {Afzal} \emph {et~al.}}]{Afzal:2024ulu}%
  \BibitemOpen
  \bibfield  {author} {\bibinfo {author} {\bibfnamefont {F.}~\bibnamefont {Afzal}} \emph {et~al.},\ }\href@noop {} {\  (\bibinfo {year} {2024})},\ \Eprint {http://arxiv.org/abs/2407.03316} {arXiv:2407.03316 [nucl-ex]} \BibitemShut {NoStop}%
\bibitem [{\citenamefont {Hergert}(2020)}]{Hergert:2020bxy}%
  \BibitemOpen
  \bibfield  {author} {\bibinfo {author} {\bibfnamefont {H.}~\bibnamefont {Hergert}},\ }\href {\doibase 10.3389/fphy.2020.00379} {\bibfield  {journal} {\bibinfo  {journal} {Front. in Phys.}\ }\textbf {\bibinfo {volume} {8}},\ \bibinfo {pages} {379} (\bibinfo {year} {2020})},\ \Eprint {http://arxiv.org/abs/2008.05061} {arXiv:2008.05061 [nucl-th]} \BibitemShut {NoStop}%
\bibitem [{\citenamefont {Aaij}\ \emph {et~al.}(2019)\citenamefont {Aaij} \emph {et~al.}}]{LHCb:2019xmb}%
  \BibitemOpen
  \bibfield  {author} {\bibinfo {author} {\bibfnamefont {R.}~\bibnamefont {Aaij}} \emph {et~al.} (\bibinfo {collaboration} {LHCb}),\ }\href {\doibase 10.1103/PhysRevLett.123.231802} {\bibfield  {journal} {\bibinfo  {journal} {Phys. Rev. Lett.}\ }\textbf {\bibinfo {volume} {123}},\ \bibinfo {pages} {231802} (\bibinfo {year} {2019})},\ \Eprint {http://arxiv.org/abs/1905.09244} {arXiv:1905.09244 [hep-ex]} \BibitemShut {NoStop}%
\bibitem [{\citenamefont {Aaij}\ \emph {et~al.}(2023)\citenamefont {Aaij} \emph {et~al.}}]{LHCb:2022fpg}%
  \BibitemOpen
  \bibfield  {author} {\bibinfo {author} {\bibfnamefont {R.}~\bibnamefont {Aaij}} \emph {et~al.} (\bibinfo {collaboration} {LHCb}),\ }\href {\doibase 10.1103/PhysRevD.108.012008} {\bibfield  {journal} {\bibinfo  {journal} {Phys. Rev. D}\ }\textbf {\bibinfo {volume} {108}},\ \bibinfo {pages} {012008} (\bibinfo {year} {2023})},\ \Eprint {http://arxiv.org/abs/2206.07622} {arXiv:2206.07622 [hep-ex]} \BibitemShut {NoStop}%
\bibitem [{\citenamefont {Aaij}\ \emph {et~al.}(2014{\natexlab{a}})\citenamefont {Aaij} \emph {et~al.}}]{LHCb:2014mir}%
  \BibitemOpen
  \bibfield  {author} {\bibinfo {author} {\bibfnamefont {R.}~\bibnamefont {Aaij}} \emph {et~al.} (\bibinfo {collaboration} {LHCb}),\ }\href {\doibase 10.1103/PhysRevD.90.112004} {\bibfield  {journal} {\bibinfo  {journal} {Phys. Rev. D}\ }\textbf {\bibinfo {volume} {90}},\ \bibinfo {pages} {112004} (\bibinfo {year} {2014}{\natexlab{a}})},\ \Eprint {http://arxiv.org/abs/1408.5373} {arXiv:1408.5373 [hep-ex]} \BibitemShut {NoStop}%
\bibitem [{\citenamefont {Aaij}\ \emph {et~al.}(2020)\citenamefont {Aaij} \emph {et~al.}}]{LHCb:2019jta}%
  \BibitemOpen
  \bibfield  {author} {\bibinfo {author} {\bibfnamefont {R.}~\bibnamefont {Aaij}} \emph {et~al.} (\bibinfo {collaboration} {LHCb}),\ }\href {\doibase 10.1103/PhysRevLett.124.031801} {\bibfield  {journal} {\bibinfo  {journal} {Phys. Rev. Lett.}\ }\textbf {\bibinfo {volume} {124}},\ \bibinfo {pages} {031801} (\bibinfo {year} {2020})},\ \Eprint {http://arxiv.org/abs/1909.05211} {arXiv:1909.05211 [hep-ex]} \BibitemShut {NoStop}%
\bibitem [{\citenamefont {Aaij}\ \emph {et~al.}(2014{\natexlab{b}})\citenamefont {Aaij} \emph {et~al.}}]{LHCb:2013lcl}%
  \BibitemOpen
  \bibfield  {author} {\bibinfo {author} {\bibfnamefont {R.}~\bibnamefont {Aaij}} \emph {et~al.} (\bibinfo {collaboration} {LHCb}),\ }\href {\doibase 10.1103/PhysRevLett.112.011801} {\bibfield  {journal} {\bibinfo  {journal} {Phys. Rev. Lett.}\ }\textbf {\bibinfo {volume} {112}},\ \bibinfo {pages} {011801} (\bibinfo {year} {2014}{\natexlab{b}})},\ \Eprint {http://arxiv.org/abs/1310.4740} {arXiv:1310.4740 [hep-ex]} \BibitemShut {NoStop}%
\bibitem [{\citenamefont {Suzuki}\ and\ \citenamefont {Wolfenstein}(1999)}]{Suzuki:1999uc}%
  \BibitemOpen
  \bibfield  {author} {\bibinfo {author} {\bibfnamefont {M.}~\bibnamefont {Suzuki}}\ and\ \bibinfo {author} {\bibfnamefont {L.}~\bibnamefont {Wolfenstein}},\ }\href {\doibase 10.1103/PhysRevD.60.074019} {\bibfield  {journal} {\bibinfo  {journal} {Phys. Rev. D}\ }\textbf {\bibinfo {volume} {60}},\ \bibinfo {pages} {074019} (\bibinfo {year} {1999})},\ \Eprint {http://arxiv.org/abs/hep-ph/9903477} {arXiv:hep-ph/9903477} \BibitemShut {NoStop}%
\bibitem [{\citenamefont {Wolfenstein}(1991)}]{Wolfenstein:1990ks}%
  \BibitemOpen
  \bibfield  {author} {\bibinfo {author} {\bibfnamefont {L.}~\bibnamefont {Wolfenstein}},\ }\href {\doibase 10.1103/PhysRevD.43.151} {\bibfield  {journal} {\bibinfo  {journal} {Phys. Rev. D}\ }\textbf {\bibinfo {volume} {43}},\ \bibinfo {pages} {151} (\bibinfo {year} {1991})}\BibitemShut {NoStop}%
\bibitem [{\citenamefont {Suzuki}(2008)}]{Suzuki:2007je}%
  \BibitemOpen
  \bibfield  {author} {\bibinfo {author} {\bibfnamefont {M.}~\bibnamefont {Suzuki}},\ }\href {\doibase 10.1103/PhysRevD.77.054021} {\bibfield  {journal} {\bibinfo  {journal} {Phys. Rev. D}\ }\textbf {\bibinfo {volume} {77}},\ \bibinfo {pages} {054021} (\bibinfo {year} {2008})},\ \Eprint {http://arxiv.org/abs/0710.5534} {arXiv:0710.5534 [hep-ph]} \BibitemShut {NoStop}%
\bibitem [{\citenamefont {Alvarenga~Nogueira}\ \emph {et~al.}(2015)\citenamefont {Alvarenga~Nogueira}, \citenamefont {Bediaga}, \citenamefont {Cavalcante}, \citenamefont {Frederico},\ and\ \citenamefont {Louren\c{c}o}}]{AlvarengaNogueira:2015wpj}%
  \BibitemOpen
  \bibfield  {author} {\bibinfo {author} {\bibfnamefont {J.~H.}\ \bibnamefont {Alvarenga~Nogueira}}, \bibinfo {author} {\bibfnamefont {I.}~\bibnamefont {Bediaga}}, \bibinfo {author} {\bibfnamefont {A.~B.~R.}\ \bibnamefont {Cavalcante}}, \bibinfo {author} {\bibfnamefont {T.}~\bibnamefont {Frederico}}, \ and\ \bibinfo {author} {\bibfnamefont {O.}~\bibnamefont {Louren\c{c}o}},\ }\href {\doibase 10.1103/PhysRevD.92.054010} {\bibfield  {journal} {\bibinfo  {journal} {Phys. Rev. D}\ }\textbf {\bibinfo {volume} {92}},\ \bibinfo {pages} {054010} (\bibinfo {year} {2015})},\ \Eprint {http://arxiv.org/abs/1506.08332} {arXiv:1506.08332 [hep-ph]} \BibitemShut {NoStop}%
\bibitem [{\citenamefont {Bediaga}\ \emph {et~al.}(2014)\citenamefont {Bediaga}, \citenamefont {Frederico},\ and\ \citenamefont {Louren\c{c}o}}]{Bediaga:2013ela}%
  \BibitemOpen
  \bibfield  {author} {\bibinfo {author} {\bibfnamefont {I.}~\bibnamefont {Bediaga}}, \bibinfo {author} {\bibfnamefont {T.}~\bibnamefont {Frederico}}, \ and\ \bibinfo {author} {\bibfnamefont {O.}~\bibnamefont {Louren\c{c}o}},\ }\href {\doibase 10.1103/PhysRevD.89.094013} {\bibfield  {journal} {\bibinfo  {journal} {Phys. Rev. D}\ }\textbf {\bibinfo {volume} {89}},\ \bibinfo {pages} {094013} (\bibinfo {year} {2014})},\ \Eprint {http://arxiv.org/abs/1307.8164} {arXiv:1307.8164 [hep-ph]} \BibitemShut {NoStop}%
\bibitem [{\citenamefont {Garrote}\ \emph {et~al.}(2023)\citenamefont {Garrote}, \citenamefont {Cuervo}, \citenamefont {Magalh\~aes},\ and\ \citenamefont {Pel\'aez}}]{Garrote:2022uub}%
  \BibitemOpen
  \bibfield  {author} {\bibinfo {author} {\bibfnamefont {R.~A.}\ \bibnamefont {Garrote}}, \bibinfo {author} {\bibfnamefont {J.}~\bibnamefont {Cuervo}}, \bibinfo {author} {\bibfnamefont {P.~C.}\ \bibnamefont {Magalh\~aes}}, \ and\ \bibinfo {author} {\bibfnamefont {J.~R.}\ \bibnamefont {Pel\'aez}},\ }\href {\doibase 10.1103/PhysRevLett.130.201901} {\bibfield  {journal} {\bibinfo  {journal} {Phys. Rev. Lett.}\ }\textbf {\bibinfo {volume} {130}},\ \bibinfo {pages} {201901} (\bibinfo {year} {2023})},\ \Eprint {http://arxiv.org/abs/2210.08354} {arXiv:2210.08354 [hep-ph]} \BibitemShut {NoStop}%
\bibitem [{\citenamefont {Jackura}\ \emph {et~al.}(2019)\citenamefont {Jackura}, \citenamefont {Fern\'andez-Ram\'\i{}rez}, \citenamefont {Mathieu}, \citenamefont {Mikhasenko}, \citenamefont {Nys}, \citenamefont {Pilloni}, \citenamefont {Salda\~na}, \citenamefont {Sherrill},\ and\ \citenamefont {Szczepaniak}}]{Jackura:2018xnx}%
  \BibitemOpen
  \bibfield  {author} {\bibinfo {author} {\bibfnamefont {A.}~\bibnamefont {Jackura}}, \bibinfo {author} {\bibfnamefont {C.}~\bibnamefont {Fern\'andez-Ram\'\i{}rez}}, \bibinfo {author} {\bibfnamefont {V.}~\bibnamefont {Mathieu}}, \bibinfo {author} {\bibfnamefont {M.}~\bibnamefont {Mikhasenko}}, \bibinfo {author} {\bibfnamefont {J.}~\bibnamefont {Nys}}, \bibinfo {author} {\bibfnamefont {A.}~\bibnamefont {Pilloni}}, \bibinfo {author} {\bibfnamefont {K.}~\bibnamefont {Salda\~na}}, \bibinfo {author} {\bibfnamefont {N.}~\bibnamefont {Sherrill}}, \ and\ \bibinfo {author} {\bibfnamefont {A.}~\bibnamefont {Szczepaniak}} (\bibinfo {collaboration} {JPAC}),\ }\href {\doibase 10.1140/epjc/s10052-019-6566-1} {\bibfield  {journal} {\bibinfo  {journal} {Eur. Phys. J. C}\ }\textbf {\bibinfo {volume} {79}},\ \bibinfo {pages} {56} (\bibinfo {year} {2019})},\ \Eprint {http://arxiv.org/abs/1809.10523} {arXiv:1809.10523 [hep-ph]} \BibitemShut {NoStop}%
\bibitem [{\citenamefont {Hansen}\ and\ \citenamefont {Sharpe}(2015)}]{Hansen:2015zga}%
  \BibitemOpen
  \bibfield  {author} {\bibinfo {author} {\bibfnamefont {M.~T.}\ \bibnamefont {Hansen}}\ and\ \bibinfo {author} {\bibfnamefont {S.~R.}\ \bibnamefont {Sharpe}},\ }\href {\doibase 10.1103/PhysRevD.92.114509} {\bibfield  {journal} {\bibinfo  {journal} {Phys. Rev.}\ }\textbf {\bibinfo {volume} {D92}},\ \bibinfo {pages} {114509} (\bibinfo {year} {2015})},\ \Eprint {http://arxiv.org/abs/1504.04248} {arXiv:1504.04248 [hep-lat]} \BibitemShut {NoStop}%
\bibitem [{\citenamefont {Jackura}\ \emph {et~al.}(2021)\citenamefont {Jackura}, \citenamefont {Brice\~no}, \citenamefont {Dawid}, \citenamefont {Islam},\ and\ \citenamefont {McCarty}}]{Jackura:2020bsk}%
  \BibitemOpen
  \bibfield  {author} {\bibinfo {author} {\bibfnamefont {A.~W.}\ \bibnamefont {Jackura}}, \bibinfo {author} {\bibfnamefont {R.~A.}\ \bibnamefont {Brice\~no}}, \bibinfo {author} {\bibfnamefont {S.~M.}\ \bibnamefont {Dawid}}, \bibinfo {author} {\bibfnamefont {M.~H.~E.}\ \bibnamefont {Islam}}, \ and\ \bibinfo {author} {\bibfnamefont {C.}~\bibnamefont {McCarty}},\ }\href {\doibase 10.1103/PhysRevD.104.014507} {\bibfield  {journal} {\bibinfo  {journal} {Phys. Rev. D}\ }\textbf {\bibinfo {volume} {104}},\ \bibinfo {pages} {014507} (\bibinfo {year} {2021})},\ \Eprint {http://arxiv.org/abs/2010.09820} {arXiv:2010.09820 [hep-lat]} \BibitemShut {NoStop}%
\bibitem [{\citenamefont {Jackura}\ and\ \citenamefont {Brice\~no}(2024)}]{Jackura:2023qtp}%
  \BibitemOpen
  \bibfield  {author} {\bibinfo {author} {\bibfnamefont {A.~W.}\ \bibnamefont {Jackura}}\ and\ \bibinfo {author} {\bibfnamefont {R.~A.}\ \bibnamefont {Brice\~no}},\ }\href {\doibase 10.1103/PhysRevD.109.096030} {\bibfield  {journal} {\bibinfo  {journal} {Phys. Rev. D}\ }\textbf {\bibinfo {volume} {109}},\ \bibinfo {pages} {096030} (\bibinfo {year} {2024})},\ \Eprint {http://arxiv.org/abs/2312.00625} {arXiv:2312.00625 [hep-ph]} \BibitemShut {NoStop}%
\bibitem [{\citenamefont {Jackura}(2023)}]{Jackura:2022gib}%
  \BibitemOpen
  \bibfield  {author} {\bibinfo {author} {\bibfnamefont {A.~W.}\ \bibnamefont {Jackura}},\ }\href {\doibase 10.1103/PhysRevD.108.034505} {\bibfield  {journal} {\bibinfo  {journal} {Phys. Rev. D}\ }\textbf {\bibinfo {volume} {108}},\ \bibinfo {pages} {034505} (\bibinfo {year} {2023})},\ \Eprint {http://arxiv.org/abs/2208.10587} {arXiv:2208.10587 [hep-lat]} \BibitemShut {NoStop}%
\bibitem [{\citenamefont {Mai}\ \emph {et~al.}(2017)\citenamefont {Mai}, \citenamefont {Hu}, \citenamefont {Doring}, \citenamefont {Pilloni},\ and\ \citenamefont {Szczepaniak}}]{Mai:2017vot}%
  \BibitemOpen
  \bibfield  {author} {\bibinfo {author} {\bibfnamefont {M.}~\bibnamefont {Mai}}, \bibinfo {author} {\bibfnamefont {B.}~\bibnamefont {Hu}}, \bibinfo {author} {\bibfnamefont {M.}~\bibnamefont {Doring}}, \bibinfo {author} {\bibfnamefont {A.}~\bibnamefont {Pilloni}}, \ and\ \bibinfo {author} {\bibfnamefont {A.}~\bibnamefont {Szczepaniak}},\ }\href {\doibase 10.1140/epja/i2017-12368-4} {\bibfield  {journal} {\bibinfo  {journal} {Eur. Phys. J. A}\ }\textbf {\bibinfo {volume} {53}},\ \bibinfo {pages} {177} (\bibinfo {year} {2017})},\ \Eprint {http://arxiv.org/abs/1706.06118} {arXiv:1706.06118 [nucl-th]} \BibitemShut {NoStop}%
\bibitem [{\citenamefont {Mikhasenko}\ \emph {et~al.}(2019)\citenamefont {Mikhasenko}, \citenamefont {Wunderlich}, \citenamefont {Jackura}, \citenamefont {Mathieu}, \citenamefont {Pilloni}, \citenamefont {Ketzer},\ and\ \citenamefont {Szczepaniak}}]{Mikhasenko:2019vhk}%
  \BibitemOpen
  \bibfield  {author} {\bibinfo {author} {\bibfnamefont {M.}~\bibnamefont {Mikhasenko}}, \bibinfo {author} {\bibfnamefont {Y.}~\bibnamefont {Wunderlich}}, \bibinfo {author} {\bibfnamefont {A.}~\bibnamefont {Jackura}}, \bibinfo {author} {\bibfnamefont {V.}~\bibnamefont {Mathieu}}, \bibinfo {author} {\bibfnamefont {A.}~\bibnamefont {Pilloni}}, \bibinfo {author} {\bibfnamefont {B.}~\bibnamefont {Ketzer}}, \ and\ \bibinfo {author} {\bibfnamefont {A.}~\bibnamefont {Szczepaniak}},\ }\href {\doibase 10.1007/JHEP08(2019)080} {\bibfield  {journal} {\bibinfo  {journal} {JHEP}\ }\textbf {\bibinfo {volume} {08}},\ \bibinfo {pages} {080} (\bibinfo {year} {2019})},\ \Eprint {http://arxiv.org/abs/1904.11894} {arXiv:1904.11894 [hep-ph]} \BibitemShut {NoStop}%
\bibitem [{\citenamefont {Dawid}\ and\ \citenamefont {Szczepaniak}(2021)}]{Dawid:2020uhn}%
  \BibitemOpen
  \bibfield  {author} {\bibinfo {author} {\bibfnamefont {S.~M.}\ \bibnamefont {Dawid}}\ and\ \bibinfo {author} {\bibfnamefont {A.~P.}\ \bibnamefont {Szczepaniak}},\ }\href {\doibase 10.1103/PhysRevD.103.014009} {\bibfield  {journal} {\bibinfo  {journal} {Phys. Rev. D}\ }\textbf {\bibinfo {volume} {103}},\ \bibinfo {pages} {014009} (\bibinfo {year} {2021})},\ \Eprint {http://arxiv.org/abs/2010.08084} {arXiv:2010.08084 [nucl-th]} \BibitemShut {NoStop}%
\bibitem [{\citenamefont {Feng}\ \emph {et~al.}(2024)\citenamefont {Feng}, \citenamefont {Gil}, \citenamefont {D\"oring}, \citenamefont {Molina}, \citenamefont {Mai}, \citenamefont {Shastry},\ and\ \citenamefont {Szczepaniak}}]{Feng:2024wyg}%
  \BibitemOpen
  \bibfield  {author} {\bibinfo {author} {\bibfnamefont {Y.}~\bibnamefont {Feng}}, \bibinfo {author} {\bibfnamefont {F.}~\bibnamefont {Gil}}, \bibinfo {author} {\bibfnamefont {M.}~\bibnamefont {D\"oring}}, \bibinfo {author} {\bibfnamefont {R.}~\bibnamefont {Molina}}, \bibinfo {author} {\bibfnamefont {M.}~\bibnamefont {Mai}}, \bibinfo {author} {\bibfnamefont {V.}~\bibnamefont {Shastry}}, \ and\ \bibinfo {author} {\bibfnamefont {A.}~\bibnamefont {Szczepaniak}},\ }\href@noop {} {\  (\bibinfo {year} {2024})},\ \Eprint {http://arxiv.org/abs/2407.08721} {arXiv:2407.08721 [nucl-th]} \BibitemShut {NoStop}%
\bibitem [{\citenamefont {Detmold}\ \emph {et~al.}(2008)\citenamefont {Detmold}, \citenamefont {Savage}, \citenamefont {Torok}, \citenamefont {Beane}, \citenamefont {Luu}, \citenamefont {Orginos},\ and\ \citenamefont {Parreno}}]{Detmold:2008fn}%
  \BibitemOpen
  \bibfield  {author} {\bibinfo {author} {\bibfnamefont {W.}~\bibnamefont {Detmold}}, \bibinfo {author} {\bibfnamefont {M.~J.}\ \bibnamefont {Savage}}, \bibinfo {author} {\bibfnamefont {A.}~\bibnamefont {Torok}}, \bibinfo {author} {\bibfnamefont {S.~R.}\ \bibnamefont {Beane}}, \bibinfo {author} {\bibfnamefont {T.~C.}\ \bibnamefont {Luu}}, \bibinfo {author} {\bibfnamefont {K.}~\bibnamefont {Orginos}}, \ and\ \bibinfo {author} {\bibfnamefont {A.}~\bibnamefont {Parreno}},\ }\href {\doibase 10.1103/PhysRevD.78.014507} {\bibfield  {journal} {\bibinfo  {journal} {Phys. Rev. D}\ }\textbf {\bibinfo {volume} {78}},\ \bibinfo {pages} {014507} (\bibinfo {year} {2008})},\ \Eprint {http://arxiv.org/abs/0803.2728} {arXiv:0803.2728 [hep-lat]} \BibitemShut {NoStop}%
\bibitem [{\citenamefont {Beane}\ \emph {et~al.}(2008)\citenamefont {Beane}, \citenamefont {Detmold}, \citenamefont {Luu}, \citenamefont {Orginos}, \citenamefont {Savage},\ and\ \citenamefont {Torok}}]{Beane:2007es}%
  \BibitemOpen
  \bibfield  {author} {\bibinfo {author} {\bibfnamefont {S.~R.}\ \bibnamefont {Beane}}, \bibinfo {author} {\bibfnamefont {W.}~\bibnamefont {Detmold}}, \bibinfo {author} {\bibfnamefont {T.~C.}\ \bibnamefont {Luu}}, \bibinfo {author} {\bibfnamefont {K.}~\bibnamefont {Orginos}}, \bibinfo {author} {\bibfnamefont {M.~J.}\ \bibnamefont {Savage}}, \ and\ \bibinfo {author} {\bibfnamefont {A.}~\bibnamefont {Torok}},\ }\href {\doibase 10.1103/PhysRevLett.100.082004} {\bibfield  {journal} {\bibinfo  {journal} {Phys. Rev. Lett.}\ }\textbf {\bibinfo {volume} {100}},\ \bibinfo {pages} {082004} (\bibinfo {year} {2008})},\ \Eprint {http://arxiv.org/abs/0710.1827} {arXiv:0710.1827 [hep-lat]} \BibitemShut {NoStop}%
\bibitem [{\citenamefont {Culver}\ \emph {et~al.}(2020)\citenamefont {Culver}, \citenamefont {Mai}, \citenamefont {Brett}, \citenamefont {Alexandru},\ and\ \citenamefont {D\"oring}}]{Culver:2019vvu}%
  \BibitemOpen
  \bibfield  {author} {\bibinfo {author} {\bibfnamefont {C.}~\bibnamefont {Culver}}, \bibinfo {author} {\bibfnamefont {M.}~\bibnamefont {Mai}}, \bibinfo {author} {\bibfnamefont {R.}~\bibnamefont {Brett}}, \bibinfo {author} {\bibfnamefont {A.}~\bibnamefont {Alexandru}}, \ and\ \bibinfo {author} {\bibfnamefont {M.}~\bibnamefont {D\"oring}},\ }\href {\doibase 10.1103/PhysRevD.101.114507} {\bibfield  {journal} {\bibinfo  {journal} {Phys. Rev. D}\ }\textbf {\bibinfo {volume} {101}},\ \bibinfo {pages} {114507} (\bibinfo {year} {2020})},\ \Eprint {http://arxiv.org/abs/1911.09047} {arXiv:1911.09047 [hep-lat]} \BibitemShut {NoStop}%
\bibitem [{\citenamefont {Alexandru}\ \emph {et~al.}(2020)\citenamefont {Alexandru}, \citenamefont {Brett}, \citenamefont {Culver}, \citenamefont {D\"oring}, \citenamefont {Guo}, \citenamefont {Lee},\ and\ \citenamefont {Mai}}]{Alexandru:2020xqf}%
  \BibitemOpen
  \bibfield  {author} {\bibinfo {author} {\bibfnamefont {A.}~\bibnamefont {Alexandru}}, \bibinfo {author} {\bibfnamefont {R.}~\bibnamefont {Brett}}, \bibinfo {author} {\bibfnamefont {C.}~\bibnamefont {Culver}}, \bibinfo {author} {\bibfnamefont {M.}~\bibnamefont {D\"oring}}, \bibinfo {author} {\bibfnamefont {D.}~\bibnamefont {Guo}}, \bibinfo {author} {\bibfnamefont {F.~X.}\ \bibnamefont {Lee}}, \ and\ \bibinfo {author} {\bibfnamefont {M.}~\bibnamefont {Mai}},\ }\href {\doibase 10.1103/PhysRevD.102.114523} {\bibfield  {journal} {\bibinfo  {journal} {Phys. Rev. D}\ }\textbf {\bibinfo {volume} {102}},\ \bibinfo {pages} {114523} (\bibinfo {year} {2020})},\ \Eprint {http://arxiv.org/abs/2009.12358} {arXiv:2009.12358 [hep-lat]} \BibitemShut {NoStop}%
\bibitem [{\citenamefont {Hansen}\ \emph {et~al.}(2021)\citenamefont {Hansen}, \citenamefont {Brice\~no}, \citenamefont {Edwards}, \citenamefont {Thomas},\ and\ \citenamefont {Wilson}}]{Hansen:2020otl}%
  \BibitemOpen
  \bibfield  {author} {\bibinfo {author} {\bibfnamefont {M.~T.}\ \bibnamefont {Hansen}}, \bibinfo {author} {\bibfnamefont {R.~A.}\ \bibnamefont {Brice\~no}}, \bibinfo {author} {\bibfnamefont {R.~G.}\ \bibnamefont {Edwards}}, \bibinfo {author} {\bibfnamefont {C.~E.}\ \bibnamefont {Thomas}}, \ and\ \bibinfo {author} {\bibfnamefont {D.~J.}\ \bibnamefont {Wilson}} (\bibinfo {collaboration} {Hadron Spectrum}),\ }\href {\doibase 10.1103/PhysRevLett.126.012001} {\bibfield  {journal} {\bibinfo  {journal} {Phys. Rev. Lett.}\ }\textbf {\bibinfo {volume} {126}},\ \bibinfo {pages} {012001} (\bibinfo {year} {2021})},\ \Eprint {http://arxiv.org/abs/2009.04931} {arXiv:2009.04931 [hep-lat]} \BibitemShut {NoStop}%
\bibitem [{\citenamefont {Draper}\ \emph {et~al.}(2023)\citenamefont {Draper}, \citenamefont {Hanlon}, \citenamefont {H\"orz}, \citenamefont {Morningstar}, \citenamefont {Romero-L\'opez},\ and\ \citenamefont {Sharpe}}]{Draper:2023boj}%
  \BibitemOpen
  \bibfield  {author} {\bibinfo {author} {\bibfnamefont {Z.~T.}\ \bibnamefont {Draper}}, \bibinfo {author} {\bibfnamefont {A.~D.}\ \bibnamefont {Hanlon}}, \bibinfo {author} {\bibfnamefont {B.}~\bibnamefont {H\"orz}}, \bibinfo {author} {\bibfnamefont {C.}~\bibnamefont {Morningstar}}, \bibinfo {author} {\bibfnamefont {F.}~\bibnamefont {Romero-L\'opez}}, \ and\ \bibinfo {author} {\bibfnamefont {S.~R.}\ \bibnamefont {Sharpe}},\ }\href {\doibase 10.1007/JHEP05(2023)137} {\bibfield  {journal} {\bibinfo  {journal} {JHEP}\ }\textbf {\bibinfo {volume} {05}},\ \bibinfo {pages} {137} (\bibinfo {year} {2023})},\ \Eprint {http://arxiv.org/abs/2302.13587} {arXiv:2302.13587 [hep-lat]} \BibitemShut {NoStop}%
\bibitem [{\citenamefont {Polejaeva}\ and\ \citenamefont {Rusetsky}(2012)}]{Polejaeva:2012ut}%
  \BibitemOpen
  \bibfield  {author} {\bibinfo {author} {\bibfnamefont {K.}~\bibnamefont {Polejaeva}}\ and\ \bibinfo {author} {\bibfnamefont {A.}~\bibnamefont {Rusetsky}},\ }\href {\doibase 10.1140/epja/i2012-12067-8} {\bibfield  {journal} {\bibinfo  {journal} {Eur. Phys. J.}\ }\textbf {\bibinfo {volume} {A48}},\ \bibinfo {pages} {67} (\bibinfo {year} {2012})},\ \Eprint {http://arxiv.org/abs/1203.1241} {arXiv:1203.1241 [hep-lat]} \BibitemShut {NoStop}%
\bibitem [{\citenamefont {Hansen}\ and\ \citenamefont {Sharpe}(2014)}]{Hansen:2014eka}%
  \BibitemOpen
  \bibfield  {author} {\bibinfo {author} {\bibfnamefont {M.~T.}\ \bibnamefont {Hansen}}\ and\ \bibinfo {author} {\bibfnamefont {S.~R.}\ \bibnamefont {Sharpe}},\ }\href {\doibase 10.1103/PhysRevD.90.116003} {\bibfield  {journal} {\bibinfo  {journal} {Phys. Rev.}\ }\textbf {\bibinfo {volume} {D90}},\ \bibinfo {pages} {116003} (\bibinfo {year} {2014})},\ \Eprint {http://arxiv.org/abs/1408.5933} {arXiv:1408.5933 [hep-lat]} \BibitemShut {NoStop}%
\bibitem [{\citenamefont {Briceno}\ \emph {et~al.}(2017)\citenamefont {Briceno}, \citenamefont {Hansen},\ and\ \citenamefont {Sharpe}}]{Briceno:2017tce}%
  \BibitemOpen
  \bibfield  {author} {\bibinfo {author} {\bibfnamefont {R.~A.}\ \bibnamefont {Briceno}}, \bibinfo {author} {\bibfnamefont {M.~T.}\ \bibnamefont {Hansen}}, \ and\ \bibinfo {author} {\bibfnamefont {S.~R.}\ \bibnamefont {Sharpe}},\ }\href {\doibase 10.1103/PhysRevD.95.074510} {\bibfield  {journal} {\bibinfo  {journal} {Phys. Rev.}\ }\textbf {\bibinfo {volume} {D95}},\ \bibinfo {pages} {074510} (\bibinfo {year} {2017})},\ \Eprint {http://arxiv.org/abs/1701.07465} {arXiv:1701.07465 [hep-lat]} \BibitemShut {NoStop}%
\bibitem [{\citenamefont {Brice\~no}\ \emph {et~al.}(2018)\citenamefont {Brice\~no}, \citenamefont {Hansen},\ and\ \citenamefont {Sharpe}}]{Briceno:2018mlh}%
  \BibitemOpen
  \bibfield  {author} {\bibinfo {author} {\bibfnamefont {R.~A.}\ \bibnamefont {Brice\~no}}, \bibinfo {author} {\bibfnamefont {M.~T.}\ \bibnamefont {Hansen}}, \ and\ \bibinfo {author} {\bibfnamefont {S.~R.}\ \bibnamefont {Sharpe}},\ }\href {\doibase 10.1103/PhysRevD.98.014506} {\bibfield  {journal} {\bibinfo  {journal} {Phys. Rev.}\ }\textbf {\bibinfo {volume} {D98}},\ \bibinfo {pages} {014506} (\bibinfo {year} {2018})},\ \Eprint {http://arxiv.org/abs/1803.04169} {arXiv:1803.04169 [hep-lat]} \BibitemShut {NoStop}%
\bibitem [{\citenamefont {Briceno}\ \emph {et~al.}(2019)\citenamefont {Briceno}, \citenamefont {Hansen},\ and\ \citenamefont {Sharpe}}]{Briceno:2018aml}%
  \BibitemOpen
  \bibfield  {author} {\bibinfo {author} {\bibfnamefont {R.~A.}\ \bibnamefont {Briceno}}, \bibinfo {author} {\bibfnamefont {M.~T.}\ \bibnamefont {Hansen}}, \ and\ \bibinfo {author} {\bibfnamefont {S.~R.}\ \bibnamefont {Sharpe}},\ }\href {\doibase 10.1103/PhysRevD.99.014516} {\bibfield  {journal} {\bibinfo  {journal} {Phys. Rev.}\ }\textbf {\bibinfo {volume} {D99}},\ \bibinfo {pages} {014516} (\bibinfo {year} {2019})},\ \Eprint {http://arxiv.org/abs/1810.01429} {arXiv:1810.01429 [hep-lat]} \BibitemShut {NoStop}%
\bibitem [{\citenamefont {Briceño}\ \emph {et~al.}(2019)\citenamefont {Briceño}, \citenamefont {Hansen}, \citenamefont {Sharpe},\ and\ \citenamefont {Szczepaniak}}]{Briceno:2019muc}%
  \BibitemOpen
  \bibfield  {author} {\bibinfo {author} {\bibfnamefont {R.~A.}\ \bibnamefont {Briceño}}, \bibinfo {author} {\bibfnamefont {M.~T.}\ \bibnamefont {Hansen}}, \bibinfo {author} {\bibfnamefont {S.~R.}\ \bibnamefont {Sharpe}}, \ and\ \bibinfo {author} {\bibfnamefont {A.~P.}\ \bibnamefont {Szczepaniak}},\ }\href {\doibase 10.1103/PhysRevD.100.054508} {\bibfield  {journal} {\bibinfo  {journal} {Phys. Rev. D}\ }\textbf {\bibinfo {volume} {100}},\ \bibinfo {pages} {054508} (\bibinfo {year} {2019})},\ \Eprint {http://arxiv.org/abs/1905.11188} {arXiv:1905.11188 [hep-lat]} \BibitemShut {NoStop}%
\bibitem [{\citenamefont {Blanton}\ \emph {et~al.}(2019)\citenamefont {Blanton}, \citenamefont {Romero-L{\'o}pez},\ and\ \citenamefont {Sharpe}}]{Blanton:2019igq}%
  \BibitemOpen
  \bibfield  {author} {\bibinfo {author} {\bibfnamefont {T.~D.}\ \bibnamefont {Blanton}}, \bibinfo {author} {\bibfnamefont {F.}~\bibnamefont {Romero-L{\'o}pez}}, \ and\ \bibinfo {author} {\bibfnamefont {S.~R.}\ \bibnamefont {Sharpe}},\ }\href {\doibase 10.1007/JHEP03(2019)106} {\bibfield  {journal} {\bibinfo  {journal} {JHEP}\ }\textbf {\bibinfo {volume} {03}},\ \bibinfo {pages} {106} (\bibinfo {year} {2019})},\ \Eprint {http://arxiv.org/abs/1901.07095} {arXiv:1901.07095 [hep-lat]} \BibitemShut {NoStop}%
\bibitem [{\citenamefont {Hansen}\ \emph {et~al.}(2020)\citenamefont {Hansen}, \citenamefont {Romero-L\'opez},\ and\ \citenamefont {Sharpe}}]{Hansen:2020zhy}%
  \BibitemOpen
  \bibfield  {author} {\bibinfo {author} {\bibfnamefont {M.~T.}\ \bibnamefont {Hansen}}, \bibinfo {author} {\bibfnamefont {F.}~\bibnamefont {Romero-L\'opez}}, \ and\ \bibinfo {author} {\bibfnamefont {S.~R.}\ \bibnamefont {Sharpe}},\ }\href {\doibase 10.1007/JHEP07(2020)047} {\bibfield  {journal} {\bibinfo  {journal} {JHEP}\ }\textbf {\bibinfo {volume} {07}},\ \bibinfo {pages} {047} (\bibinfo {year} {2020})},\ \bibinfo {note} {[Erratum: JHEP 02, 014 (2021)]},\ \Eprint {http://arxiv.org/abs/2003.10974} {arXiv:2003.10974 [hep-lat]} \BibitemShut {NoStop}%
\bibitem [{\citenamefont {Blanton}\ and\ \citenamefont {Sharpe}(2020{\natexlab{a}})}]{Blanton:2020gha}%
  \BibitemOpen
  \bibfield  {author} {\bibinfo {author} {\bibfnamefont {T.~D.}\ \bibnamefont {Blanton}}\ and\ \bibinfo {author} {\bibfnamefont {S.~R.}\ \bibnamefont {Sharpe}},\ }\href {\doibase 10.1103/PhysRevD.102.054520} {\bibfield  {journal} {\bibinfo  {journal} {Phys. Rev. D}\ }\textbf {\bibinfo {volume} {102}},\ \bibinfo {pages} {054520} (\bibinfo {year} {2020}{\natexlab{a}})},\ \Eprint {http://arxiv.org/abs/2007.16188} {arXiv:2007.16188 [hep-lat]} \BibitemShut {NoStop}%
\bibitem [{\citenamefont {Hammer}\ \emph {et~al.}(2017{\natexlab{a}})\citenamefont {Hammer}, \citenamefont {Pang},\ and\ \citenamefont {Rusetsky}}]{Hammer:2017uqm}%
  \BibitemOpen
  \bibfield  {author} {\bibinfo {author} {\bibfnamefont {H.-W.}\ \bibnamefont {Hammer}}, \bibinfo {author} {\bibfnamefont {J.-Y.}\ \bibnamefont {Pang}}, \ and\ \bibinfo {author} {\bibfnamefont {A.}~\bibnamefont {Rusetsky}},\ }\href {\doibase 10.1007/JHEP09(2017)109} {\bibfield  {journal} {\bibinfo  {journal} {JHEP}\ }\textbf {\bibinfo {volume} {09}},\ \bibinfo {pages} {109} (\bibinfo {year} {2017}{\natexlab{a}})},\ \Eprint {http://arxiv.org/abs/1706.07700} {arXiv:1706.07700 [hep-lat]} \BibitemShut {NoStop}%
\bibitem [{\citenamefont {Hammer}\ \emph {et~al.}(2017{\natexlab{b}})\citenamefont {Hammer}, \citenamefont {Pang},\ and\ \citenamefont {Rusetsky}}]{Hammer:2017kms}%
  \BibitemOpen
  \bibfield  {author} {\bibinfo {author} {\bibfnamefont {H.~W.}\ \bibnamefont {Hammer}}, \bibinfo {author} {\bibfnamefont {J.~Y.}\ \bibnamefont {Pang}}, \ and\ \bibinfo {author} {\bibfnamefont {A.}~\bibnamefont {Rusetsky}},\ }\href {\doibase 10.1007/JHEP10(2017)115} {\bibfield  {journal} {\bibinfo  {journal} {JHEP}\ }\textbf {\bibinfo {volume} {10}},\ \bibinfo {pages} {115} (\bibinfo {year} {2017}{\natexlab{b}})},\ \Eprint {http://arxiv.org/abs/1707.02176} {arXiv:1707.02176 [hep-lat]} \BibitemShut {NoStop}%
\bibitem [{\citenamefont {Meng}\ \emph {et~al.}(2018)\citenamefont {Meng}, \citenamefont {Liu}, \citenamefont {Mei\ss{}ner},\ and\ \citenamefont {Rusetsky}}]{Meng:2017jgx}%
  \BibitemOpen
  \bibfield  {author} {\bibinfo {author} {\bibfnamefont {Y.}~\bibnamefont {Meng}}, \bibinfo {author} {\bibfnamefont {C.}~\bibnamefont {Liu}}, \bibinfo {author} {\bibfnamefont {U.-G.}\ \bibnamefont {Mei\ss{}ner}}, \ and\ \bibinfo {author} {\bibfnamefont {A.}~\bibnamefont {Rusetsky}},\ }\href {\doibase 10.1103/PhysRevD.98.014508} {\bibfield  {journal} {\bibinfo  {journal} {Phys. Rev. D}\ }\textbf {\bibinfo {volume} {98}},\ \bibinfo {pages} {014508} (\bibinfo {year} {2018})},\ \Eprint {http://arxiv.org/abs/1712.08464} {arXiv:1712.08464 [hep-lat]} \BibitemShut {NoStop}%
\bibitem [{\citenamefont {Pang}\ \emph {et~al.}(2019)\citenamefont {Pang}, \citenamefont {Wu}, \citenamefont {Hammer}, \citenamefont {Mei\ss{}ner},\ and\ \citenamefont {Rusetsky}}]{Pang:2019dfe}%
  \BibitemOpen
  \bibfield  {author} {\bibinfo {author} {\bibfnamefont {J.-Y.}\ \bibnamefont {Pang}}, \bibinfo {author} {\bibfnamefont {J.-J.}\ \bibnamefont {Wu}}, \bibinfo {author} {\bibfnamefont {H.~W.}\ \bibnamefont {Hammer}}, \bibinfo {author} {\bibfnamefont {U.-G.}\ \bibnamefont {Mei\ss{}ner}}, \ and\ \bibinfo {author} {\bibfnamefont {A.}~\bibnamefont {Rusetsky}},\ }\href {\doibase 10.1103/PhysRevD.99.074513} {\bibfield  {journal} {\bibinfo  {journal} {Phys. Rev. D}\ }\textbf {\bibinfo {volume} {99}},\ \bibinfo {pages} {074513} (\bibinfo {year} {2019})},\ \Eprint {http://arxiv.org/abs/1902.01111} {arXiv:1902.01111 [hep-lat]} \BibitemShut {NoStop}%
\bibitem [{\citenamefont {M\"uller}\ \emph {et~al.}(2022)\citenamefont {M\"uller}, \citenamefont {Pang}, \citenamefont {Rusetsky},\ and\ \citenamefont {Wu}}]{Muller:2021uur}%
  \BibitemOpen
  \bibfield  {author} {\bibinfo {author} {\bibfnamefont {F.}~\bibnamefont {M\"uller}}, \bibinfo {author} {\bibfnamefont {J.-Y.}\ \bibnamefont {Pang}}, \bibinfo {author} {\bibfnamefont {A.}~\bibnamefont {Rusetsky}}, \ and\ \bibinfo {author} {\bibfnamefont {J.-J.}\ \bibnamefont {Wu}},\ }\href {\doibase 10.1007/JHEP02(2022)158} {\bibfield  {journal} {\bibinfo  {journal} {JHEP}\ }\textbf {\bibinfo {volume} {02}},\ \bibinfo {pages} {158} (\bibinfo {year} {2022})},\ \Eprint {http://arxiv.org/abs/2110.09351} {arXiv:2110.09351 [hep-lat]} \BibitemShut {NoStop}%
\bibitem [{\citenamefont {Blanton}\ and\ \citenamefont {Sharpe}(2021{\natexlab{a}})}]{Blanton:2020gmf}%
  \BibitemOpen
  \bibfield  {author} {\bibinfo {author} {\bibfnamefont {T.~D.}\ \bibnamefont {Blanton}}\ and\ \bibinfo {author} {\bibfnamefont {S.~R.}\ \bibnamefont {Sharpe}},\ }\href {\doibase 10.1103/PhysRevD.103.054503} {\bibfield  {journal} {\bibinfo  {journal} {Phys. Rev. D}\ }\textbf {\bibinfo {volume} {103}},\ \bibinfo {pages} {054503} (\bibinfo {year} {2021}{\natexlab{a}})},\ \Eprint {http://arxiv.org/abs/2011.05520} {arXiv:2011.05520 [hep-lat]} \BibitemShut {NoStop}%
\bibitem [{\citenamefont {Blanton}\ and\ \citenamefont {Sharpe}(2021{\natexlab{b}})}]{Blanton:2021mih}%
  \BibitemOpen
  \bibfield  {author} {\bibinfo {author} {\bibfnamefont {T.~D.}\ \bibnamefont {Blanton}}\ and\ \bibinfo {author} {\bibfnamefont {S.~R.}\ \bibnamefont {Sharpe}},\ }\href {\doibase 10.1103/PhysRevD.104.034509} {\bibfield  {journal} {\bibinfo  {journal} {Phys. Rev. D}\ }\textbf {\bibinfo {volume} {104}},\ \bibinfo {pages} {034509} (\bibinfo {year} {2021}{\natexlab{b}})},\ \Eprint {http://arxiv.org/abs/2105.12094} {arXiv:2105.12094 [hep-lat]} \BibitemShut {NoStop}%
\bibitem [{\citenamefont {Luscher}(1986{\natexlab{a}})}]{Luscher:1985dn}%
  \BibitemOpen
  \bibfield  {author} {\bibinfo {author} {\bibfnamefont {M.}~\bibnamefont {Luscher}},\ }\href {\doibase 10.1007/BF01211589} {\bibfield  {journal} {\bibinfo  {journal} {Commun. Math. Phys.}\ }\textbf {\bibinfo {volume} {104}},\ \bibinfo {pages} {177} (\bibinfo {year} {1986}{\natexlab{a}})}\BibitemShut {NoStop}%
\bibitem [{\citenamefont {Luscher}(1986{\natexlab{b}})}]{Luscher:1986n2}%
  \BibitemOpen
  \bibfield  {author} {\bibinfo {author} {\bibfnamefont {M.}~\bibnamefont {Luscher}},\ }\href {\doibase 10.1007/BF01211097} {\bibfield  {journal} {\bibinfo  {journal} {Commun.Math.Phys.}\ }\textbf {\bibinfo {volume} {105}},\ \bibinfo {pages} {153} (\bibinfo {year} {1986}{\natexlab{b}})}\BibitemShut {NoStop}%
\bibitem [{\citenamefont {Luscher}(1991)}]{Luscher:1990ux}%
  \BibitemOpen
  \bibfield  {author} {\bibinfo {author} {\bibfnamefont {M.}~\bibnamefont {Luscher}},\ }\href {\doibase 10.1016/0550-3213(91)90366-6} {\bibfield  {journal} {\bibinfo  {journal} {Nucl. Phys.}\ }\textbf {\bibinfo {volume} {B354}},\ \bibinfo {pages} {531} (\bibinfo {year} {1991})}\BibitemShut {NoStop}%
\bibitem [{\citenamefont {Rummukainen}\ and\ \citenamefont {Gottlieb}(1995)}]{Rummukainen:1995vs}%
  \BibitemOpen
  \bibfield  {author} {\bibinfo {author} {\bibfnamefont {K.}~\bibnamefont {Rummukainen}}\ and\ \bibinfo {author} {\bibfnamefont {S.~A.}\ \bibnamefont {Gottlieb}},\ }\href {\doibase 10.1016/0550-3213(95)00313-H} {\bibfield  {journal} {\bibinfo  {journal} {Nucl. Phys.}\ }\textbf {\bibinfo {volume} {B450}},\ \bibinfo {pages} {397} (\bibinfo {year} {1995})},\ \Eprint {http://arxiv.org/abs/hep-lat/9503028} {arXiv:hep-lat/9503028 [hep-lat]} \BibitemShut {NoStop}%
\bibitem [{\citenamefont {Kim}\ \emph {et~al.}(2005)\citenamefont {Kim}, \citenamefont {Sachrajda},\ and\ \citenamefont {Sharpe}}]{Kim:2005gf}%
  \BibitemOpen
  \bibfield  {author} {\bibinfo {author} {\bibfnamefont {C.~h.}\ \bibnamefont {Kim}}, \bibinfo {author} {\bibfnamefont {C.~T.}\ \bibnamefont {Sachrajda}}, \ and\ \bibinfo {author} {\bibfnamefont {S.~R.}\ \bibnamefont {Sharpe}},\ }\href {\doibase 10.1016/j.nuclphysb.2005.08.029} {\bibfield  {journal} {\bibinfo  {journal} {Nucl. Phys.}\ }\textbf {\bibinfo {volume} {B727}},\ \bibinfo {pages} {218} (\bibinfo {year} {2005})},\ \Eprint {http://arxiv.org/abs/hep-lat/0507006} {arXiv:hep-lat/0507006 [hep-lat]} \BibitemShut {NoStop}%
\bibitem [{\citenamefont {He}\ \emph {et~al.}(2005)\citenamefont {He}, \citenamefont {Feng},\ and\ \citenamefont {Liu}}]{He:2005ey}%
  \BibitemOpen
  \bibfield  {author} {\bibinfo {author} {\bibfnamefont {S.}~\bibnamefont {He}}, \bibinfo {author} {\bibfnamefont {X.}~\bibnamefont {Feng}}, \ and\ \bibinfo {author} {\bibfnamefont {C.}~\bibnamefont {Liu}},\ }\href {\doibase 10.1088/1126-6708/2005/07/011} {\bibfield  {journal} {\bibinfo  {journal} {JHEP}\ }\textbf {\bibinfo {volume} {07}},\ \bibinfo {pages} {011} (\bibinfo {year} {2005})},\ \Eprint {http://arxiv.org/abs/hep-lat/0504019} {arXiv:hep-lat/0504019 [hep-lat]} \BibitemShut {NoStop}%
\bibitem [{\citenamefont {Hansen}\ and\ \citenamefont {Sharpe}(2012)}]{Hansen:2012tf}%
  \BibitemOpen
  \bibfield  {author} {\bibinfo {author} {\bibfnamefont {M.~T.}\ \bibnamefont {Hansen}}\ and\ \bibinfo {author} {\bibfnamefont {S.~R.}\ \bibnamefont {Sharpe}},\ }\href {\doibase 10.1103/PhysRevD.86.016007} {\bibfield  {journal} {\bibinfo  {journal} {Phys. Rev.}\ }\textbf {\bibinfo {volume} {D86}},\ \bibinfo {pages} {016007} (\bibinfo {year} {2012})},\ \Eprint {http://arxiv.org/abs/1204.0826} {arXiv:1204.0826 [hep-lat]} \BibitemShut {NoStop}%
\bibitem [{\citenamefont {Brice\~no}\ and\ \citenamefont {Davoudi}(2013)}]{Briceno:2012yi}%
  \BibitemOpen
  \bibfield  {author} {\bibinfo {author} {\bibfnamefont {R.~A.}\ \bibnamefont {Brice\~no}}\ and\ \bibinfo {author} {\bibfnamefont {Z.}~\bibnamefont {Davoudi}},\ }\href {\doibase 10.1103/PhysRevD.88.094507} {\bibfield  {journal} {\bibinfo  {journal} {Phys. Rev.}\ }\textbf {\bibinfo {volume} {D88}},\ \bibinfo {pages} {094507} (\bibinfo {year} {2013})},\ \Eprint {http://arxiv.org/abs/1204.1110} {arXiv:1204.1110 [hep-lat]} \BibitemShut {NoStop}%
\bibitem [{\citenamefont {Brice\~no}\ \emph {et~al.}(2013)\citenamefont {Brice\~no}, \citenamefont {Davoudi},\ and\ \citenamefont {Luu}}]{Briceno:2013lba}%
  \BibitemOpen
  \bibfield  {author} {\bibinfo {author} {\bibfnamefont {R.~A.}\ \bibnamefont {Brice\~no}}, \bibinfo {author} {\bibfnamefont {Z.}~\bibnamefont {Davoudi}}, \ and\ \bibinfo {author} {\bibfnamefont {T.~C.}\ \bibnamefont {Luu}},\ }\href {\doibase 10.1103/PhysRevD.88.034502} {\bibfield  {journal} {\bibinfo  {journal} {Phys. Rev.}\ }\textbf {\bibinfo {volume} {D88}},\ \bibinfo {pages} {034502} (\bibinfo {year} {2013})},\ \Eprint {http://arxiv.org/abs/1305.4903} {arXiv:1305.4903 [hep-lat]} \BibitemShut {NoStop}%
\bibitem [{\citenamefont {Brice\~no}(2014)}]{Briceno:2014oea}%
  \BibitemOpen
  \bibfield  {author} {\bibinfo {author} {\bibfnamefont {R.~A.}\ \bibnamefont {Brice\~no}},\ }\href {\doibase 10.1103/PhysRevD.89.074507} {\bibfield  {journal} {\bibinfo  {journal} {Phys. Rev.}\ }\textbf {\bibinfo {volume} {D89}},\ \bibinfo {pages} {074507} (\bibinfo {year} {2014})},\ \Eprint {http://arxiv.org/abs/1401.3312} {arXiv:1401.3312 [hep-lat]} \BibitemShut {NoStop}%
\bibitem [{\citenamefont {Dudek}\ \emph {et~al.}(2011)\citenamefont {Dudek}, \citenamefont {Edwards}, \citenamefont {Peardon}, \citenamefont {Richards},\ and\ \citenamefont {Thomas}}]{Dudek:2010ew}%
  \BibitemOpen
  \bibfield  {author} {\bibinfo {author} {\bibfnamefont {J.~J.}\ \bibnamefont {Dudek}}, \bibinfo {author} {\bibfnamefont {R.~G.}\ \bibnamefont {Edwards}}, \bibinfo {author} {\bibfnamefont {M.~J.}\ \bibnamefont {Peardon}}, \bibinfo {author} {\bibfnamefont {D.~G.}\ \bibnamefont {Richards}}, \ and\ \bibinfo {author} {\bibfnamefont {C.~E.}\ \bibnamefont {Thomas}},\ }\href {\doibase 10.1103/PhysRevD.83.071504} {\bibfield  {journal} {\bibinfo  {journal} {Phys. Rev.}\ }\textbf {\bibinfo {volume} {D83}},\ \bibinfo {pages} {071504} (\bibinfo {year} {2011})},\ \Eprint {http://arxiv.org/abs/1011.6352} {arXiv:1011.6352 [hep-ph]} \BibitemShut {NoStop}%
\bibitem [{\citenamefont {Pelissier}\ and\ \citenamefont {Alexandru}(2013)}]{Pelissier:2012pi}%
  \BibitemOpen
  \bibfield  {author} {\bibinfo {author} {\bibfnamefont {C.}~\bibnamefont {Pelissier}}\ and\ \bibinfo {author} {\bibfnamefont {A.}~\bibnamefont {Alexandru}},\ }\href {\doibase 10.1103/PhysRevD.87.014503} {\bibfield  {journal} {\bibinfo  {journal} {Phys. Rev.}\ }\textbf {\bibinfo {volume} {D87}},\ \bibinfo {pages} {014503} (\bibinfo {year} {2013})},\ \Eprint {http://arxiv.org/abs/1211.0092} {arXiv:1211.0092 [hep-lat]} \BibitemShut {NoStop}%
\bibitem [{\citenamefont {Dudek}\ \emph {et~al.}(2013)\citenamefont {Dudek}, \citenamefont {Edwards},\ and\ \citenamefont {Thomas}}]{Dudek:2012xn}%
  \BibitemOpen
  \bibfield  {author} {\bibinfo {author} {\bibfnamefont {J.~J.}\ \bibnamefont {Dudek}}, \bibinfo {author} {\bibfnamefont {R.~G.}\ \bibnamefont {Edwards}}, \ and\ \bibinfo {author} {\bibfnamefont {C.~E.}\ \bibnamefont {Thomas}} (\bibinfo {collaboration} {Hadron Spectrum}),\ }\href {\doibase 10.1103/PhysRevD.87.034505, 10.1103/PhysRevD.90.099902} {\bibfield  {journal} {\bibinfo  {journal} {Phys. Rev.}\ }\textbf {\bibinfo {volume} {D87}},\ \bibinfo {pages} {034505} (\bibinfo {year} {2013})},\ \bibinfo {note} {[Erratum: Phys. Rev.D90,no.9,099902(2014)]},\ \Eprint {http://arxiv.org/abs/1212.0830} {arXiv:1212.0830 [hep-ph]} \BibitemShut {NoStop}%
\bibitem [{\citenamefont {Liu}\ \emph {et~al.}(2013)\citenamefont {Liu}, \citenamefont {Orginos}, \citenamefont {Guo}, \citenamefont {Hanhart},\ and\ \citenamefont {Meissner}}]{Liu:2012zya}%
  \BibitemOpen
  \bibfield  {author} {\bibinfo {author} {\bibfnamefont {L.}~\bibnamefont {Liu}}, \bibinfo {author} {\bibfnamefont {K.}~\bibnamefont {Orginos}}, \bibinfo {author} {\bibfnamefont {F.-K.}\ \bibnamefont {Guo}}, \bibinfo {author} {\bibfnamefont {C.}~\bibnamefont {Hanhart}}, \ and\ \bibinfo {author} {\bibfnamefont {U.-G.}\ \bibnamefont {Meissner}},\ }\href {\doibase 10.1103/PhysRevD.87.014508} {\bibfield  {journal} {\bibinfo  {journal} {Phys. Rev.}\ }\textbf {\bibinfo {volume} {D87}},\ \bibinfo {pages} {014508} (\bibinfo {year} {2013})},\ \Eprint {http://arxiv.org/abs/1208.4535} {arXiv:1208.4535 [hep-lat]} \BibitemShut {NoStop}%
\bibitem [{\citenamefont {Wilson}\ \emph {et~al.}(2015{\natexlab{a}})\citenamefont {Wilson}, \citenamefont {Dudek}, \citenamefont {Edwards},\ and\ \citenamefont {Thomas}}]{Wilson:2014cna}%
  \BibitemOpen
  \bibfield  {author} {\bibinfo {author} {\bibfnamefont {D.~J.}\ \bibnamefont {Wilson}}, \bibinfo {author} {\bibfnamefont {J.~J.}\ \bibnamefont {Dudek}}, \bibinfo {author} {\bibfnamefont {R.~G.}\ \bibnamefont {Edwards}}, \ and\ \bibinfo {author} {\bibfnamefont {C.~E.}\ \bibnamefont {Thomas}},\ }\href {\doibase 10.1103/PhysRevD.91.054008} {\bibfield  {journal} {\bibinfo  {journal} {Phys. Rev.}\ }\textbf {\bibinfo {volume} {D91}},\ \bibinfo {pages} {054008} (\bibinfo {year} {2015}{\natexlab{a}})},\ \Eprint {http://arxiv.org/abs/1411.2004} {arXiv:1411.2004 [hep-ph]} \BibitemShut {NoStop}%
\bibitem [{\citenamefont {Dudek}\ \emph {et~al.}(2014)\citenamefont {Dudek}, \citenamefont {Edwards}, \citenamefont {Thomas},\ and\ \citenamefont {Wilson}}]{Dudek:2014qha}%
  \BibitemOpen
  \bibfield  {author} {\bibinfo {author} {\bibfnamefont {J.~J.}\ \bibnamefont {Dudek}}, \bibinfo {author} {\bibfnamefont {R.~G.}\ \bibnamefont {Edwards}}, \bibinfo {author} {\bibfnamefont {C.~E.}\ \bibnamefont {Thomas}}, \ and\ \bibinfo {author} {\bibfnamefont {D.~J.}\ \bibnamefont {Wilson}} (\bibinfo {collaboration} {Hadron Spectrum}),\ }\href {\doibase 10.1103/PhysRevLett.113.182001} {\bibfield  {journal} {\bibinfo  {journal} {Phys. Rev. Lett.}\ }\textbf {\bibinfo {volume} {113}},\ \bibinfo {pages} {182001} (\bibinfo {year} {2014})},\ \Eprint {http://arxiv.org/abs/1406.4158} {arXiv:1406.4158 [hep-ph]} \BibitemShut {NoStop}%
\bibitem [{\citenamefont {Lang}\ \emph {et~al.}(2015)\citenamefont {Lang}, \citenamefont {Mohler}, \citenamefont {Prelovsek},\ and\ \citenamefont {Woloshyn}}]{Lang:2015hza}%
  \BibitemOpen
  \bibfield  {author} {\bibinfo {author} {\bibfnamefont {C.~B.}\ \bibnamefont {Lang}}, \bibinfo {author} {\bibfnamefont {D.}~\bibnamefont {Mohler}}, \bibinfo {author} {\bibfnamefont {S.}~\bibnamefont {Prelovsek}}, \ and\ \bibinfo {author} {\bibfnamefont {R.~M.}\ \bibnamefont {Woloshyn}},\ }\href {\doibase 10.1016/j.physletb.2015.08.038} {\bibfield  {journal} {\bibinfo  {journal} {Phys. Lett.}\ }\textbf {\bibinfo {volume} {B750}},\ \bibinfo {pages} {17} (\bibinfo {year} {2015})},\ \Eprint {http://arxiv.org/abs/1501.01646} {arXiv:1501.01646 [hep-lat]} \BibitemShut {NoStop}%
\bibitem [{\citenamefont {Wilson}\ \emph {et~al.}(2015{\natexlab{b}})\citenamefont {Wilson}, \citenamefont {Brice\~no}, \citenamefont {Dudek}, \citenamefont {Edwards},\ and\ \citenamefont {Thomas}}]{Wilson:2015dqa}%
  \BibitemOpen
  \bibfield  {author} {\bibinfo {author} {\bibfnamefont {D.~J.}\ \bibnamefont {Wilson}}, \bibinfo {author} {\bibfnamefont {R.~A.}\ \bibnamefont {Brice\~no}}, \bibinfo {author} {\bibfnamefont {J.~J.}\ \bibnamefont {Dudek}}, \bibinfo {author} {\bibfnamefont {R.~G.}\ \bibnamefont {Edwards}}, \ and\ \bibinfo {author} {\bibfnamefont {C.~E.}\ \bibnamefont {Thomas}},\ }\href {\doibase 10.1103/PhysRevD.92.094502} {\bibfield  {journal} {\bibinfo  {journal} {Phys. Rev.}\ }\textbf {\bibinfo {volume} {D92}},\ \bibinfo {pages} {094502} (\bibinfo {year} {2015}{\natexlab{b}})},\ \Eprint {http://arxiv.org/abs/1507.02599} {arXiv:1507.02599 [hep-ph]} \BibitemShut {NoStop}%
\bibitem [{\citenamefont {Dudek}\ \emph {et~al.}(2016)\citenamefont {Dudek}, \citenamefont {Edwards},\ and\ \citenamefont {Wilson}}]{Dudek:2016cru}%
  \BibitemOpen
  \bibfield  {author} {\bibinfo {author} {\bibfnamefont {J.~J.}\ \bibnamefont {Dudek}}, \bibinfo {author} {\bibfnamefont {R.~G.}\ \bibnamefont {Edwards}}, \ and\ \bibinfo {author} {\bibfnamefont {D.~J.}\ \bibnamefont {Wilson}} (\bibinfo {collaboration} {Hadron Spectrum}),\ }\href {\doibase 10.1103/PhysRevD.93.094506} {\bibfield  {journal} {\bibinfo  {journal} {Phys. Rev.}\ }\textbf {\bibinfo {volume} {D93}},\ \bibinfo {pages} {094506} (\bibinfo {year} {2016})},\ \Eprint {http://arxiv.org/abs/1602.05122} {arXiv:1602.05122 [hep-ph]} \BibitemShut {NoStop}%
\bibitem [{\citenamefont {Brice\~no}\ \emph {et~al.}(2017)\citenamefont {Brice\~no}, \citenamefont {Dudek}, \citenamefont {Edwards},\ and\ \citenamefont {Wilson}}]{Briceno:2016mjc}%
  \BibitemOpen
  \bibfield  {author} {\bibinfo {author} {\bibfnamefont {R.~A.}\ \bibnamefont {Brice\~no}}, \bibinfo {author} {\bibfnamefont {J.~J.}\ \bibnamefont {Dudek}}, \bibinfo {author} {\bibfnamefont {R.~G.}\ \bibnamefont {Edwards}}, \ and\ \bibinfo {author} {\bibfnamefont {D.~J.}\ \bibnamefont {Wilson}},\ }\href {\doibase 10.1103/PhysRevLett.118.022002} {\bibfield  {journal} {\bibinfo  {journal} {Phys. Rev. Lett.}\ }\textbf {\bibinfo {volume} {118}},\ \bibinfo {pages} {022002} (\bibinfo {year} {2017})},\ \Eprint {http://arxiv.org/abs/1607.05900} {arXiv:1607.05900 [hep-ph]} \BibitemShut {NoStop}%
\bibitem [{\citenamefont {Moir}\ \emph {et~al.}(2016)\citenamefont {Moir}, \citenamefont {Peardon}, \citenamefont {Ryan}, \citenamefont {Thomas},\ and\ \citenamefont {Wilson}}]{Moir:2016srx}%
  \BibitemOpen
  \bibfield  {author} {\bibinfo {author} {\bibfnamefont {G.}~\bibnamefont {Moir}}, \bibinfo {author} {\bibfnamefont {M.}~\bibnamefont {Peardon}}, \bibinfo {author} {\bibfnamefont {S.~M.}\ \bibnamefont {Ryan}}, \bibinfo {author} {\bibfnamefont {C.~E.}\ \bibnamefont {Thomas}}, \ and\ \bibinfo {author} {\bibfnamefont {D.~J.}\ \bibnamefont {Wilson}},\ }\href {\doibase 10.1007/JHEP10(2016)011} {\bibfield  {journal} {\bibinfo  {journal} {JHEP}\ }\textbf {\bibinfo {volume} {10}},\ \bibinfo {pages} {011} (\bibinfo {year} {2016})},\ \Eprint {http://arxiv.org/abs/1607.07093} {arXiv:1607.07093 [hep-lat]} \BibitemShut {NoStop}%
\bibitem [{\citenamefont {Bulava}\ \emph {et~al.}(2016)\citenamefont {Bulava}, \citenamefont {Fahy}, \citenamefont {Horz}, \citenamefont {Juge}, \citenamefont {Morningstar},\ and\ \citenamefont {Wong}}]{Bulava:2016mks}%
  \BibitemOpen
  \bibfield  {author} {\bibinfo {author} {\bibfnamefont {J.}~\bibnamefont {Bulava}}, \bibinfo {author} {\bibfnamefont {B.}~\bibnamefont {Fahy}}, \bibinfo {author} {\bibfnamefont {B.}~\bibnamefont {Horz}}, \bibinfo {author} {\bibfnamefont {K.~J.}\ \bibnamefont {Juge}}, \bibinfo {author} {\bibfnamefont {C.}~\bibnamefont {Morningstar}}, \ and\ \bibinfo {author} {\bibfnamefont {C.~H.}\ \bibnamefont {Wong}},\ }\href {\doibase 10.1016/j.nuclphysb.2016.07.024} {\bibfield  {journal} {\bibinfo  {journal} {Nucl. Phys.}\ }\textbf {\bibinfo {volume} {B910}},\ \bibinfo {pages} {842} (\bibinfo {year} {2016})},\ \Eprint {http://arxiv.org/abs/1604.05593} {arXiv:1604.05593 [hep-lat]} \BibitemShut {NoStop}%
\bibitem [{\citenamefont {Hu}\ \emph {et~al.}(2016)\citenamefont {Hu}, \citenamefont {Molina}, \citenamefont {Doring},\ and\ \citenamefont {Alexandru}}]{Hu:2016shf}%
  \BibitemOpen
  \bibfield  {author} {\bibinfo {author} {\bibfnamefont {B.}~\bibnamefont {Hu}}, \bibinfo {author} {\bibfnamefont {R.}~\bibnamefont {Molina}}, \bibinfo {author} {\bibfnamefont {M.}~\bibnamefont {Doring}}, \ and\ \bibinfo {author} {\bibfnamefont {A.}~\bibnamefont {Alexandru}},\ }\href {\doibase 10.1103/PhysRevLett.117.122001} {\bibfield  {journal} {\bibinfo  {journal} {Phys. Rev. Lett.}\ }\textbf {\bibinfo {volume} {117}},\ \bibinfo {pages} {122001} (\bibinfo {year} {2016})},\ \Eprint {http://arxiv.org/abs/1605.04823} {arXiv:1605.04823 [hep-lat]} \BibitemShut {NoStop}%
\bibitem [{\citenamefont {Alexandrou}\ \emph {et~al.}(2017)\citenamefont {Alexandrou}, \citenamefont {Leskovec}, \citenamefont {Meinel}, \citenamefont {Negele}, \citenamefont {Paul}, \citenamefont {Petschlies}, \citenamefont {Pochinsky}, \citenamefont {Rendon},\ and\ \citenamefont {Syritsyn}}]{Alexandrou:2017mpi}%
  \BibitemOpen
  \bibfield  {author} {\bibinfo {author} {\bibfnamefont {C.}~\bibnamefont {Alexandrou}}, \bibinfo {author} {\bibfnamefont {L.}~\bibnamefont {Leskovec}}, \bibinfo {author} {\bibfnamefont {S.}~\bibnamefont {Meinel}}, \bibinfo {author} {\bibfnamefont {J.}~\bibnamefont {Negele}}, \bibinfo {author} {\bibfnamefont {S.}~\bibnamefont {Paul}}, \bibinfo {author} {\bibfnamefont {M.}~\bibnamefont {Petschlies}}, \bibinfo {author} {\bibfnamefont {A.}~\bibnamefont {Pochinsky}}, \bibinfo {author} {\bibfnamefont {G.}~\bibnamefont {Rendon}}, \ and\ \bibinfo {author} {\bibfnamefont {S.}~\bibnamefont {Syritsyn}},\ }\href {\doibase 10.1103/PhysRevD.96.034525} {\bibfield  {journal} {\bibinfo  {journal} {Phys. Rev.}\ }\textbf {\bibinfo {volume} {D96}},\ \bibinfo {pages} {034525} (\bibinfo {year} {2017})},\ \Eprint {http://arxiv.org/abs/1704.05439} {arXiv:1704.05439 [hep-lat]} \BibitemShut {NoStop}%
\bibitem [{\citenamefont {Bali}\ \emph {et~al.}(2017)\citenamefont {Bali}, \citenamefont {Collins}, \citenamefont {Cox},\ and\ \citenamefont {Schäfer}}]{Bali:2017pdv}%
  \BibitemOpen
  \bibfield  {author} {\bibinfo {author} {\bibfnamefont {G.~S.}\ \bibnamefont {Bali}}, \bibinfo {author} {\bibfnamefont {S.}~\bibnamefont {Collins}}, \bibinfo {author} {\bibfnamefont {A.}~\bibnamefont {Cox}}, \ and\ \bibinfo {author} {\bibfnamefont {A.}~\bibnamefont {Schäfer}},\ }\href {\doibase 10.1103/PhysRevD.96.074501} {\bibfield  {journal} {\bibinfo  {journal} {Phys. Rev.}\ }\textbf {\bibinfo {volume} {D96}},\ \bibinfo {pages} {074501} (\bibinfo {year} {2017})},\ \Eprint {http://arxiv.org/abs/1706.01247} {arXiv:1706.01247 [hep-lat]} \BibitemShut {NoStop}%
\bibitem [{\citenamefont {Wagman}\ \emph {et~al.}(2017)\citenamefont {Wagman}, \citenamefont {Winter}, \citenamefont {Chang}, \citenamefont {Davoudi}, \citenamefont {Detmold}, \citenamefont {Orginos}, \citenamefont {Savage},\ and\ \citenamefont {Shanahan}}]{Wagman:2017tmp}%
  \BibitemOpen
  \bibfield  {author} {\bibinfo {author} {\bibfnamefont {M.~L.}\ \bibnamefont {Wagman}}, \bibinfo {author} {\bibfnamefont {F.}~\bibnamefont {Winter}}, \bibinfo {author} {\bibfnamefont {E.}~\bibnamefont {Chang}}, \bibinfo {author} {\bibfnamefont {Z.}~\bibnamefont {Davoudi}}, \bibinfo {author} {\bibfnamefont {W.}~\bibnamefont {Detmold}}, \bibinfo {author} {\bibfnamefont {K.}~\bibnamefont {Orginos}}, \bibinfo {author} {\bibfnamefont {M.~J.}\ \bibnamefont {Savage}}, \ and\ \bibinfo {author} {\bibfnamefont {P.~E.}\ \bibnamefont {Shanahan}},\ }\href {\doibase 10.1103/PhysRevD.96.114510} {\bibfield  {journal} {\bibinfo  {journal} {Phys. Rev.}\ }\textbf {\bibinfo {volume} {D96}},\ \bibinfo {pages} {114510} (\bibinfo {year} {2017})},\ \Eprint {http://arxiv.org/abs/1706.06550} {arXiv:1706.06550 [hep-lat]} \BibitemShut {NoStop}%
\bibitem [{\citenamefont {Andersen}\ \emph {et~al.}(2018)\citenamefont {Andersen}, \citenamefont {Bulava}, \citenamefont {Horz},\ and\ \citenamefont {Morningstar}}]{Andersen:2017una}%
  \BibitemOpen
  \bibfield  {author} {\bibinfo {author} {\bibfnamefont {C.~W.}\ \bibnamefont {Andersen}}, \bibinfo {author} {\bibfnamefont {J.}~\bibnamefont {Bulava}}, \bibinfo {author} {\bibfnamefont {B.}~\bibnamefont {Horz}}, \ and\ \bibinfo {author} {\bibfnamefont {C.}~\bibnamefont {Morningstar}},\ }\href {\doibase 10.1103/PhysRevD.97.014506} {\bibfield  {journal} {\bibinfo  {journal} {Phys. Rev.}\ }\textbf {\bibinfo {volume} {D97}},\ \bibinfo {pages} {014506} (\bibinfo {year} {2018})},\ \Eprint {http://arxiv.org/abs/1710.01557} {arXiv:1710.01557 [hep-lat]} \BibitemShut {NoStop}%
\bibitem [{\citenamefont {Briceno}\ \emph {et~al.}(2018{\natexlab{a}})\citenamefont {Briceno}, \citenamefont {Dudek}, \citenamefont {Edwards},\ and\ \citenamefont {Wilson}}]{Briceno:2017qmb}%
  \BibitemOpen
  \bibfield  {author} {\bibinfo {author} {\bibfnamefont {R.~A.}\ \bibnamefont {Briceno}}, \bibinfo {author} {\bibfnamefont {J.~J.}\ \bibnamefont {Dudek}}, \bibinfo {author} {\bibfnamefont {R.~G.}\ \bibnamefont {Edwards}}, \ and\ \bibinfo {author} {\bibfnamefont {D.~J.}\ \bibnamefont {Wilson}},\ }\href {\doibase 10.1103/PhysRevD.97.054513} {\bibfield  {journal} {\bibinfo  {journal} {Phys. Rev.}\ }\textbf {\bibinfo {volume} {D97}},\ \bibinfo {pages} {054513} (\bibinfo {year} {2018}{\natexlab{a}})},\ \Eprint {http://arxiv.org/abs/1708.06667} {arXiv:1708.06667 [hep-lat]} \BibitemShut {NoStop}%
\bibitem [{\citenamefont {Woss}\ \emph {et~al.}(2018)\citenamefont {Woss}, \citenamefont {Thomas}, \citenamefont {Dudek}, \citenamefont {Edwards},\ and\ \citenamefont {Wilson}}]{Woss:2018irj}%
  \BibitemOpen
  \bibfield  {author} {\bibinfo {author} {\bibfnamefont {A.}~\bibnamefont {Woss}}, \bibinfo {author} {\bibfnamefont {C.~E.}\ \bibnamefont {Thomas}}, \bibinfo {author} {\bibfnamefont {J.~J.}\ \bibnamefont {Dudek}}, \bibinfo {author} {\bibfnamefont {R.~G.}\ \bibnamefont {Edwards}}, \ and\ \bibinfo {author} {\bibfnamefont {D.~J.}\ \bibnamefont {Wilson}},\ }\href {\doibase 10.1007/JHEP07(2018)043} {\bibfield  {journal} {\bibinfo  {journal} {JHEP}\ }\textbf {\bibinfo {volume} {07}},\ \bibinfo {pages} {043} (\bibinfo {year} {2018})},\ \Eprint {http://arxiv.org/abs/1802.05580} {arXiv:1802.05580 [hep-lat]} \BibitemShut {NoStop}%
\bibitem [{\citenamefont {Brett}\ \emph {et~al.}(2018)\citenamefont {Brett}, \citenamefont {Bulava}, \citenamefont {Fallica}, \citenamefont {Hanlon}, \citenamefont {Horz},\ and\ \citenamefont {Morningstar}}]{Brett:2018jqw}%
  \BibitemOpen
  \bibfield  {author} {\bibinfo {author} {\bibfnamefont {R.}~\bibnamefont {Brett}}, \bibinfo {author} {\bibfnamefont {J.}~\bibnamefont {Bulava}}, \bibinfo {author} {\bibfnamefont {J.}~\bibnamefont {Fallica}}, \bibinfo {author} {\bibfnamefont {A.}~\bibnamefont {Hanlon}}, \bibinfo {author} {\bibfnamefont {B.}~\bibnamefont {Horz}}, \ and\ \bibinfo {author} {\bibfnamefont {C.}~\bibnamefont {Morningstar}},\ }\href {\doibase 10.1016/j.nuclphysb.2018.05.008} {\bibfield  {journal} {\bibinfo  {journal} {Nucl. Phys.}\ }\textbf {\bibinfo {volume} {B932}},\ \bibinfo {pages} {29} (\bibinfo {year} {2018})},\ \Eprint {http://arxiv.org/abs/1802.03100} {arXiv:1802.03100 [hep-lat]} \BibitemShut {NoStop}%
\bibitem [{\citenamefont {Mai}\ \emph {et~al.}(2019)\citenamefont {Mai}, \citenamefont {Culver}, \citenamefont {Alexandru}, \citenamefont {D\"oring},\ and\ \citenamefont {Lee}}]{Mai:2019pqr}%
  \BibitemOpen
  \bibfield  {author} {\bibinfo {author} {\bibfnamefont {M.}~\bibnamefont {Mai}}, \bibinfo {author} {\bibfnamefont {C.}~\bibnamefont {Culver}}, \bibinfo {author} {\bibfnamefont {A.}~\bibnamefont {Alexandru}}, \bibinfo {author} {\bibfnamefont {M.}~\bibnamefont {D\"oring}}, \ and\ \bibinfo {author} {\bibfnamefont {F.~X.}\ \bibnamefont {Lee}},\ }\href {\doibase 10.1103/PhysRevD.100.114514} {\bibfield  {journal} {\bibinfo  {journal} {Phys. Rev. D}\ }\textbf {\bibinfo {volume} {100}},\ \bibinfo {pages} {114514} (\bibinfo {year} {2019})},\ \Eprint {http://arxiv.org/abs/1908.01847} {arXiv:1908.01847 [hep-lat]} \BibitemShut {NoStop}%
\bibitem [{\citenamefont {Woss}\ \emph {et~al.}(2019)\citenamefont {Woss}, \citenamefont {Thomas}, \citenamefont {Dudek}, \citenamefont {Edwards},\ and\ \citenamefont {Wilson}}]{Woss:2019hse}%
  \BibitemOpen
  \bibfield  {author} {\bibinfo {author} {\bibfnamefont {A.~J.}\ \bibnamefont {Woss}}, \bibinfo {author} {\bibfnamefont {C.~E.}\ \bibnamefont {Thomas}}, \bibinfo {author} {\bibfnamefont {J.~J.}\ \bibnamefont {Dudek}}, \bibinfo {author} {\bibfnamefont {R.~G.}\ \bibnamefont {Edwards}}, \ and\ \bibinfo {author} {\bibfnamefont {D.~J.}\ \bibnamefont {Wilson}},\ }\href {\doibase 10.1103/PhysRevD.100.054506} {\bibfield  {journal} {\bibinfo  {journal} {Phys. Rev. D}\ }\textbf {\bibinfo {volume} {100}},\ \bibinfo {pages} {054506} (\bibinfo {year} {2019})},\ \Eprint {http://arxiv.org/abs/1904.04136} {arXiv:1904.04136 [hep-lat]} \BibitemShut {NoStop}%
\bibitem [{\citenamefont {Wilson}\ \emph {et~al.}(2019)\citenamefont {Wilson}, \citenamefont {Briceno}, \citenamefont {Dudek}, \citenamefont {Edwards},\ and\ \citenamefont {Thomas}}]{Wilson:2019wfr}%
  \BibitemOpen
  \bibfield  {author} {\bibinfo {author} {\bibfnamefont {D.~J.}\ \bibnamefont {Wilson}}, \bibinfo {author} {\bibfnamefont {R.~A.}\ \bibnamefont {Briceno}}, \bibinfo {author} {\bibfnamefont {J.~J.}\ \bibnamefont {Dudek}}, \bibinfo {author} {\bibfnamefont {R.~G.}\ \bibnamefont {Edwards}}, \ and\ \bibinfo {author} {\bibfnamefont {C.~E.}\ \bibnamefont {Thomas}},\ }\href {\doibase 10.1103/PhysRevLett.123.042002} {\bibfield  {journal} {\bibinfo  {journal} {Phys. Rev. Lett.}\ }\textbf {\bibinfo {volume} {123}},\ \bibinfo {pages} {042002} (\bibinfo {year} {2019})},\ \Eprint {http://arxiv.org/abs/1904.03188} {arXiv:1904.03188 [hep-lat]} \BibitemShut {NoStop}%
\bibitem [{\citenamefont {Cheung}\ \emph {et~al.}(2021)\citenamefont {Cheung}, \citenamefont {Thomas}, \citenamefont {Wilson}, \citenamefont {Moir}, \citenamefont {Peardon},\ and\ \citenamefont {Ryan}}]{Cheung:2020mql}%
  \BibitemOpen
  \bibfield  {author} {\bibinfo {author} {\bibfnamefont {G.~K.~C.}\ \bibnamefont {Cheung}}, \bibinfo {author} {\bibfnamefont {C.~E.}\ \bibnamefont {Thomas}}, \bibinfo {author} {\bibfnamefont {D.~J.}\ \bibnamefont {Wilson}}, \bibinfo {author} {\bibfnamefont {G.}~\bibnamefont {Moir}}, \bibinfo {author} {\bibfnamefont {M.}~\bibnamefont {Peardon}}, \ and\ \bibinfo {author} {\bibfnamefont {S.~M.}\ \bibnamefont {Ryan}} (\bibinfo {collaboration} {Hadron Spectrum}),\ }\href {\doibase 10.1007/JHEP02(2021)100} {\bibfield  {journal} {\bibinfo  {journal} {JHEP}\ }\textbf {\bibinfo {volume} {02}},\ \bibinfo {pages} {100} (\bibinfo {year} {2021})},\ \Eprint {http://arxiv.org/abs/2008.06432} {arXiv:2008.06432 [hep-lat]} \BibitemShut {NoStop}%
\bibitem [{\citenamefont {Rendon}\ \emph {et~al.}(2020)\citenamefont {Rendon}, \citenamefont {Leskovec}, \citenamefont {Meinel}, \citenamefont {Negele}, \citenamefont {Paul}, \citenamefont {Petschlies}, \citenamefont {Pochinsky}, \citenamefont {Silvi},\ and\ \citenamefont {Syritsyn}}]{Rendon:2020rtw}%
  \BibitemOpen
  \bibfield  {author} {\bibinfo {author} {\bibfnamefont {G.}~\bibnamefont {Rendon}}, \bibinfo {author} {\bibfnamefont {L.}~\bibnamefont {Leskovec}}, \bibinfo {author} {\bibfnamefont {S.}~\bibnamefont {Meinel}}, \bibinfo {author} {\bibfnamefont {J.}~\bibnamefont {Negele}}, \bibinfo {author} {\bibfnamefont {S.}~\bibnamefont {Paul}}, \bibinfo {author} {\bibfnamefont {M.}~\bibnamefont {Petschlies}}, \bibinfo {author} {\bibfnamefont {A.}~\bibnamefont {Pochinsky}}, \bibinfo {author} {\bibfnamefont {G.}~\bibnamefont {Silvi}}, \ and\ \bibinfo {author} {\bibfnamefont {S.}~\bibnamefont {Syritsyn}},\ }\href {\doibase 10.1103/PhysRevD.102.114520} {\bibfield  {journal} {\bibinfo  {journal} {Phys. Rev. D}\ }\textbf {\bibinfo {volume} {102}},\ \bibinfo {pages} {114520} (\bibinfo {year} {2020})},\ \Eprint {http://arxiv.org/abs/2006.14035} {arXiv:2006.14035 [hep-lat]} \BibitemShut {NoStop}%
\bibitem [{\citenamefont {Woss}\ \emph {et~al.}(2021)\citenamefont {Woss}, \citenamefont {Dudek}, \citenamefont {Edwards}, \citenamefont {Thomas},\ and\ \citenamefont {Wilson}}]{Woss:2020ayi}%
  \BibitemOpen
  \bibfield  {author} {\bibinfo {author} {\bibfnamefont {A.~J.}\ \bibnamefont {Woss}}, \bibinfo {author} {\bibfnamefont {J.~J.}\ \bibnamefont {Dudek}}, \bibinfo {author} {\bibfnamefont {R.~G.}\ \bibnamefont {Edwards}}, \bibinfo {author} {\bibfnamefont {C.~E.}\ \bibnamefont {Thomas}}, \ and\ \bibinfo {author} {\bibfnamefont {D.~J.}\ \bibnamefont {Wilson}} (\bibinfo {collaboration} {Hadron Spectrum}),\ }\href {\doibase 10.1103/PhysRevD.103.054502} {\bibfield  {journal} {\bibinfo  {journal} {Phys. Rev. D}\ }\textbf {\bibinfo {volume} {103}},\ \bibinfo {pages} {054502} (\bibinfo {year} {2021})},\ \Eprint {http://arxiv.org/abs/2009.10034} {arXiv:2009.10034 [hep-lat]} \BibitemShut {NoStop}%
\bibitem [{\citenamefont {H\"orz}\ \emph {et~al.}(2021)\citenamefont {H\"orz} \emph {et~al.}}]{Horz:2020zvv}%
  \BibitemOpen
  \bibfield  {author} {\bibinfo {author} {\bibfnamefont {B.}~\bibnamefont {H\"orz}} \emph {et~al.},\ }\href {\doibase 10.1103/PhysRevC.103.014003} {\bibfield  {journal} {\bibinfo  {journal} {Phys. Rev. C}\ }\textbf {\bibinfo {volume} {103}},\ \bibinfo {pages} {014003} (\bibinfo {year} {2021})},\ \Eprint {http://arxiv.org/abs/2009.11825} {arXiv:2009.11825 [hep-lat]} \BibitemShut {NoStop}%
\bibitem [{\citenamefont {Briceno}\ \emph {et~al.}(2018{\natexlab{b}})\citenamefont {Briceno}, \citenamefont {Dudek},\ and\ \citenamefont {Young}}]{Briceno:2017max}%
  \BibitemOpen
  \bibfield  {author} {\bibinfo {author} {\bibfnamefont {R.~A.}\ \bibnamefont {Briceno}}, \bibinfo {author} {\bibfnamefont {J.~J.}\ \bibnamefont {Dudek}}, \ and\ \bibinfo {author} {\bibfnamefont {R.~D.}\ \bibnamefont {Young}},\ }\href {\doibase 10.1103/RevModPhys.90.025001} {\bibfield  {journal} {\bibinfo  {journal} {Rev. Mod. Phys.}\ }\textbf {\bibinfo {volume} {90}},\ \bibinfo {pages} {025001} (\bibinfo {year} {2018}{\natexlab{b}})},\ \Eprint {http://arxiv.org/abs/1706.06223} {arXiv:1706.06223 [hep-lat]} \BibitemShut {NoStop}%
\bibitem [{\citenamefont {Hansen}\ and\ \citenamefont {Sharpe}(2019)}]{Hansen:2019nir}%
  \BibitemOpen
  \bibfield  {author} {\bibinfo {author} {\bibfnamefont {M.~T.}\ \bibnamefont {Hansen}}\ and\ \bibinfo {author} {\bibfnamefont {S.~R.}\ \bibnamefont {Sharpe}},\ }\href {\doibase 10.1146/annurev-nucl-101918-023723} {\bibfield  {journal} {\bibinfo  {journal} {Annual Review of Nuclear and Particle Science}\ }\textbf {\bibinfo {volume} {69}},\ \bibinfo {pages} {null} (\bibinfo {year} {2019})},\ \Eprint {http://arxiv.org/abs/1901.00483} {arXiv:1901.00483 [hep-lat]} \BibitemShut {NoStop}%
\bibitem [{\citenamefont {Dawid}\ \emph {et~al.}(2023{\natexlab{a}})\citenamefont {Dawid}, \citenamefont {Islam},\ and\ \citenamefont {Brice\~no}}]{Dawid:2023jrj}%
  \BibitemOpen
  \bibfield  {author} {\bibinfo {author} {\bibfnamefont {S.~M.}\ \bibnamefont {Dawid}}, \bibinfo {author} {\bibfnamefont {M.~H.~E.}\ \bibnamefont {Islam}}, \ and\ \bibinfo {author} {\bibfnamefont {R.~A.}\ \bibnamefont {Brice\~no}},\ }\href {\doibase 10.1103/PhysRevD.108.034016} {\bibfield  {journal} {\bibinfo  {journal} {Phys. Rev. D}\ }\textbf {\bibinfo {volume} {108}},\ \bibinfo {pages} {034016} (\bibinfo {year} {2023}{\natexlab{a}})},\ \Eprint {http://arxiv.org/abs/2303.04394} {arXiv:2303.04394 [nucl-th]} \BibitemShut {NoStop}%
\bibitem [{\citenamefont {Brice\~no}\ \emph {et~al.}(2024)\citenamefont {Brice\~no}, \citenamefont {Jackura}, \citenamefont {Pefkou},\ and\ \citenamefont {Romero-L\'opez}}]{Briceno:2024txg}%
  \BibitemOpen
  \bibfield  {author} {\bibinfo {author} {\bibfnamefont {R.~A.}\ \bibnamefont {Brice\~no}}, \bibinfo {author} {\bibfnamefont {A.~W.}\ \bibnamefont {Jackura}}, \bibinfo {author} {\bibfnamefont {D.~A.}\ \bibnamefont {Pefkou}}, \ and\ \bibinfo {author} {\bibfnamefont {F.}~\bibnamefont {Romero-L\'opez}},\ }\href {\doibase 10.1007/JHEP05(2024)279} {\bibfield  {journal} {\bibinfo  {journal} {JHEP}\ }\textbf {\bibinfo {volume} {05}},\ \bibinfo {pages} {279} (\bibinfo {year} {2024})},\ \Eprint {http://arxiv.org/abs/2402.12167} {arXiv:2402.12167 [hep-lat]} \BibitemShut {NoStop}%
\bibitem [{\citenamefont {Romero-L\'opez}\ \emph {et~al.}(2019)\citenamefont {Romero-L\'opez}, \citenamefont {Sharpe}, \citenamefont {Blanton}, \citenamefont {Brice\~no},\ and\ \citenamefont {Hansen}}]{Romero-Lopez:2019qrt}%
  \BibitemOpen
  \bibfield  {author} {\bibinfo {author} {\bibfnamefont {F.}~\bibnamefont {Romero-L\'opez}}, \bibinfo {author} {\bibfnamefont {S.~R.}\ \bibnamefont {Sharpe}}, \bibinfo {author} {\bibfnamefont {T.~D.}\ \bibnamefont {Blanton}}, \bibinfo {author} {\bibfnamefont {R.~A.}\ \bibnamefont {Brice\~no}}, \ and\ \bibinfo {author} {\bibfnamefont {M.~T.}\ \bibnamefont {Hansen}},\ }\href {\doibase 10.1007/JHEP10(2019)007} {\bibfield  {journal} {\bibinfo  {journal} {JHEP}\ }\textbf {\bibinfo {volume} {10}},\ \bibinfo {pages} {007} (\bibinfo {year} {2019})},\ \Eprint {http://arxiv.org/abs/1908.02411} {arXiv:1908.02411 [hep-lat]} \BibitemShut {NoStop}%
\bibitem [{\citenamefont {Blanton}\ and\ \citenamefont {Sharpe}(2020{\natexlab{b}})}]{Blanton:2020jnm}%
  \BibitemOpen
  \bibfield  {author} {\bibinfo {author} {\bibfnamefont {T.~D.}\ \bibnamefont {Blanton}}\ and\ \bibinfo {author} {\bibfnamefont {S.~R.}\ \bibnamefont {Sharpe}},\ }\href {\doibase 10.1103/PhysRevD.102.054515} {\bibfield  {journal} {\bibinfo  {journal} {Phys. Rev. D}\ }\textbf {\bibinfo {volume} {102}},\ \bibinfo {pages} {054515} (\bibinfo {year} {2020}{\natexlab{b}})},\ \Eprint {http://arxiv.org/abs/2007.16190} {arXiv:2007.16190 [hep-lat]} \BibitemShut {NoStop}%
\bibitem [{\citenamefont {Varshalovich}\ \emph {et~al.}(1988)\citenamefont {Varshalovich}, \citenamefont {Moskalev},\ and\ \citenamefont {Khersonskii}}]{VMK}%
  \BibitemOpen
  \bibfield  {author} {\bibinfo {author} {\bibfnamefont {D.~A.}\ \bibnamefont {Varshalovich}}, \bibinfo {author} {\bibfnamefont {A.~N.}\ \bibnamefont {Moskalev}}, \ and\ \bibinfo {author} {\bibfnamefont {V.~K.}\ \bibnamefont {Khersonskii}},\ }\href@noop {} {\emph {\bibinfo {title} {Quantum Theory of Angular Momentum}}}\ (\bibinfo  {publisher} {World Scientific},\ \bibinfo {address} {Singapore},\ \bibinfo {year} {1988})\BibitemShut {NoStop}%
\bibitem [{\citenamefont {Peardon}\ \emph {et~al.}(2009)\citenamefont {Peardon}, \citenamefont {Bulava}, \citenamefont {Foley}, \citenamefont {Morningstar}, \citenamefont {Dudek}, \citenamefont {Edwards}, \citenamefont {Joo}, \citenamefont {Lin}, \citenamefont {Richards},\ and\ \citenamefont {Juge}}]{HadronSpectrum:2009krc}%
  \BibitemOpen
  \bibfield  {author} {\bibinfo {author} {\bibfnamefont {M.}~\bibnamefont {Peardon}}, \bibinfo {author} {\bibfnamefont {J.}~\bibnamefont {Bulava}}, \bibinfo {author} {\bibfnamefont {J.}~\bibnamefont {Foley}}, \bibinfo {author} {\bibfnamefont {C.}~\bibnamefont {Morningstar}}, \bibinfo {author} {\bibfnamefont {J.}~\bibnamefont {Dudek}}, \bibinfo {author} {\bibfnamefont {R.~G.}\ \bibnamefont {Edwards}}, \bibinfo {author} {\bibfnamefont {B.}~\bibnamefont {Joo}}, \bibinfo {author} {\bibfnamefont {H.-W.}\ \bibnamefont {Lin}}, \bibinfo {author} {\bibfnamefont {D.~G.}\ \bibnamefont {Richards}}, \ and\ \bibinfo {author} {\bibfnamefont {K.~J.}\ \bibnamefont {Juge}} (\bibinfo {collaboration} {Hadron Spectrum}),\ }\href {\doibase 10.1103/PhysRevD.80.054506} {\bibfield  {journal} {\bibinfo  {journal} {Phys. Rev. D}\ }\textbf {\bibinfo {volume} {80}},\ \bibinfo {pages} {054506} (\bibinfo {year} {2009})},\ \Eprint {http://arxiv.org/abs/0905.2160} {arXiv:0905.2160 [hep-lat]} \BibitemShut {NoStop}%
\bibitem [{\citenamefont {Rodas}\ \emph {et~al.}(2023)\citenamefont {Rodas}, \citenamefont {Dudek},\ and\ \citenamefont {Edwards}}]{Rodas:2023gma}%
  \BibitemOpen
  \bibfield  {author} {\bibinfo {author} {\bibfnamefont {A.}~\bibnamefont {Rodas}}, \bibinfo {author} {\bibfnamefont {J.~J.}\ \bibnamefont {Dudek}}, \ and\ \bibinfo {author} {\bibfnamefont {R.~G.}\ \bibnamefont {Edwards}} (\bibinfo {collaboration} {Hadron Spectrum}),\ }\href {\doibase 10.1103/PhysRevD.108.034513} {\bibfield  {journal} {\bibinfo  {journal} {Phys. Rev. D}\ }\textbf {\bibinfo {volume} {108}},\ \bibinfo {pages} {034513} (\bibinfo {year} {2023})},\ \Eprint {http://arxiv.org/abs/2303.10701} {arXiv:2303.10701 [hep-lat]} \BibitemShut {NoStop}%
\bibitem [{\citenamefont {Delves}\ and\ \citenamefont {Mohamed}(1985)}]{Delves_Mohamed_1985}%
  \BibitemOpen
  \bibfield  {author} {\bibinfo {author} {\bibfnamefont {L.~M.}\ \bibnamefont {Delves}}\ and\ \bibinfo {author} {\bibfnamefont {J.~L.}\ \bibnamefont {Mohamed}},\ }\href@noop {} {\emph {\bibinfo {title} {Computational Methods for Integral Equations}}}\ (\bibinfo  {publisher} {Cambridge University Press},\ \bibinfo {year} {1985})\BibitemShut {NoStop}%
\bibitem [{\citenamefont {Dawid}\ \emph {et~al.}(2023{\natexlab{b}})\citenamefont {Dawid}, \citenamefont {Islam}, \citenamefont {Brice\~no},\ and\ \citenamefont {Jackura}}]{Dawid:2023kxu}%
  \BibitemOpen
  \bibfield  {author} {\bibinfo {author} {\bibfnamefont {S.~M.}\ \bibnamefont {Dawid}}, \bibinfo {author} {\bibfnamefont {M.~H.~E.}\ \bibnamefont {Islam}}, \bibinfo {author} {\bibfnamefont {R.~A.}\ \bibnamefont {Brice\~no}}, \ and\ \bibinfo {author} {\bibfnamefont {A.~W.}\ \bibnamefont {Jackura}},\ }\href@noop {} {\  (\bibinfo {year} {2023}{\natexlab{b}})},\ \Eprint {http://arxiv.org/abs/2309.01732} {arXiv:2309.01732 [nucl-th]} \BibitemShut {NoStop}%
\bibitem [{\citenamefont {Dudek}\ \emph {et~al.}(2012)\citenamefont {Dudek}, \citenamefont {Edwards},\ and\ \citenamefont {Thomas}}]{Dudek:2012gj}%
  \BibitemOpen
  \bibfield  {author} {\bibinfo {author} {\bibfnamefont {J.~J.}\ \bibnamefont {Dudek}}, \bibinfo {author} {\bibfnamefont {R.~G.}\ \bibnamefont {Edwards}}, \ and\ \bibinfo {author} {\bibfnamefont {C.~E.}\ \bibnamefont {Thomas}},\ }\href {\doibase 10.1103/PhysRevD.86.034031} {\bibfield  {journal} {\bibinfo  {journal} {Phys. Rev.}\ }\textbf {\bibinfo {volume} {D86}},\ \bibinfo {pages} {034031} (\bibinfo {year} {2012})},\ \Eprint {http://arxiv.org/abs/1203.6041} {arXiv:1203.6041 [hep-ph]} \BibitemShut {NoStop}%
\bibitem [{\citenamefont {Beane}\ \emph {et~al.}(2012)\citenamefont {Beane}, \citenamefont {Chang}, \citenamefont {Detmold}, \citenamefont {Lin}, \citenamefont {Luu}, \citenamefont {Orginos}, \citenamefont {Parreno}, \citenamefont {Savage}, \citenamefont {Torok},\ and\ \citenamefont {Walker-Loud}}]{NPLQCD:2011htk}%
  \BibitemOpen
  \bibfield  {author} {\bibinfo {author} {\bibfnamefont {S.~R.}\ \bibnamefont {Beane}}, \bibinfo {author} {\bibfnamefont {E.}~\bibnamefont {Chang}}, \bibinfo {author} {\bibfnamefont {W.}~\bibnamefont {Detmold}}, \bibinfo {author} {\bibfnamefont {H.~W.}\ \bibnamefont {Lin}}, \bibinfo {author} {\bibfnamefont {T.~C.}\ \bibnamefont {Luu}}, \bibinfo {author} {\bibfnamefont {K.}~\bibnamefont {Orginos}}, \bibinfo {author} {\bibfnamefont {A.}~\bibnamefont {Parreno}}, \bibinfo {author} {\bibfnamefont {M.~J.}\ \bibnamefont {Savage}}, \bibinfo {author} {\bibfnamefont {A.}~\bibnamefont {Torok}}, \ and\ \bibinfo {author} {\bibfnamefont {A.}~\bibnamefont {Walker-Loud}} (\bibinfo {collaboration} {NPLQCD}),\ }\href {\doibase 10.1103/PhysRevD.85.034505} {\bibfield  {journal} {\bibinfo  {journal} {Phys. Rev. D}\ }\textbf {\bibinfo {volume} {85}},\ \bibinfo {pages} {034505} (\bibinfo {year} {2012})},\ \Eprint {http://arxiv.org/abs/1107.5023} {arXiv:1107.5023 [hep-lat]} \BibitemShut {NoStop}%
\bibitem [{\citenamefont {Jacob}\ and\ \citenamefont {Wick}(1959)}]{Jacob:1959at}%
  \BibitemOpen
  \bibfield  {author} {\bibinfo {author} {\bibfnamefont {M.}~\bibnamefont {Jacob}}\ and\ \bibinfo {author} {\bibfnamefont {G.~C.}\ \bibnamefont {Wick}},\ }\href {\doibase 10.1016/0003-4916(59)90051-X} {\bibfield  {journal} {\bibinfo  {journal} {Annals Phys.}\ }\textbf {\bibinfo {volume} {7}},\ \bibinfo {pages} {404} (\bibinfo {year} {1959})}\BibitemShut {NoStop}%
\end{thebibliography}%

\end{document}